\documentclass{article}
\usepackage{authblk}
\usepackage[square, numbers]{natbib}
\usepackage{amsmath}
\usepackage{graphicx}
\usepackage{color}
\usepackage{cases}
\usepackage{float}
\usepackage{amssymb}
\usepackage{subcaption}
\newcommand{\sbt}{\,\begin{picture}(-1,1)(-1,-3)\circle*{3}\end{picture}\;\,}
\newcommand{\trasp}[1]{{#1}^{\mathsf T}}
\newcommand{\vett}[1]{\boldsymbol{#1}}
\DeclareMathOperator{\Tr}{Tr}
\newcommand{\tens}[1]{\boldsymbol{\mathsf{#1}}}
\usepackage{mathtools}

\begin{document}

\title{\large \textbf{ANN-aided incremental multiscale-remodelling-based finite strain poroelasticity\vspace{1cm}}}

\author{ Hamidreza Dehghani \thanks{hamidreza.dehghani@uni.lu} }
\author{Andreas Zilian \vspace{1cm}}

\affil{\textit{\small Institute of Computational Engineering and Sciences, Department of Engineering, Faculty of Science, Technology and Medicine, University of Luxembourg}} 

\date{}
\maketitle

\begin{abstract}
Mechanical modelling of poroelastic media under finite strain is usually carried out via phenomenological models neglecting complex micro-macro scales interdependency. One reason is that the mathematical two-scale analysis is only straightforward assuming infinitesimal strain theory. Exploiting the potential of ANNs for fast and reliable upscaling and localisation procedures, we propose an incremental numerical approach that considers rearrangement of the cell properties based on its current deformation, which leads to the remodelling of the macroscopic model after each time increment. This computational framework is valid for finite strain and large deformation problems while it ensures infinitesimal strain increments within time steps. The full effects of the interdependency between the properties and response of macro and micro scales are considered for the first time providing more accurate predictive analysis of fluid-saturated porous media which is studied via a numerical consolidation example. Furthermore, the (nonlinear) deviation from Darcy's law is captured in fluid filtration numerical analyses. Finally, the brain tissue mechanical response under uniaxial cyclic test is simulated and studied.
\end{abstract}
\vspace{1cm}
\newpage
\section{Introduction}
Poroelasticity deals with the mechanical and hydraulic responses of the materials consisting of an elastic porous solid matrix interacting with viscous fluid percolating its pores. This field covers a wide range of applications from soil and rock mechanics to soft biological tissues such as the brain and cancerous ones \cite{Biot1941,FRANCESCHINI20062592}.
 There are different theories developed for poroelastic problems, including the phenomenological Biot's theory \cite{Biot1941} and microstructure-derived equations in \cite{burrigekeller}. The former considers only the homogenised form to be fitted into experiments while the latter is based on asymptotic homogenisation providing analytical relationships, in the form of PDEs to be solved in the cell domain, between the microscale and effective properties. The cell problems, then, can be solved using numerical tools such as Finite Element (FE) method as in \cite{PENTA201580, Hdehghani, HdehghaniThesis} which, later on, improved by means of Artificial Intelligence (AI) in \cite{HDAZ2020}.
 
One considerable limitation of using asymptotic homogenisation is that the upscaling procedure is only straightforward for infinitesimal deformations which restricts us to the choice of linear poroelasticity (in the sense of linear constitutive equations). On the other hand, the phenomenological approaches such as Biot's poroelastic theory \cite{Biot1941} and biphasic mixture theory \cite{BOWEN19801129} have been developed to predict the nonlinear poroelastic behaviour of soft tissues \cite{simon1983poroelastic, Oomens1985, Mow1986, Simon1991, Simon1992, alfio_grillo}. These theories, however, do not provide sufficient links between micro and macro scales, which considering the complexity and multiscale and multiphysics nature of poroelastic problems, is crucial for a realistic analysis of the scenarios of interest.
 The fluid flow in poroelastic media is usually described via linear Darcy's law \cite{Darcy1856}, which again is valid in the cases with infinitesimal fluid velocity and solid deformation.
 Nonlinear fluid flow in poroelastic media has been considered as another challenge which, so far, is addressed via cubic and quadratic corrections (in terms of averaged velocity or Reynolds number) to linear Darcy's law in a rigid porous media \cite{forchheimer1901, mei_auriault_1991, Wodie1991CorrectionNL, Firdaouss1997}. However, not only another parameter (apart from hydraulic conductivity) is to be assumed/determined for the correction term but also the solid deformation does not play any role in the fluid flow nonlinearity using such models.

In this study, we aim at overcoming the mentioned hurdles to achieve a more accurate model describing mechanical and hydraulic behaviour of poroelastic media by means of an incremental remodelling-based nonlinear method for numerical analysis of such complex problems. 
Incremental analysis of structures has long been used in the numerical analysis of "path-dependent" problems such as soil mechanics, plasticity, large (geometric) deformations, etc. \cite{Roger1958, Yarimci1966, Christian1977, YAGHMAI19711375, Sharifi1971NonlinearBA, David1969}. In such frameworks, three configurations are considered in the deformation history, namely, \emph{reference configuration} which is the original state at the beginning of the analysis, the current \emph{deformed configuration} at time $t + \Delta t$, and the \emph{intermediate configuration} at time $t$ just before the deformed configuration. In other words, the intermediate configuration is at the beginning of the time increment, which ends at the deformed configuration. The path from the former to the latter is usually linear, while the transformation from the reference configuration to the deformed configuration can be nonlinear. Although, in linear poroelastic problems, the constitutive equations are linear the nonlinearity of mechanical and hydraulic response arises from solid-fluid interaction. In fact, as a poroelastic problem is generally history/path-dependent, it is usually performed in an incremental numerical framework regardless of being linear \cite{Hdehghani1} or nonlinear \cite{Simon1991}, however, the full potential of such frameworks have not been exploited. In particular, this framework allows the rearrangement of material properties in each time increment (remodelling) without considerable loss of efficiency if the updated properties are available in real-time. A similar approach is exploited in poroplastic models in \cite{alfio_grillo} in which the structural reorganisation is assumed to obey a phenomenological flow rule driven by stress. In the field of heterogeneous media, the microstructural evolution is captured based on asymptotic homogenisation in \cite{RAMIREZTORRES2018245} which is limited to solving the cell problems at each time increment and spatial point to update the macroscopic properties imposing a high computational cost. A similar framework is also applicable in the field of poroelasticity. However, due to the inefficiency imposed by the upscaling and localisation, it is not feasible for real-world problems.  

The key to overcoming this problem could be found in Artificial Intelligence (AI). Data-driven modelling has been successfully applied in a wide range of problems from marketing \cite{Malhotra1999} to computational analysis of mechanics problems. The well known Artificial Neural Networks (ANNs) \cite{Rosenblatt58theperceptron} has been successfully applied in the field of computational mechanics by providing a powerful interpolation mean. This method is inspired by human brain architecture, acting as a transfer function by providing accurate outputs from given inputs. An ANN consists of some hidden layers each adopting a number of neurones (see, Figure \ref{fig_ANN_Parallel} for ANN architecture) which includes a weight and a bias to be tuned during an optimisation procedure called ANN training. The number of hidden layers and the neurones inside each are chosen based on the user's experience. Having a large number of neurones and layers could lead to inefficiency of training and output calculation (so-called feed-forward procedure) while having a small number of neurones can lead to a loss of accuracy. The training procedure, which determines the network's accuracy, is based on an optimisation procedure minimising a distance function from, in the simplest case, \emph{a priori} provided training dataset consisting of a number of \emph{exact} outputs with their corresponding inputs.
 In computational mechanics, ANNs are employed to complement standard approaches such as FE \cite{HDAZ2020} or, in specific cases, to carry out the whole computations \cite{Raissi2018, dehghani2020hybrid}.
  In poroelasticity, ANNs are employed for effective model parameter identification in order to efficiently solve complex problems with spatially dependent porosity and solid matrix properties by the present authors in \cite{HDAZ2020}. The training dataset in the latter study was acquired by solving a certain number of cell problems. We employ ANNs for fast computation of both effective model parameters and coefficients required for localisation procedure. 
  
Here, we integrate the mentioned techniques and develop an ANN-informed incremental computational methodology for numerical analysis of poroelastic media under finite deformation considering rearrangement of microstructural properties (porosity and material properties of the solid matrix) and their effects on the effective coefficients of homogenised system of PDEs. The following steps are taken for this purpose:
\begin{itemize}
\item First, a certain number of cell problems are solved in order to provide a training dataset with inputs being the microscopic properties and the outputs being the tensors required for effective properties calculation as well as localisation procedure.

\item A suitable ANN is designed and trained via the provided training dataset replacing the time-consuming FE analysis of the cell problems for the calculation of the effective properties of the medium. 

\item Then, the time domain is discretised, ensuring infinitesimal deformation in each time increment followed by the space discretisation of macroscale domain for FE analysis.

\item The homogenised system of PDEs obtained via asymptotic homogenisation multiscale analysis for poroelastic media \cite{burrigekeller} (linear poroelasticity) is employed for numerical analysis of macroscopic response in one increment. 

\item With the help of ANN, this is followed by localisation analysis which calculates the average microscale solid matrix deformation due to the macroscale mechanical and hydraulic response. 

\item The porosity and solid matrix material properties, using a specific strain energy function (here, neo-Hookean), are updated based on the current microscopic deformation (remodelling). 

\item The ANN, again, is employed to carry out upscaling for rearranged effective coefficients identification based on the current (updated) microscopic properties. 

\item The new properties of macroscale system of PDEs are used for the numerical analysis of the next time increment starting from the fourth step.
 \end{itemize}
  
 The governing equations of the mentioned framework are presented in details in Section \ref{sec:2}. In Section \ref{Sec_numericalExamples}, the described methodology is employed, first, for the analysis of a simple poroelastic problem in order to study the micro-macro interdependency of deformation and properties highlighting the feasibility and importance of the present methodology. Then, Darcy's experiment is simulated showing that employing this framework, there is no need for any correction to Darcy's law as the nonlinear fluid flow is captured automatically by updating the hydraulic conductivity. Last but not least, a cyclic uniaxial test, similar to the experiments on the brain tissue in \cite{FRANCESCHINI20062592}, is carried out showing the viability of the simulation of complex real-world problems by the present computational framework. The promising results of the numerical examples open several directions for the future works, which together with discussion and conclusions are provided in Section \ref{Sec_conclusions}. Furthermore, a part of the previously developed homogenisation procedure is placed in Appendix \ref{Homo_app}.

\section{Governing equations}
\label{sec:2}
Let us consider an arbitrary poroelastic domain $\Omega \in \mathbb R^3$ which consists of the solid matrix $\Omega_s$ and interstitial fluid $\Omega_f$ such that $\Omega = \Omega_s \cup \Omega_f$.
We assume that the poroelastic medium has the average pore size $d$ which is infinitesimal compared to the medium size $L$ (e.g. as shown in Figure \ref{fig:lenghtscale}). As the exploited analytical homogenisation and localisation are valid for infinitesimal strain, the time discretisation is the first matter to consider, clarifying the applicability of the multiscale methodology. 

The time-dependent state $\vett U_t$ (at time $t$) of the problem is advanced to the time instant $t+\Delta t$ as follows:
In one increment, $\vett U_{t+\Delta t}$ is calculated from the intermediate configuration (configuration at time $t$), the time increment $\Delta t$, and solution increment $\Delta \vett U$. As the intermediate configuration and $\Delta t$ are fixed, the problem decreases to finding $\Delta \vett U$ and consequently $\vett U_{t+\Delta t} $ via
\begin{equation}
\vett U_{t+\Delta t} = \vett U_t + \Delta \vett U \label{timedisc}
\end{equation}

At this stage, for the sake of simplicity, we drop $\Delta$ notation, however, we emphasise that the following equations and mechanical relationships are considered to be within one increment (i.e. from time $t$ to $t + \Delta t$), thus, applying the infinitesimal deformation assumptions. 

 Considering the mentioned condition, in one increment, the solid compartment is linear elastically interacting with incompressible Newtonian fluid percolating its pores with no-slip boundary condition on the solid-fluid interface. The corresponding equations are
\begin{align}
          \nabla \cdot \tens{\tau}&=0 &\textrm{in}\,\,\Omega_s\label{eq:solid} \\
          \tens{\tau}&=\mathbb{C} \xi(\vett u) &\textrm{in}\,\,\Omega_s \label{eq:solidstress} \\ \nonumber \\ \nonumber \\
\label{eq:fluido}
     \nabla \cdot \tens{\sigma}&=0  &\textrm{in}\,\,\Omega_f\\
     \tens{\sigma}&=-p\tens{I}+2\mu \xi(\vett{v})  &\textrm{in}\,\,\Omega_f \label{fluidstress}\\
     \nabla \cdot \vett{v}&=0 &\textrm{in}\,\,\Omega_f \\ \nonumber \\ \nonumber \\
\label{eq:stress}
    \tens{\tau}\vett{n}&=\tens{\sigma}\vett{n} &\textrm{on}\,\,\Gamma\,\,\\
    \dot{\vett{u}}&=\vett{v} &\textrm{on}\,\,\Gamma\,\,  \label{eq:tangente}
\end{align}
for solid, fluid, and interface conditions, respectively, where
$\vett u$ indicates the displacement of solid phase, $\vett{v}$ represents the fluid velocity, $\mathbb{C}$ is the fourth rank solid matrix elasticity tensor, and $\mu$ is the interstitial fluid dynamic viscosity. $\tau$ and $\sigma$ are the solid Cauchy and fluid viscous stress tensors, respectively.
Furthermore, the symmetric gradient operator $\xi$ is defined as
\begin{equation}
\label{eq:symgrad21}
\xi{(\sbt)} = \frac{1}{2}\left[ \nabla(\sbt)+[\nabla(\sbt)]^{\tens T}\right].
\end{equation}


\subsection{Non-dimensionalisation}
Poroelasticity has a wide range of applications from soil and rock mechanics to different biological tissues (such as brain tissue). It also has a complex interdependency between the units of measurement due to its multiscale and multiphysics nature which is a significant source of ambiguity leading to a misunderstanding of the model parameters and response \cite{Firdaouss1997}. In this study, before we proceed further, we take advantage of Non-dimensionalisation procedure which is similar to the one in \cite{growing}. For each example, we adopt four formally independent characteristic values for macroscale length $L$, average microscale cell dimension $d$, fluid viscosity $\mu_c$, and force $f_c$ such that
\begin{equation}
 \vett{x}=L\vett{x}',\quad \vett y = d \vett y', \quad \vett F = f_c \vett F' \quad  \mu=\mu_c \mu', \label{nondim2}
 \end{equation}
where $\vett x$ and $\vett y$ are macroscale and microscale spatial variables, $\vett F$ is (any) force, and $\mu$ is the interstitial fluid dynamic viscosity. All other parameters in this study are non-dimensionalised with respect to the mentioned independent characteristic values as
\begin{align}
\nonumber  &\vett u = L \vett u',\quad \vett{v}=\frac{f_cd^2}{L^3\mu_c}\vett{v}', \quad \mathbb{C}=\frac{f_c}{L^2 }\mathbb{C}', \quad t = \frac{L^4 \mu_c}{f_c d^2} t',\\ \nonumber & \quad \quad\quad  \vett{\tens K} = \frac{d^2}{\mu_c}\vett{\tens K}', \quad M = \frac{f_c}{L^2 } M', \quad \tilde{\mathbb{C}}=\frac{f_c}{L^2 }\tilde{\mathbb{C}'}
\end{align}
where $\sbt'$ indicates the non-dimensional parameter. $\vett{\tens K}$, $M$, and $\tilde{\mathbb C}$ are the hydraulic conductivity, Biot modulus, and effective (drained) elasticity tensor which are to be defined in Section \ref{physical}.

\subsection{Homogenisation} \label{physical}
In this section, we outline the mathematical two-scale homogenisation based on non-dimensionalised parameters which is developed and employed in several studies in the literature \cite{burrigekeller, HdehghaniThesis, growing}. For the reader's convenience and as we will refer to the notations later on, we provide more details of this procedure in Appendix \ref{Homo_app}.

Removing the prime symbol (indicating non-dimens- ionalised variables), for the sake of simplicity, all the Equations \eqref{eq:solid}-\eqref{eq:tangente} will be written the same except for Equation \eqref{fluidstress} which adopts the form 
\begin{equation}
\tens{\sigma}=-p\tens{I}+\epsilon^22\mu \xi(\vett{v}) \quad \quad\quad\quad\,\,\,\textrm{in}\,\,\Omega_f \label{Nonfluidstress}
\end{equation}
where $
\epsilon=d/L \ll 1
$ is the length scale separation parameter. The coefficient $\epsilon^2$ appears due to the scaling of Stokes flow in porous media (see e.g. \cite{burrigekeller, growing}).

An infinitesimal length scale separation parameter $\epsilon$
 allows to make use of asymptotic homogenisation technique. We assume that $\vett{x}$ and $\vett{y}$  ($ 
\vett{y}=\vett{x}/\epsilon
$) indicate two formally independent macroscale and microscale spatial variables, respectively.
\begin{figure*}
\centering
\includegraphics[width=12cm]{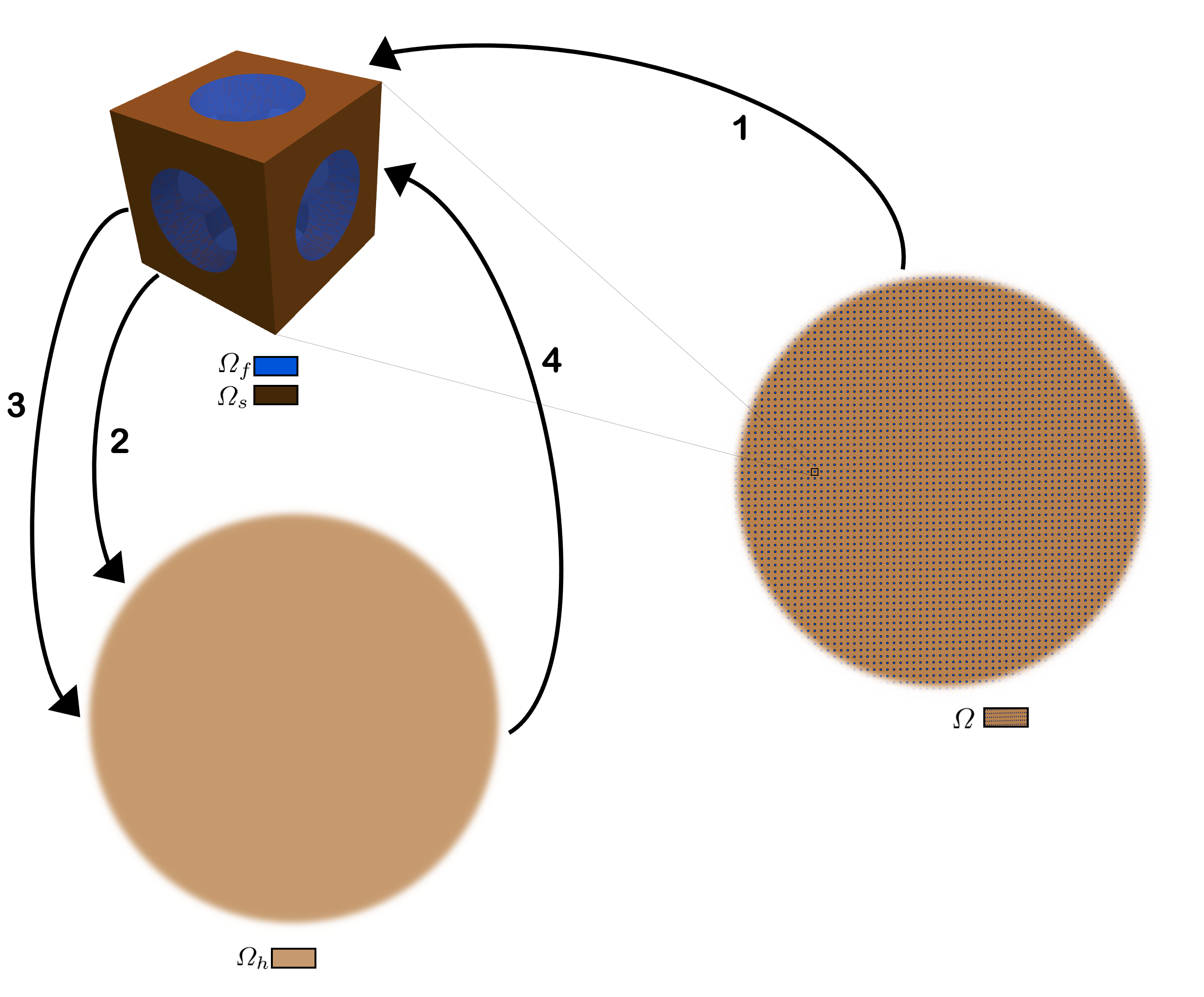}
\label{fig:lenghtscale}
\centering
\caption{A periodic multiphysics cell is assumed based on the initial peoperties of the poroelastic medium via Path 1.
The governing equations of homogenised medium (in $\Omega_h$) are constructed (Path 2) and characterised (via auxiliary cell problems considered as Path 3) based on the assumed cell properties (upscaling or homogenisation). Based on the response of the homogenised model and identified parameters in Path 3, the deformation of solid matrix ($\Omega_s$) is determined via localisation (Path 4). The fluid volume fraction (porosity) and solid material properties are modified via solid phase kinematics and (arbitrary) material model. Note that Path 1 and Path 2 are only considered once for each problem (which are equivalent for a wide range of problems) but the loop, Path 3--macroscale analysis--Path 4--cell remodelling, is performed at each time increment. The ANN is employed in Path 3 and Path 4.
}
\label{fig:lenghtscale}
\end{figure*}
Then, the differential operators transform into two independent ones for macroscale and microscale as
\begin{equation}
\nabla \rightarrow \nabla_{\vett{x}}+\frac{1}{\epsilon} \nabla_{\vett{y}} \label{eq:gradnew}
\end{equation}
and, similarly,
\begin{equation}
\xi \rightarrow \xi_{\vett{x}}+\frac{1}{\epsilon} \xi_{\vett{y}} \label{eq:xiTransform}
\end{equation}
which separates the problem into two-scales.

Applying the transformation \eqref{eq:gradnew} and \eqref{eq:xiTransform} into Equations \eqref{eq:solid}-\eqref{eq:tangente} yields the multiscale equations (Equations \eqref{eq:hstruttura}-\eqref{eq_dispinterface} in Appendix \ref{Homo_app}).

Equating the coefficients of $\epsilon^0$ and $\epsilon^1$, by replacing every field (stress, displacements etc.) by its power of series representation, $\psi_{\epsilon}(\vett{x},\vett{y})=\sum_{l=0}^{\infty}\psi^{(l)}(\vett{x},\vett{y})\epsilon^ l$, yields zero-th and first order equations (i.e. by choosing $l=0,1$) (Equations \eqref{eq:zstruttura}-\eqref{eq:utangente} in Appendix \ref{Homo_app})  which are the principal ones of the upscaling process based on which the leading order solid displacement $\vett{u}^{(0)}$ and hydrostatic pressure $p^{(0)}$ are microscale independent (locally constant) and could be referred to as macroscale variables. The zero-th (leading) and first order variables are shown by $\sbt^{(0)}$ and $\sbt^{(1)}$, respectively.

As the relative fluid velocity and the solutions of the cell problems (will be defined later) are not constant at the microscale, we apply the integral average
\begin{equation}
\label{eq:intavg}
         \left\langle \psi \right\rangle_k= 
         \frac{1}{|\Omega|}\int_{\Omega_k}\psi(\vett{x},\vett{y})\,\textrm{d}\vett{y} \quad k=f,s,
\end{equation}
 to exploit the acquired equations in the homogenisation process and macroscale.
Assuming $\vett{y}$-periodicity, no-growth limit \cite{growing}, and the following ansatzes
\begin{align}
\vett{u}^{(1)}&=\mathcal{A}\xi_{\vett{x}}(\vett u^{(0)}) +\vett{a}p^{(0)} \label{ansatzsolid}\\
\vett{v}_{rf}&=-\vett{\tens{W}}\nabla_{\vett{x}}p^{(0)},\label{eq:vzero}\\
p^{(1)}&=-\vett{P}\cdot\nabla_{\vett{x}}p^{(0)},\label{eq:puno}
\end{align}

the system of PDEs governing the macroscale problems (in homogenised domain $\Omega_h$) in a time increment is obtained as
\begin{align}
\displaystyle \nabla_{\vett{x}}\cdot \boldsymbol{\boldsymbol{\tens{\tau}}}_E&=\boldsymbol 0, &\textrm{in}\,\,\Omega_h   \label{macroequi}\\
\boldsymbol{\boldsymbol{\tens{\tau}}}_E &\coloneqq \mathbb{\tilde{C}}\, {:}\, \vett \varepsilon^{(0)}-\tens{\tilde{\vett{\alpha}}}p^{(0)} &\textrm{in}\,\,\Omega_h \label{solidCons}  \\
\dot{p}^{(0)}&=-M[\,\tens{\tilde{\vett{\alpha}}}\, {:}\, \dot{\vett \varepsilon}^{(0)} + \nabla_{\vett{x}}\cdot  \vett{v}_{rf}]  &\textrm{in}\,\,\Omega_h\label{eqMassC} \\ 
\vett{v}_{rf}&=-\boldsymbol{\tens{K}} \nabla_{\vett{x}}p^{(0)}.  &\textrm{in}\,\,\Omega_h \label{eqDarcy}
\intertext{where,}
\boldsymbol{\boldsymbol{\tens{\tau}}}_E &= \langle \boldsymbol{\boldsymbol{\tens{\tau}}}^{(0)} \rangle - \phi p^{(0)} \boldsymbol{\boldsymbol{\tens{I}}}\\
 \vett \varepsilon^{(0)}&= \xi_{\vett{x}}(\vett u^{(0)})\\
 \vett{v}_{rf} &= \left\langle \vett{v}^{(0)}\right\rangle_f-\phi\dot{\vett{u}}^{(0)} \label{eq_vfr}
 \end{align}
indicate effective stress, macroscopic strain tensor,  and relative fluid velocity.

This system of PDEs is characterised by some coefficients, namely, effective elasticity tensor, Biot coefficient and modulus, and hydraulic conductivity, that are defined, respectively, as
\begin{align}
\tilde{\mathbb{C}}\coloneqq &\left\langle \mathbb{C}+\mathbb{C}\mathbb{M}\right \rangle_s, \label{eq_TildeC} \\
 \tens{\tilde{\vett{\alpha}}}\coloneqq &\phi \boldsymbol{\tens{I}} - \Tr{\left\langle \mathbb{M}\right\rangle_s}, \\ 
 M\coloneqq & -\frac{1}{\left\langle \Tr{\boldsymbol{\tens{Q}} }\right\rangle_s}, \\
  \boldsymbol{\tens{K}} \coloneqq &\left\langle \boldsymbol{\tens{W} } \right\rangle_f, \label{eq_HydraulicConductivity}
\end{align}
 where, $\mathbb{C}$ and $\phi$ indicate the elasticity tensor of solid matrix and porosity (i.e. volume fraction of fluid phase), respectively.

The fourth rank tensor $\mathbb{M}$ and the second rank tensors $\boldsymbol{\tens{Q}} $ and $\boldsymbol{\tens{W}} $ are the solutions of three systems of PDEs resulted from the multiscale procedure (for more details regarding the auxiliary cell problems please see Appendix \ref{Homo_app} and \cite{Hdehghani}).

\paragraph{Remark.} \label{Qremark1}
Although the solution of the auxiliary cell problems are obtained by solving a linear elastic type problem via FEM, considering Equation \eqref{eq:unoutside21}, the second rank tensor $\xi_{\vett{y}} (\vett{a})$ has the dimension $\left[\frac{L^2}{f_c}\right]$ meaning that Biot modulus $M$ has the same dimension as $\mathbb{C}$ which is correct according to Equation \eqref{eqMassC}. We highlight that, consequently, the vector $\vett a$ has the characteristic dimension $\left[\frac{dL^2}{f_c}\right]$ which agrees with the solid ansatz \eqref{ansatzsolid}.

\subsection{Localisation}
In this section, based on the asymptotic homogenisation, we develop an analytical localisation and properties rearrangement procedure in order to obtain the microscopic deformation due to the macroscale mechanical and hydraulic response.
The Ansatz \eqref{ansatzsolid}-\eqref{eq:puno}, in fact, are the bridge between microscale and macroscale. As we mentioned earlier (and from Equation \eqref{eq:zsolidstress} in Appendix \ref{Homo_app}) $\vett{u}^{(0)}$ is locally-constant meaning that it can only cause rigid body motions which does not affect the microscale deformation gradient tensor. The same condition holds for $p^{(0)}$ (see Equations \eqref{eq:zfluido} and \eqref{eq:zfluidstress} in Appendix \ref{Homo_app}).

Taking the microscopic spatial gradient of Equation \eqref{ansatzsolid} reads
\begin{equation}
\xi_{\vett{y}}(\vett{u}^{(1)}) = \xi_{\vett{y}}(\mathcal{A} \vett \varepsilon^{(0)}) + \xi_{\vett{y}}(\vett{a}p^{(0)})
\end{equation}
which, taking into account that $\varepsilon^{(0)}$, and $p^{(0)}$ are microscale spatially independent (locally constant), reduces to
\begin{align}
\xi_{\vett{y}}(\vett{u}^{(1)}) = \xi_{\vett{y}}(\mathcal{A}) \vett \varepsilon^{(0)} + \xi_{\vett{y}}(\vett{a})p^{(0)} \label{eq_micro_disp}
\end{align}
Thus, taking the integral average from both sides yields
\begin{equation}
 \langle\vett{\varepsilon}^{(1)} \rangle=  \langle\mathbb{M}\rangle \vett{\varepsilon}^{(0)} +  \langle\boldsymbol{\tens{Q}}\rangle p^{(0)} \label{eqMicroStrain}
\end{equation}
where, $ \langle\vett{\varepsilon}^{(1)}\rangle =  \langle\xi_{\vett{y}} \vett{u}^{(1)}\rangle$ is exploited to update solid matrix Young's modulus and Poisson's ratio, as well as porosity. 


The coefficients of macroscale system of PDEs are computed based on the averaged values of stain/velocity components over their corresponding phase in the cell problem. Taking the integral average of Equation \eqref{eq:puno} we can write
\begin{equation}
\left\langle p^{(1)} \right\rangle_f = - \left\langle \vett{P} \right\rangle_f \cdot \phi \nabla_{\vett{x}}p^{(0)},
\end{equation}
which, considering the uniqueness condition \eqref{equnique}, results in $ \left\langle p^{(1)} \right\rangle_f = 0$ and highlights that $p^{(1)}$ does not play a role in material properties update as they are considered in an average sense. Consequently, $\langle\vett{\varepsilon}^{(1)} \rangle$ is the only parameter that plays a role in this procedure so we call it \emph{microscopic strain tensor}.

\subsubsection{Microscopic properties update}

At this stage, we flash back to the increment notation $\Delta$ (so $\langle\vett{\varepsilon}^{(1)} \rangle$ becomes $\langle\Delta\vett{ \varepsilon}^{(1)} \rangle$) for clarification in updating the stiffness/tangent tensor of the solid phase based on the current strain tensor $\langle\vett{\varepsilon}^{(1)}_{t+\Delta t} \rangle$. The latter can be calculated via
\begin{equation}
\langle\vett{\varepsilon}^{(1)}_{t+\Delta t} \rangle = \langle\vett{\varepsilon}^{(1)}_{t} \rangle + \langle\Delta \vett{\varepsilon}^{(1)} \rangle.
\end{equation}
We make use of the Lagrangian-based formulation to update the solid matrix mechanical properties. For the sake of clarification, we highlight that this formulation is only used for determining the updated stiffness tensor and can be replaced by other methods.

Assuming that the pore section remains circular (so one can consider that the average microscopic displacement gradient tensor is symmetric), the deformation gradient tensor $\boldsymbol{\tens F}$ and Green strain tensor $\boldsymbol{\tens E}$ can be calculated via
\begin{align}
\boldsymbol{\tens F} &=  \langle\vett{\varepsilon}^{(1)}_{t+\Delta t} \rangle/(1-\phi) + \boldsymbol{\tens I} \\
\boldsymbol{\tens E} &= \frac{1}{2} \left(\boldsymbol{\tens F}^T \cdot \boldsymbol{\tens F} - \boldsymbol{\tens I} \right),
\end{align}
where $\boldsymbol{\tens I}$ is the second rank identity tensor. We highlight that $ \langle\vett{\varepsilon}^{(1)} \rangle$ is the integral average over the entire cell which could be transformed to the integral average over the solid matrix by using the coefficient $\frac{|\Omega|}{|\Omega_s|}$ which is equal to $\frac{1}{1-\phi}$ in the case of the assumed unit cube cell.

As the fluid part is assumed incompressible, then $\phi$, $V_s^i$, and $\phi_i$ are related by
\begin{align}
\phi &= 1 - (V_s^iJ) \label{phiupdate}
\intertext{and}
\phi &= 1- (1- \phi_i)J \label{phiupdate1},
\end{align}
where, $J \coloneqq  \det(\boldsymbol{\tens F})$ is the total solid matrix volume change at the point. 
 $V_s^i$ and $\phi_i$ are, respectively, the solid and fluid phase volume fractions in the reference configuration. 

In this study, we consider isotropic compressible neo-Hookean strain energy function, which,
can be written as \cite{bonet_wood_2008}
\begin{equation}
W(I_1, J) = C_{10}(I_1 - 3) - 2C_{10} \ln(J) + \frac{1}{D_1} \ln(J)^2. \label{Neo_hookean}
\end{equation}
where $C_{10}$ and $D_1$ are material parameters. We determine the latter parameters based on initial Young's modulus $E_i$ and Poisson's ratio $\nu_i$ via
%

 \begin{align}
 C_{10} &= \frac{E_i}{4(1+\nu_i)} \label{eqC10},\\
 D_1&=\frac{6(1-2\nu_i)}{E_i} \label{eqD1}.
 \end{align}
 
The components of the deformation dependent tangent matrix for this material are given by
\begin{equation}
C_{ijkl}^t= \frac{\partial W}{\partial{E_{ij}} \partial{E_{kl}}}.
\label{eqTangent}
\end{equation}
The fourth rank tensor $C_{ijkl}^t$ (note that the superscript $t$, here, does not indicate the time) is the deformation dependent tangent matrix calculated based on $\langle\vett{\varepsilon}^{(1)}_{t+\Delta t} \rangle$. We take this tensor as the current stiffness tensor for the next time increment. In other words, for each increment, we use the rearranged  stiffness tensor based on the intermediate configuration.
Having sufficiently small strain increments one can assume $C_{ijkl}^t \approx C_{ijkl}$, where $C_{ijkl}^t = \frac{\partial S_{ij}}{\partial E_{kl}}$ and $C_{ijkl} = \frac{ \Delta \sigma_{ij}}{ \Delta \varepsilon_{kl}}$ for the next increment.
 
In this study, for the sake of simplicity, we assume that the solid matrix remains isotropic under the deformation so that its Young's modulus and Poisson's ratio can be updated by
\begin{align}
E&=\frac{{C}_{11}({C}_{11}+{C}_{12})-2{C}^2_{12}}{{C}_{11}+{C}_{12}}, \label{effectiveE} \\
\nu&=\frac{{C}_{12}}{{C}_{11}+{C}_{12}}. \label{effectiveNu}
\end{align}

\subsection{ANN application for localisation and homogenisation}
In order to obtain the coefficients of the system of PDEs governing the macroscale mechanical and hydraulic response (e.g. Biot modulus and coefficient), the tensors $\mathbb M$, $\boldsymbol{\tens{Q}}$, and $\boldsymbol{\tens{K}}$ are required which could be determined by solving the auxiliary cell problems (defined via Equations \eqref{equnique}-\eqref{6solid21} in Appendix \ref{Homo_app}). On the other hand, $\mathbb M$ and $\boldsymbol{\tens{Q}}$ are necessary to carry out the localisation to update the solid matrix elasticity tensor and the cell porosity. We notice that due to the inhomogeneous/une- qual strain and pore pressure distribution in the space and time the mentioned cell properties are spatially and temporally dependent imposing the need to solve the cell problems at each spatial point and time increment which is not feasible (or too expensive). 

In order to address above mentioned issue, we employ Artificial Neural Networks (ANNs) \cite{Rosenblatt58theperceptron} in a similar way that is used in a former study of the current authors \cite{HDAZ2020}. However, here, instead of computing the coefficients of macroscale system of PDEs, we make use of the capacity of ANNs to calculate $\mathbb M$, $\boldsymbol{\tens{Q}}$, and $\boldsymbol{\tens{K}}$ so that we carry out the localisation procedure using the same outputs which means increasing the efficiency of the framework. This network can be written as a micro-macro scale transfer function i.e.
\begin{equation}
(\mathbb M, \boldsymbol{\tens{Q}}, \boldsymbol{\tens{K}}) = \mathit{ANN}(\mathbb C, \mu, \phi)
\end{equation}
which, assuming solid matrix isotropy and geometric rotational invariance with respect to three orthogonal axes, reduces to \cite{Hdehghani1, HdehghaniThesis}
\begin{equation}
(M_{11},M_{12},M_{44}, Q_{11}, K_{11}) = \mathit{ANN}(E, \nu, \mu, \phi)
\end{equation}
Considering that throughout the homogenisation procedure the parameters $E$ and $\mu$ are non-dimensionalised and the problems are linear elastic solid and newtonian fluid types one can reduce the input size of the ANNs to
\begin{equation}
(M_{11},M_{12},M_{44}, Q_{11}, K_{11}) = \mathit{ANN}(\nu, \phi)
\end{equation}

At this point we highlight that, according to Equation \eqref{eqMassC} and Remark \ref{Qremark1}, if the nondimensional solid matrix Young's modulus $E$ is different from its initial value (which is the case when including material remodelling), as it is excluded from ANN inputs, one should consider its effects on $Q_{11}$ multiplying it by $\frac{E_i}{E}$.

\begin{figure*}
\centering
\includegraphics[width=12cm]{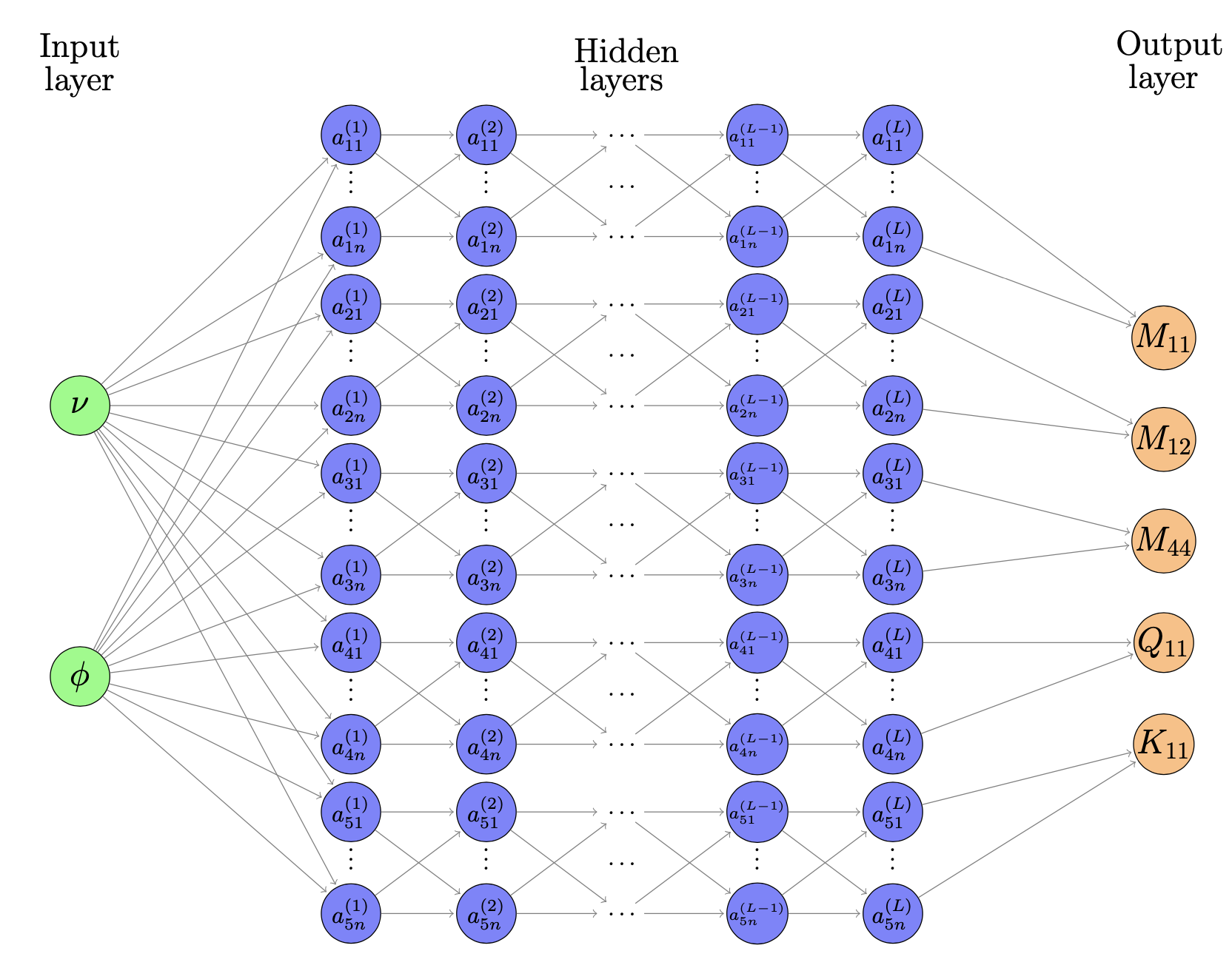}
\caption{Schematic representation of feed-forward procedure via Parallel ANNs for upscaling, as well as downscaling, providing real-time solution to auxiliary cell problems. Note that each parallel network can have distinct number of neurones $n$ and hidden layers $L$. In this study, respectively from upper network to the bottom, the parameters are $n=50$ $L = 3$, $n=50$ $L=3$, $n=20$ $L=3$, $n=50$ $L=3$, and $n=10$ $L=3$. We note that, in total, the network size is even smaller than the one in \cite{HDAZ2020} but, although the relationship is more complex here, the results are more accurate.}
\label{fig_ANN_Parallel}
\end{figure*}

In this study, we make use of five parallel networks each responsible for only one of the outputs in order to decrease the complexity while increase the accuracy of the training procedure. Figure \ref{fig_ANN_Parallel} schematically shows the configuration of this type of ANNs while, mathematically, the \emph{feed-forward} process which calculates the output based on the inputs through hidden layers can be expressed via
  \begin{align}
 z^{(i)}_k&=w^{(i)}_{kj}a^{(i-1)}_j+b^{(i)}_k \\
 a^{(i)}_k&=ReLU(z^{(i)}_k)
 \end{align}
where $a^{(i)}_k$ is the value of the $k$ -$th$ neurone in the $(i)$ -$th$ layer, $(i)\quad i \in(0,1,..,L)$ is the number of hidden layers, $k$ and $j$ indicate the number of the neurones in the $(i)$ -$th$ and $(i-1)$ -$th$ layers, respectively. The \emph{activation function} $ReLU(x) = \max (0,x)$ determines whether a neurone is active (adopts nonzero value).

At this stage, in order to achieve accurate outputs, we need to \emph{train} the networks which is viable, in the simplest case, by means of a training dataset consisting of sufficiently rich "exact" outputs each corresponding to distinct inputs. The later information, in this case, is provided by solving the auxiliary cell problems  via FE method. We solve the cell problems for 50 different porosities $\phi$ covering the interval from 0.082 to 0.783 with fine steps and for each porosity 50 different Poisson's ratio $\nu$ are considered so that the role of both variables are well reflected. 

In the training procedure, the ANNs are fed with the same inputs ($\phi$ and $\nu$) as the one in the FE problems (the target ones). Then, based on the later target, the $L^2$ error is computed and an optimiser (in this case Adam optimiser is chosen \cite{kingma2014adam}) to minimise the difference between the ANNs output and the target values (more details are available widely in the literature, e.g. \cite{HDAZ2020, OISHI2017327, KIRCHDOERFER201681}).

\subsection{Incremental nonlinear algorithm}
In strongly coupled solid-fluid problems such as the poroelastic models, fine discretisation of both space and time is required even assuming infinitesimal deformations. The latter is achieved by means of the incremental solution via Finite Element Method in which the total time of the problem is divided into sufficiently small increments $\Delta t$ so that accurate numerical temporal integration/differentiation can be achieved. Here, the simple linear time integration scheme is used. 
The time discretisation described via Equation \eqref{timedisc} can be rewritten for the macroscopic poroelastic case as 
\begin{align}
\vett u^{(0)}_{t+\Delta t} &= \vett u^{(0)}_t + \Delta \vett u^{(0)} \label{timediscu}\\
p^{(0)}_{t+\Delta t} &= p^{(0)}_{t} + \Delta p^{(0)} \label{timediscp}
\end{align}
In other words, the force vector and fluid relative velocity (which are history dependent) are calculated based on the intermediate configuration (configuration at time $t$), the time increment $\Delta t$, and displacement and pore pressure increment $\Delta \vett u^{(0)}$ and $\Delta p^{(0)}$, respectively. As the intermediate configuration and $\Delta t$ are fixed, the problem decreases to finding $\Delta \vett u^{(0)}$ and $\Delta p^{(0)}$ for which the infinitesimal deformation assumption is valid, thus, the employed mathematical homogenisation and localisation are applicable.

Consider a vector of DOF increments that must yield balance of linear momentum and the mass conservation provided, respectively, by Equations \eqref{macroequi} and \eqref{eqMassC} which can be approximated using the weak formulation. The latter can be obtained by integrating mentioned equations with respect to the corresponding volume and multiplying them with arbitrary test functions. In this case, two test functions $\delta\vett{u}^{(0)}$ and $\delta p^{(0)}$ which represent the variation of the macro\-scale displacement and pore pressure are required. Using Equations \eqref{timediscu} and \eqref{timediscp} and applying the divergence theorem the weak form of the governing equations to be solved in each increment is

\begin{align}
{0} =&  \int_{\partial\mathcal{B}} (\vett t_s^t + \Delta \vett t_s) \cdot \delta\vett{u}^{(0)} \, \mathrm{d}S
   - \int_{\mathcal{B}} (\boldsymbol{\tens{\tau}}_E^t + \Delta \boldsymbol{\tens{\tau}}_E) {:} \nabla_{\vett{x}} \, \delta\vett{u}^{(0)}\, \mathrm{d}V \nonumber \\
   &+ \int_{\mathcal{B}} \frac{1}{M_t} \, (\dot p^{(0)}_t + \Delta \dot p^{(0)}) \, \delta p^{(0)} \, \mathrm{d} V 
   + \int_{\partial\mathcal{ B}} (\vett{v}_{rf}^t + \Delta \vett{v}_{rf}) \cdot \vett n \, \delta p^{(0)} \, \mathrm{d}S  \nonumber \\
	 &- \int_{\mathcal{B}}  (\vett{v}_{rf}^t + \Delta \vett{v}_{rf}) \cdot  \nabla_{\vett{x}}  \, \delta p^{(0)} \, \mathrm{d} V + \int_{\mathcal{B}} \tilde{\boldsymbol{\alpha}}: (\dot{\vett \varepsilon}_t + \Delta \dot{\vett \varepsilon}) \, \delta p^{(0)} \, \mathrm{d} V \nonumber\\
	 &\quad \forall\;\delta\vett{u}^{(0)},\,\delta p^{(0)}.  \label{eq:weak_continuity_Delta}
\end{align}

which $\sbt^t$ or $\sbt_t$ show the situation of $\sbt$ at the intermediate configuration (at time $t$).
We highlight that the first terms inside each integral are fixed and known which form the residuals of the intermediate configuration and could be factorised. However, in this study, in order to avoid the accumulation of the residuals (numerical errors), we consider them in the formulation and implementation \cite{HOFMEISTER}.

Then substituting the constitutive equations \eqref{solidCons} and \eqref{eqDarcy}  into \eqref{eq:weak_continuity_Delta} yields
\begin{align}
{0} =&  \int_{\partial\mathcal{B}} (\vett t_s^t + \Delta \vett t_s) \cdot \delta\vett{u}^{(0)} \, \mathrm{d}S - \int_{\mathcal{B}} \left(\boldsymbol{\tens{\tau}}_E^t + (\mathbb{\tilde{C}}_t\, {:}\, \Delta \vett \varepsilon^{(0)}-\tens{\tilde{\vett{\alpha}}}_t \Delta p^{(0)})\right) {:} \nabla_{\vett{x}} \, \delta\vett{u}^{(0)}\, \mathrm{d}V \nonumber \\
   &+ \int_{\mathcal{B}} \left(\frac{1}{M_{t-\Delta t}} \dot p^{(0)}_t + \frac{1}{M_t} \Delta \dot p^{(0)}\right) \, \delta p^{(0)} \, \mathrm{d} V + \int_{\partial\mathcal{ B}} (\vett{v}_{rf}^t -\boldsymbol{\tens{K}}_t \Delta \nabla_{\vett{x}}p^{(0)}) \cdot \vett n \, \delta p^{(0)} \, \mathrm{d}S  \nonumber \\
	 &- \int_{\mathcal{B}}  (\vett{v}_{rf}^t - \boldsymbol{\tens{K}}_t \Delta \nabla_{\vett{x}}p^{(0)}) \cdot  \nabla_{\vett{x}}  \, \delta p^{(0)} \, \mathrm{d} V + \int_{\mathcal{B}} \left(\tilde{\boldsymbol{\alpha}}_{t-\Delta t}: \dot{\vett \varepsilon}_t + \tilde{\boldsymbol{\alpha}}_t: \Delta \dot{\vett \varepsilon}\right) \, \delta p^{(0)} \, \mathrm{d} V \nonumber\\
	 &\quad \forall\;\delta\vett{u}^{(0)},\,\delta p^{(0)}.  \label{eq:weak_continuity_Delta_1}
\end{align}
The coefficients $\mathbb{\tilde{C}}_t$, $\tens{\tilde{\vett{\alpha}}}_t$, $M_t$, and $\boldsymbol{\tens{K}}_t$ are the updated ones based on the updated microscopic material and geometrical parameters on the intermediate configuration. 

As in \cite{HDAZ2020}, free drainage can be introduced on a surface by enforcing 
\begin{equation}
\vett{v}_{rf} = \boldsymbol{\tens{K}} \frac{\Delta p^{(0)}}{\Delta x} \quad \quad \textrm{on}\quad \quad S_{fd}
\end{equation}
as a part of the fourth term on the right hand side of Equation \eqref{eq:weak_continuity_Delta_1}.
Here, $\Delta p$ and $\Delta x$ are the differences in pressure and distance between the surface of free drainage ($S_{fd}$) and the environment. If this condition is not introduced, an impermeable boundary condition is imposed.

\subsection{Implementation verification}
Due to the complexity of the methodology, the verification procedure is not a straight forward task. In this case, we divide the framework into simpler cases and verify them step by step. 
\begin{itemize}
\item The upscaling at the beginning of each iteration is carried out via ANN which is verified by producing random test dataset and observing high accuracy shown in Figure \ref{fig_ANNtest} and as is done in \cite{HDAZ2020}.
  \begin{figure*}
  \centering
  \begin{subfigure}{5.8cm}
  \includegraphics[width = 5.8cm]{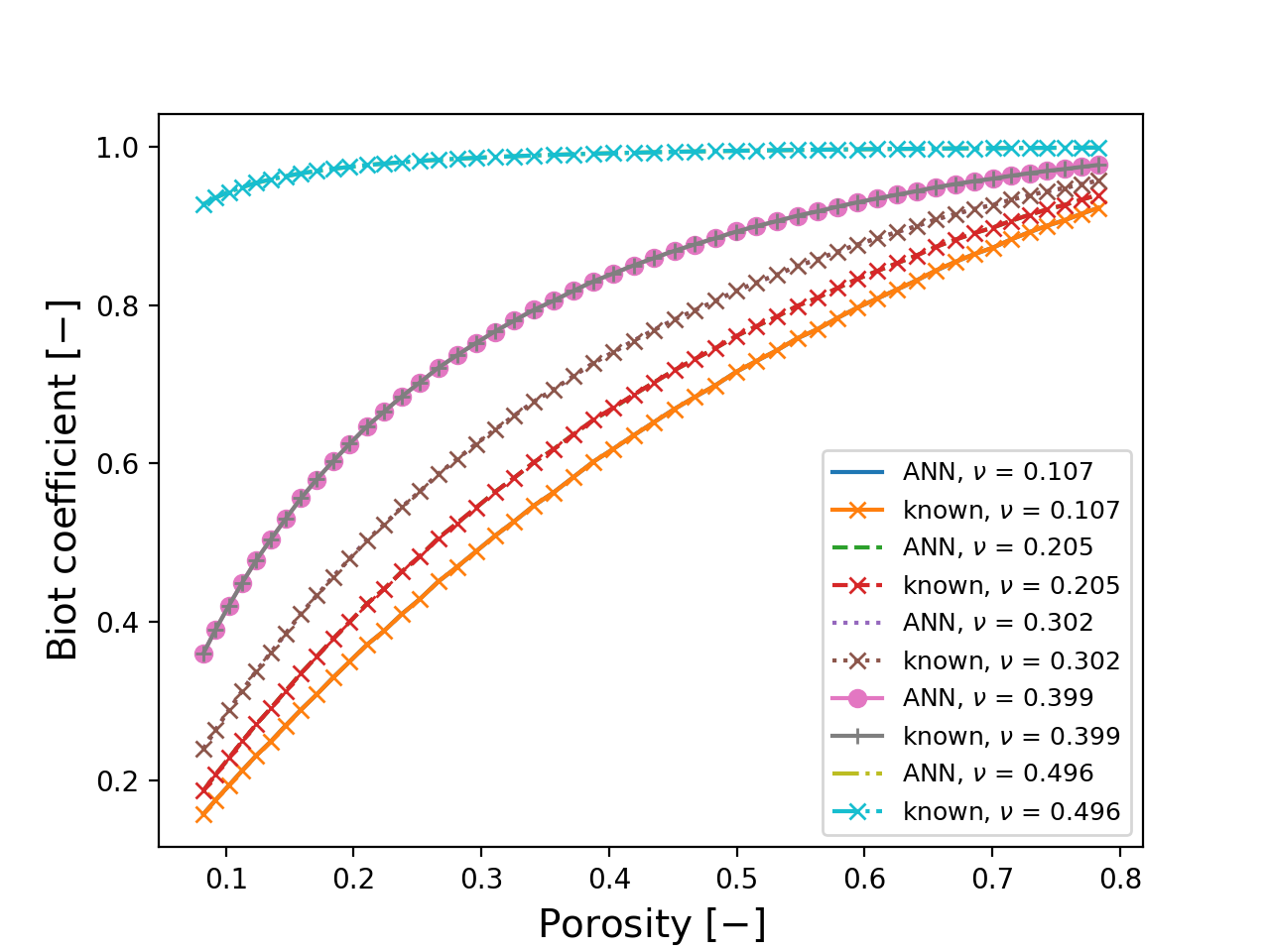}
  \caption{The trained ANN predicts Biot coefficients corresponding to a wide range of porosities and solid matrix Poisson's ratio.\vspace{0.38cm}}
  \label{fig_ANNalpha}
  \end{subfigure} \quad
  \begin{subfigure}{5.8cm}
  \includegraphics[width = 5.8cm]{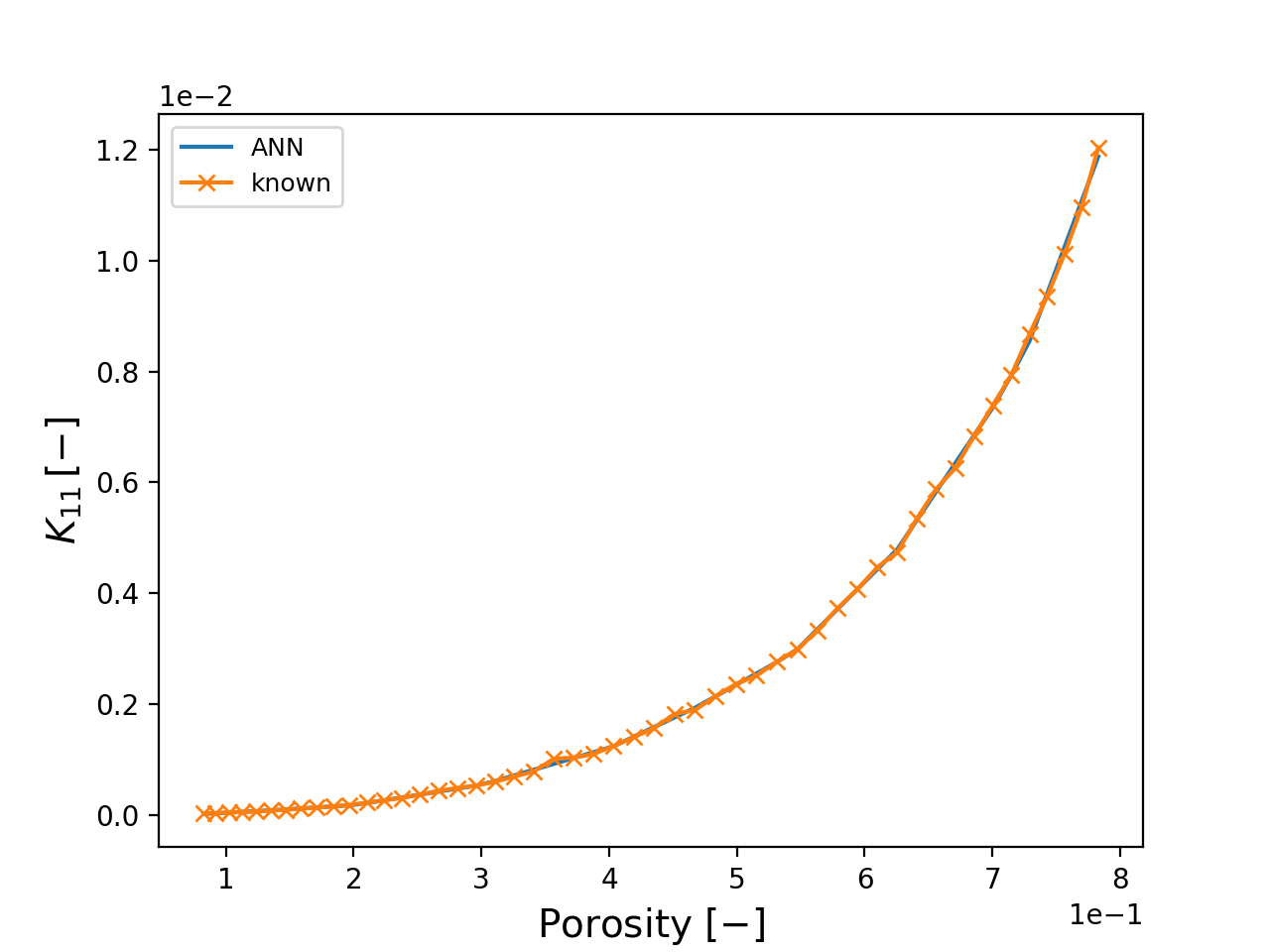}
  \caption{The hydraulic conductivity at different porosities, which is a source of nonlinearity in our methodology, can be approximated accurately in real-time via the trained ANN.}
  \label{fig_ANNK11}
  \end{subfigure}
  \caption{The trained ANN is tested, as usual, by means of a test dataset. The latter consists of the known results of the cell problems with properties different from the training dataset. Figures \ref{fig_ANNalpha} and \ref{fig_ANNK11} are two examples of this procedure.}
  \label{fig_ANNtest}
  \end{figure*}
\item At each iteration a linear consolidation problem is solved. This part is verified agreeing Terzaghi's analytical solution for a column of poroelastic material as in \cite{Hdehghani1, HdehghaniThesis}.
\item In order to verify the general incremental nonlinear analysis based on the specified strain energy function we perform a uniaxial test a linear elastic model, a typical nonlinear model, and the remodelling based incremental nonlinear analysis. The results are provided in Figure \ref{fig_Hyper_test}.
\begin{figure}
 \centering
\includegraphics[width=8cm]{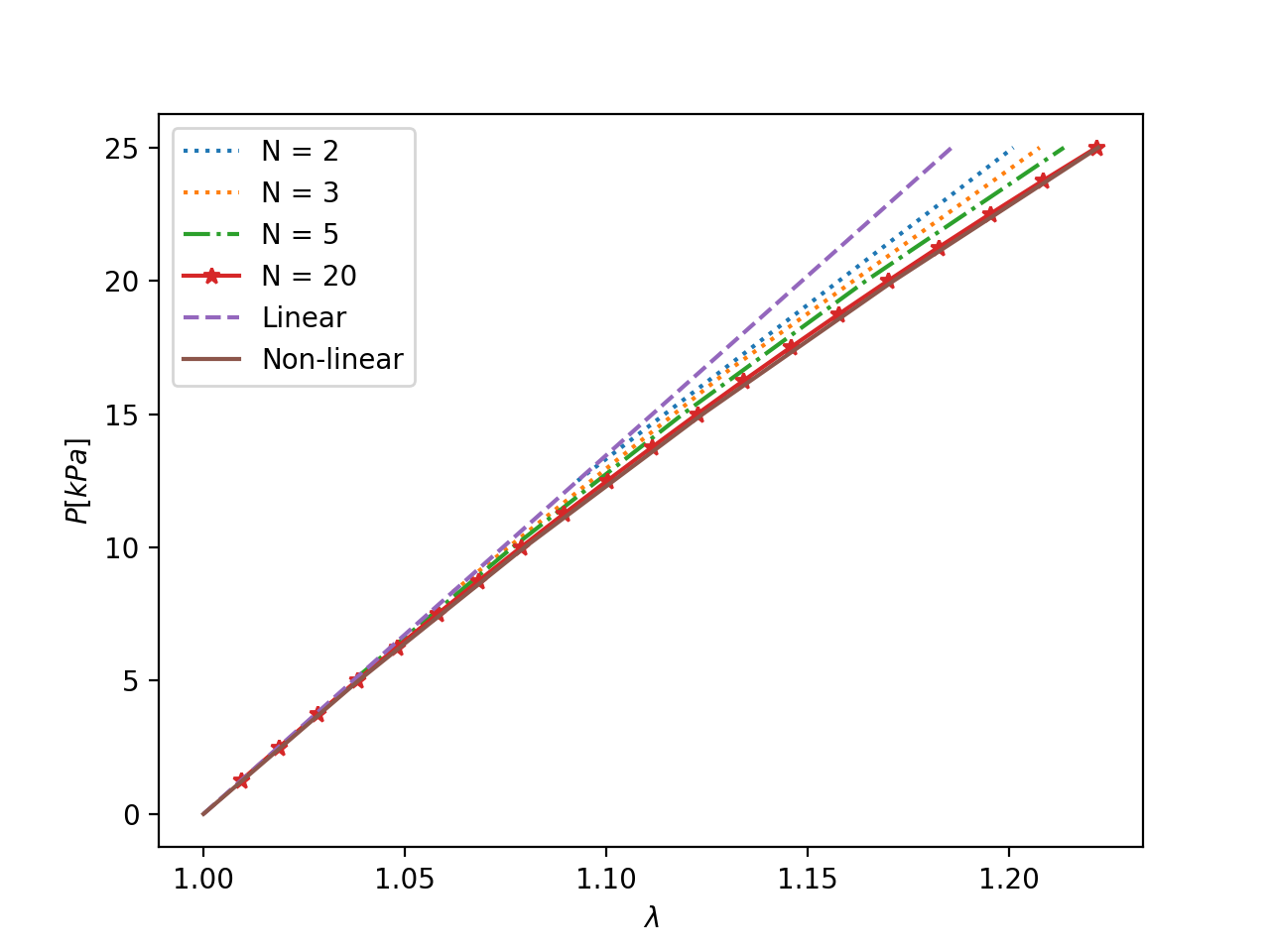}
\caption{The incremental nonlinear analysis procedure is tested and verified via an elastic problem. The dashed line shows the Hookean linear elastic response, the solid line is the results of the Neo-Hookean nonlinear elastic material (Lagrangian formulation), and the curves in between are the results of the described incremental nonlinear analysis procedure with different increment numbers (N).}
\label{fig_Hyper_test}
\end{figure}

\item Finally, in order to verify the integrity of the presented incremental nonlinear poroelastic framework we compare the maximum settlement of a column of poroelastic material under compression computed by the presented incremental nonlinear model with the linear one shown in Figure \ref{fig_MaxSettlement_T}. Furthermore, the results provided in the next section are also helpful for verification purposes.

  The associated model geometry, material properties, BCs etc. is the same as the first numerical example.
\begin{figure}
	\includegraphics[width=8cm]{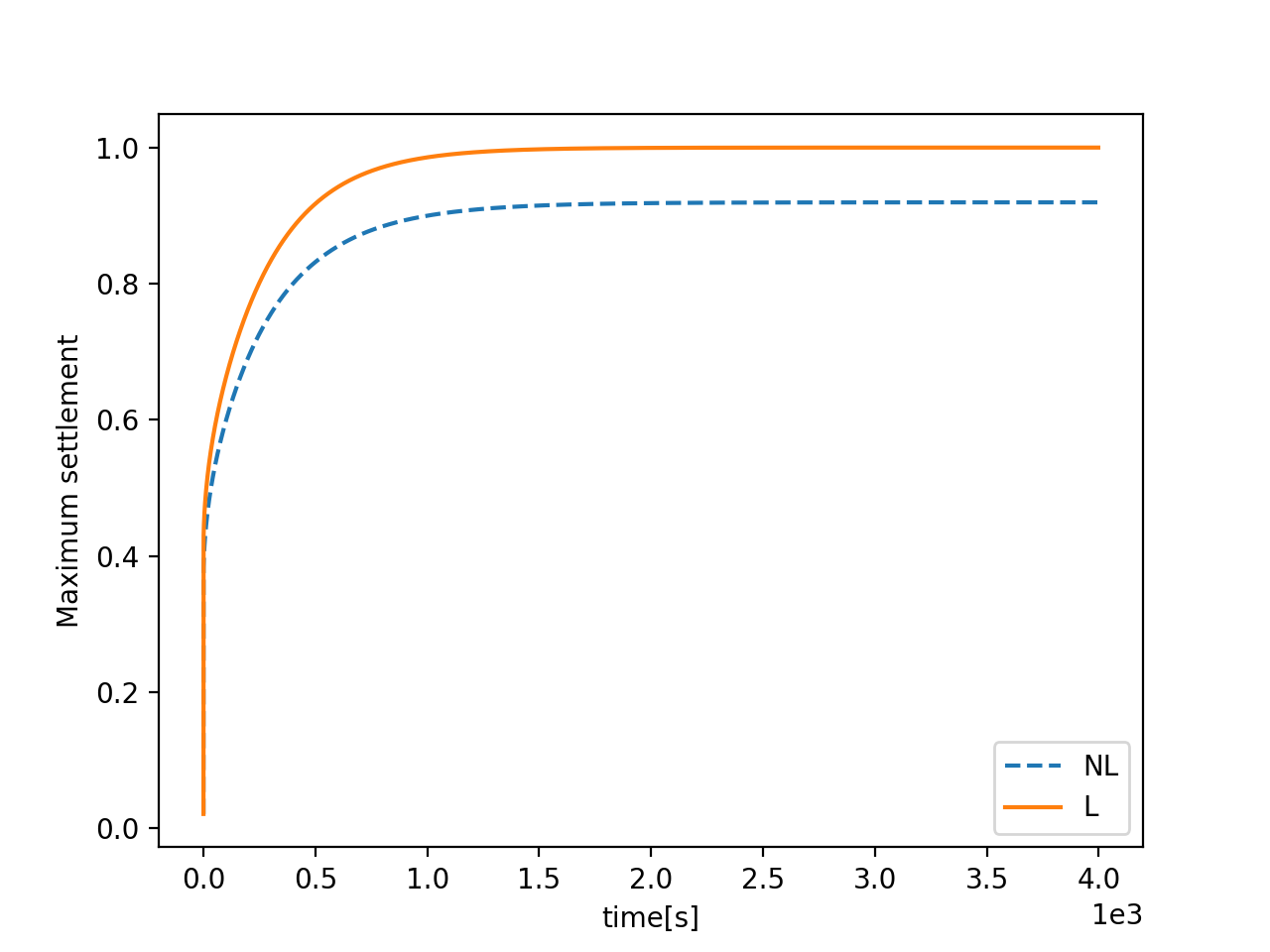}
	\centering
	\caption{The smaller settlement computed via our novel incremental nonlinear analysis for poroelastic media (dashed line) compared with the standard linear poroelasticity (solid line) is due to the strain stiffening imposed by the chosen material model which verifies the overall integrity of the model.}
	\label{fig_MaxSettlement_T}
\end{figure}
\end{itemize}

\section{Numerical examples} \label{Sec_numericalExamples}
We have, so far, dedicated our attention to developing an AI-assisted incremental computational method based on remodelling and asymptotic homogenisation/- localisation able to analyse nonlinear poroelastic problems. One major advantage of this method over the other methods in the field of nonlinear poroelasticity is that this framework, for the first time, considers the full macro and micro scale response/properties interdependencies. Hence it responds to several questions, in this study, including how are the microscale, as well as effective, properties affected by the macroscale mechanical and hydraulic response? Is an appropriate deviation from the overall linear Darcy's law for fluid flow in porous media (under finite deformation) enforced or fractional/exponential Darcy's law is needed? Why the preconditioning effects (residual strains), hysteresis response, and Mullins effects appear in poroelastic media such as brain tissue under cyclic loading? And, finally, how important is (or when is it important) to employ such a nonlinear method?
The numerical examples performed in this section respond to the above-mentioned questions.


\subsection{Uni-axial consolidation test}
Let us start from a simple example which has been exploited many times through the history of fully or partially fluid-saturated porous media. In this test, a column of poroelastic material is compressed by a constant mechanical pressure with free drainage BC on top. The model is given sufficient time to approach the steady-state where the time-dependent variations become negligible. Although this test is very simple, it is helpful to gain a better insight and to observe basic facts during the consolidation procedure. 

Let us assume initial isotropic solid matrix Young's modulus and Poisson's ratio, respectively, equal to $E_i=15e{6} [Pa]$ and $\nu_i=0.3$ with initial porosity of $\phi_i = 0.3$. The mechanical pressure is assumed $P=0.2E_i$ and the fluid is free to drain only from the top of the column. Prior to load application, the length of the tube is $L=7.5 [m]$ and the width is $B=0.1 [m]$. Zero displacements in all directions are applied at the bottom and the body is able to move only in the axial direction on the sides (the width remains constant). The load is applied in a very short time but in several increments ensuring the applicability of linear response in each increment (similar to the conditions of the hyperelastic test performed as a verification tool).

\begin{figure*}
  \centering
  \begin{subfigure}{5.8cm}
  \includegraphics[width = 5.8cm]{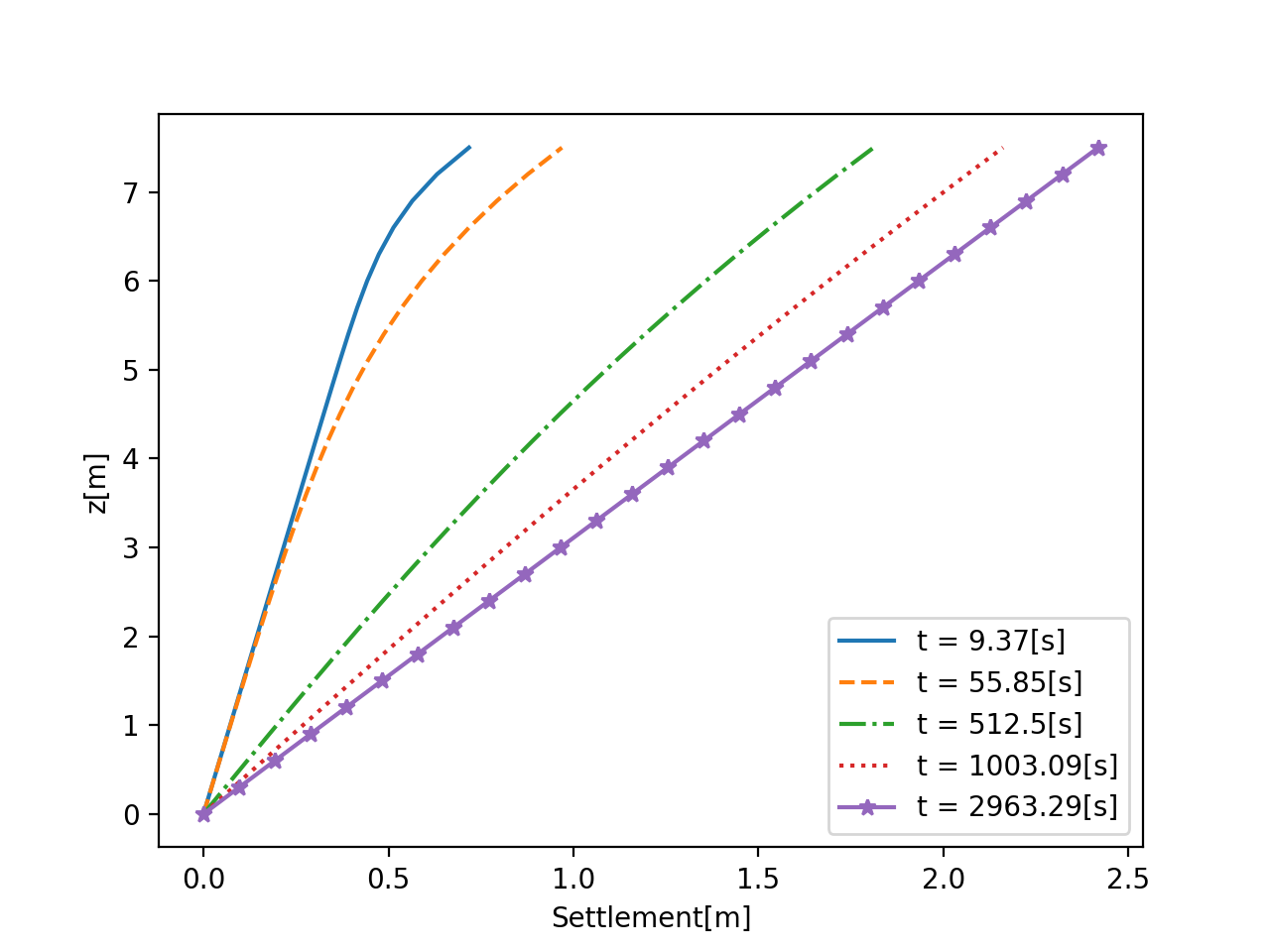}
  \caption{Shortly after the load application the settlement varies nonlinearly along the depth while the degree of nonlinearity decreases at greater times until it reaches linear spatial profile at steady-state. \vspace{0.38cm}}
  \label{fig_SettlementNL_Z}
  \end{subfigure}\quad 
  \begin{subfigure}{5.8cm}
  \includegraphics[width = 5.8cm]{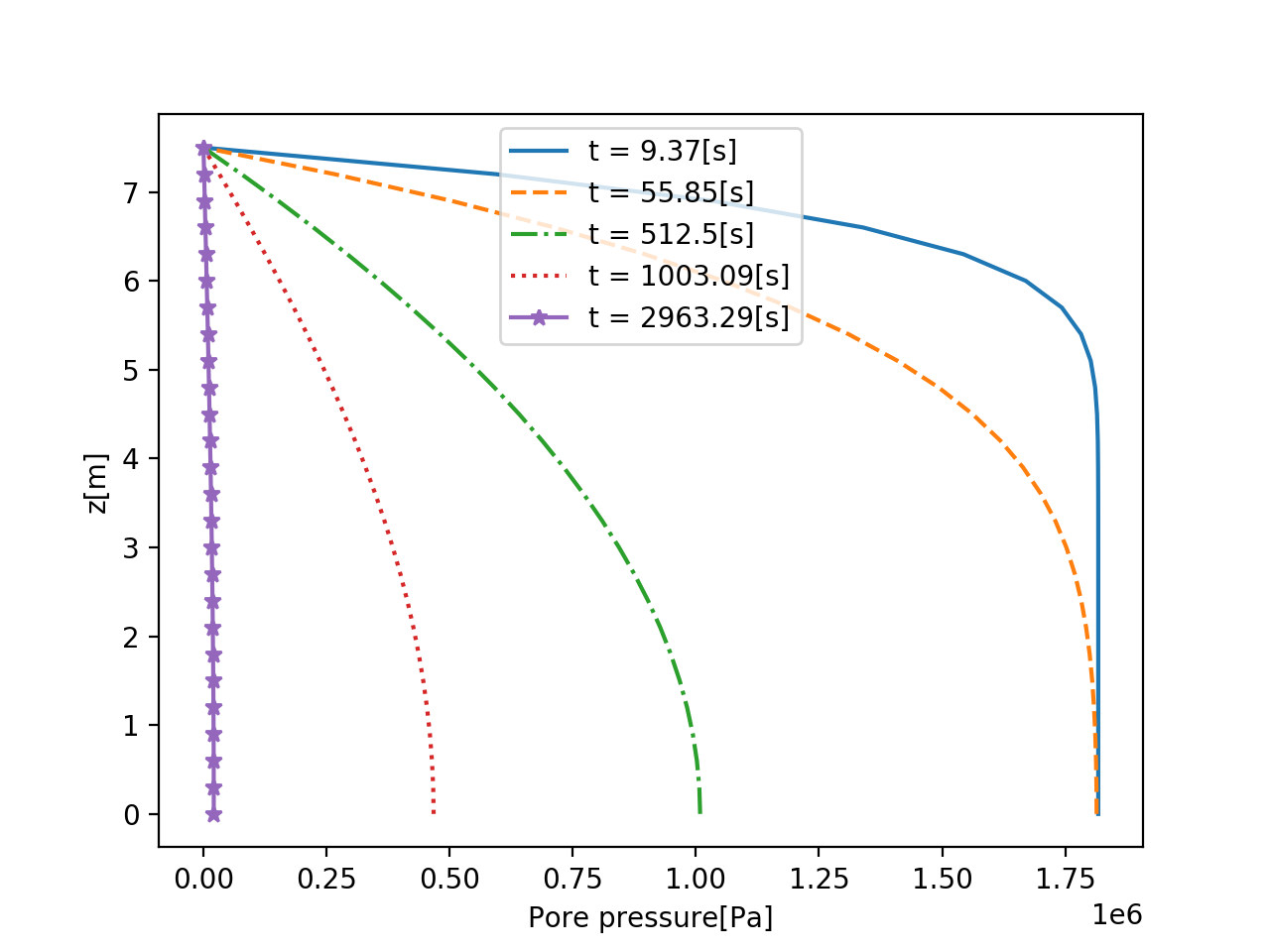}
  \caption{Pore pressure starts from a high value with great spatial gradient near the free drainage surface after the loading. The pore pressure spatial gradient drives the fluid flow approaching zero with pore pressure equal to the ambient pressure.}
  \label{fig_PPNL_Z}
  \end{subfigure}
  \caption{A representation of the spatial profile of DOFs at different times which determines the variation of material properties.}
  \label{fig_macro_response}
  \end{figure*}

We, first, visualise the macroscale response together with microscale and effective properties in order to understand their interdependency. Figure \ref{fig_macro_response} shows the mac- roscale response of the model along the axial axis (length) at different times with Figure \ref{fig_SettlementNL_Z} and Figure \ref{fig_PPNL_Z} being the settlement and interstitial pore pressure, respectively. The study of the spatial profile of settlement (which is the major displacement component in this study) and pore pressure are important and the starting point to understand the profile of micro and macro scale properties during the analysis. 
Considering Equation \eqref{eqMicroStrain}, we notice that shortly after the load application, in the major part of the length, pore pressure profile is the variable enforcing the variation of the micro and macro scale properties, while the spatial gradient of displacement affects only the part near the top of the model. This is due to the sharp decrease in pore pressure which is a consequence of the free drainage BC at top. In fact, the profiles of the properties at short times, mostly, should follow the variations of a high pore pressure while at longer times they are dominated by the solid strain.

  \begin{figure*}
  \centering
  \begin{subfigure}{5.8cm}
  \includegraphics[width = 5.8cm]{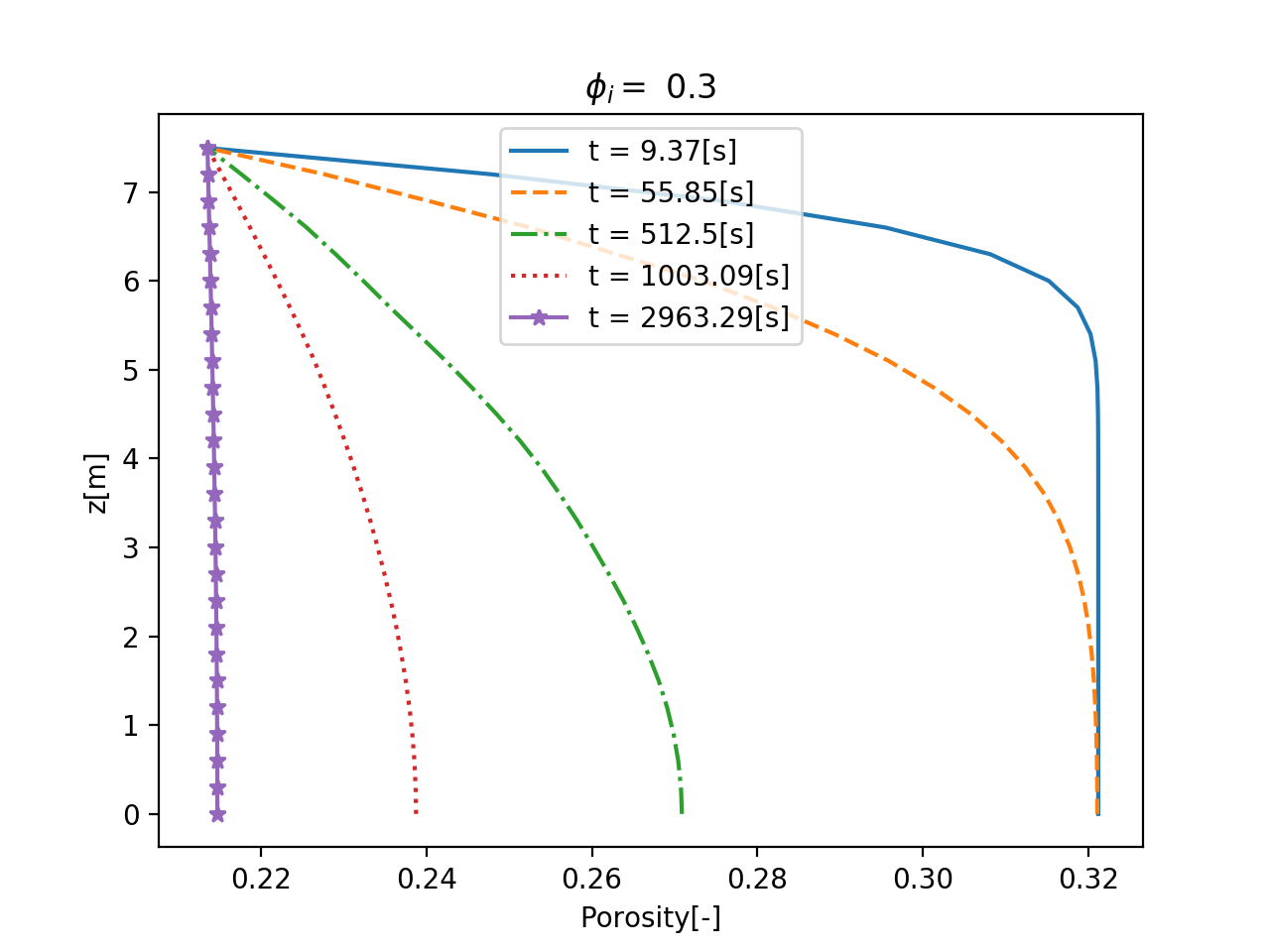}
  \caption{Although the final value of porosity at steady-state is determined by the solid phase deformation, its variation during the transient state is similar to the pore pressure spatial and temporal profile.}
  \label{fig_PhiNL_Z}
  \end{subfigure}\quad
  \begin{subfigure}{5.8cm}
  \includegraphics[width = 5.8cm]{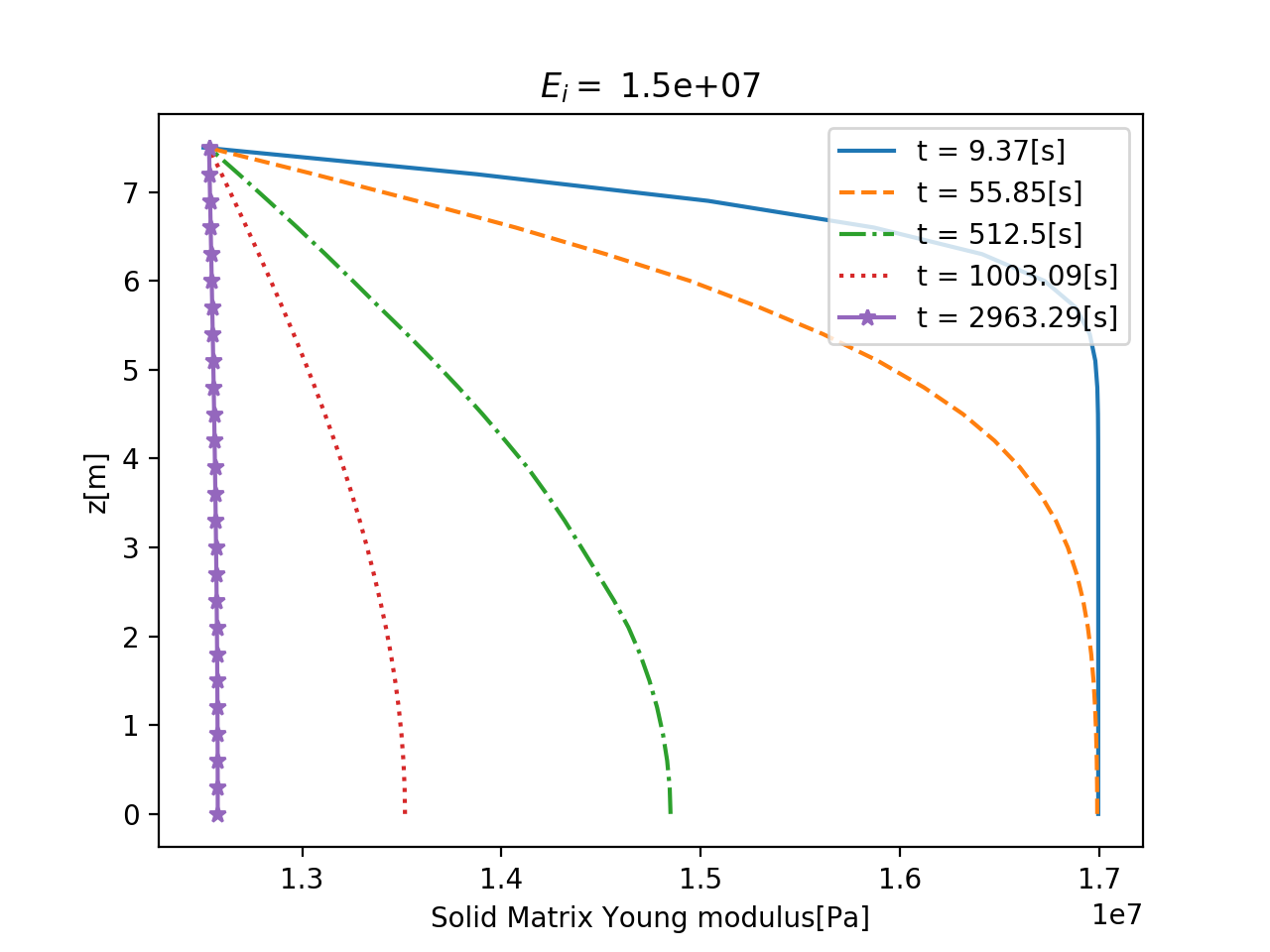}
  \caption{Similar to the porosity in Figure \ref{fig_PhiNL_Z} is the Solid Matrix Young's modulus profile.\vspace{0.38cm}\vspace{0.38cm}\vspace{0.38cm}}
  \label{fig_ENL_Z}
  \end{subfigure}
  
  \begin{subfigure}{5.8cm}
  \includegraphics[width = 5.8cm]{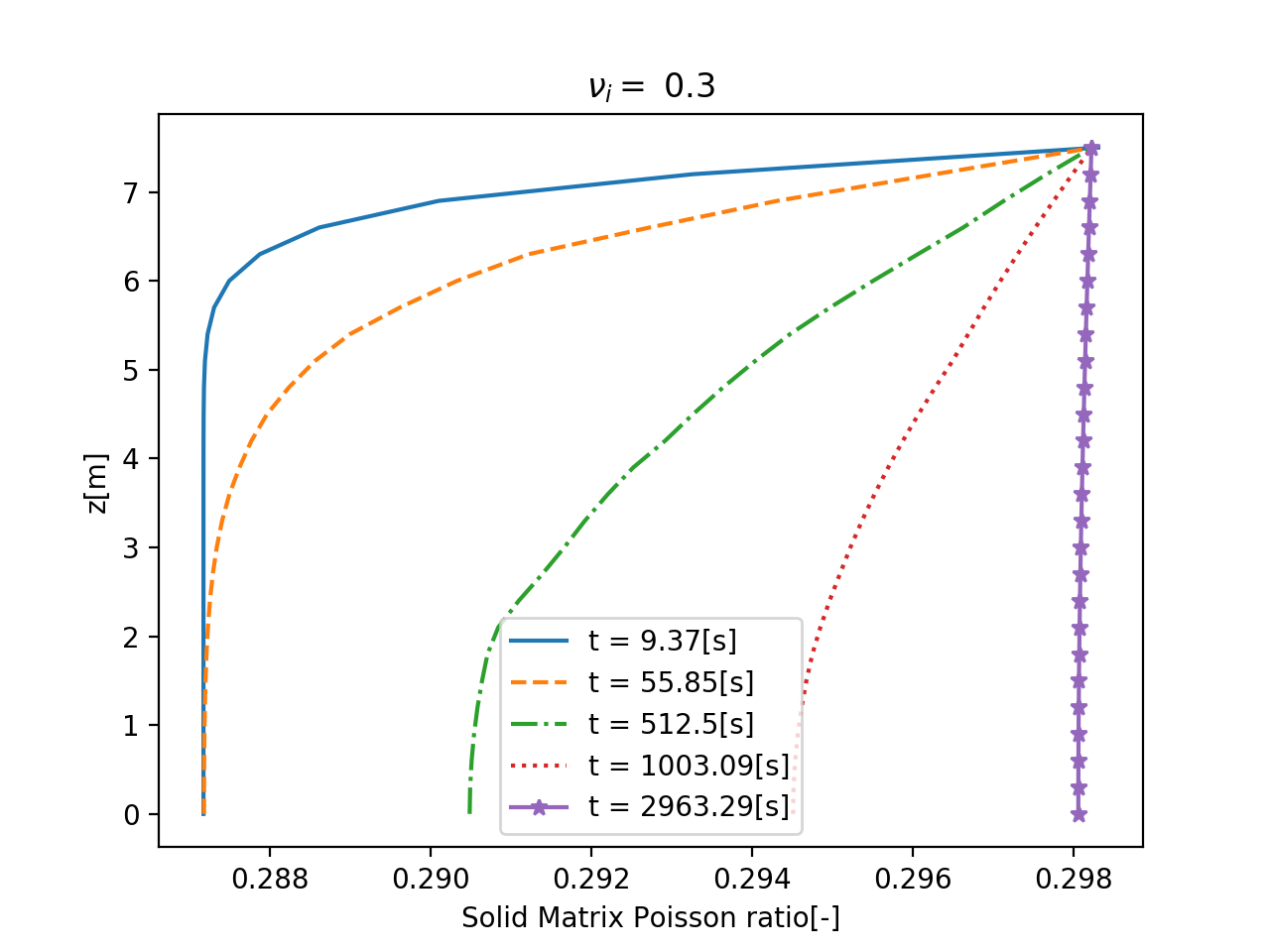}
  \caption{Solid Matrix Poisson's ratio shows inverse relationship with both pore pressure and settlement so it is below the initial value at all the times.\vspace{0.38cm}\vspace{0.38cm}}
  \label{fig_NuNL_Z}
  \end{subfigure}\quad
  \begin{subfigure}{5.8cm}
  \includegraphics[width = 5.8cm]{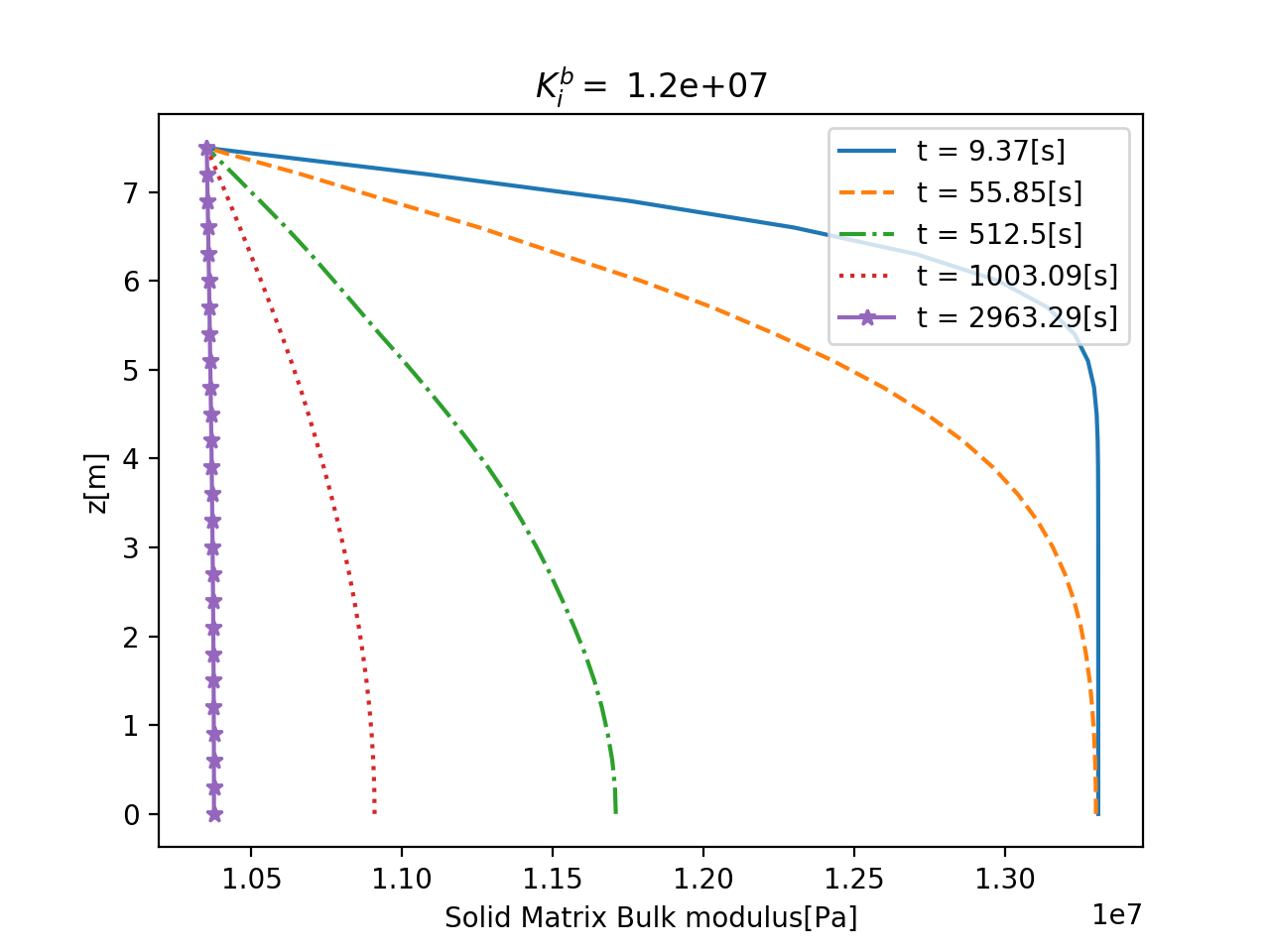}
  \caption{The matrix bulk modulus as a measure of compressibility depicts the expected behaviour. A contraction in solid matrix makes it less compressible while inverse effect is shown under the expansion.}
  \label{fig_KbulkNL_Z}
  \end{subfigure}
  \caption{The effect of macroscopic mechanical and hydraulic response on the spatial profile of microscopic properties at different times based on neo-Hookean material model.}
  \label{fig_micro_prop}
  \end{figure*}
  
At small times, the porosity distribution shown in Figure \ref{fig_PhiNL_Z} (which is similar to the studies focused on the porosity variation such as \cite{Lee1981}) reveals sharp spatial variations starting from the values below the initial porosity at the top of the model due to the solid strain (which shows a volume increase in the solid matrix) reaching to the values greater than the initial one under the influence of pore pressure. The latter highlights the fact that under the positive pore pressure, due to external compressive load, the porosity increases, causing a contraction in the solid matrix.  
However, as time passes, the pore pressure decreases and the porosity profile follows the macroscale solid strain showing an obvious effect of consolidation, which means higher volume fraction of the solid matter. We highlight that the direction of a positive pore fluid hydrostatic pressure acting on the solid-fluid interface is from the centre of the pore towards the solid domain.

The solid matrix material properties (Young's modulus and Poisson's ratio) are updated by the linearisation of the strain energy function given by Equations \eqref{Neo_hookean} and \eqref{eqTangent} based on the microscale kinematics. Figure \ref{fig_ENL_Z} shows that considering the porosity variation, the solid matrix Young's modulus is smaller than its initial value when there is an expansion in the solid domain showing strain-softening, while it is greater than the initial value in the area undergoing solid matrix contraction following the strain-hardening property of the neo-Hookean material model in compression.

On the other hand, this profile is not the same in the case of Poisson's ratio shown in Figure \ref{fig_NuNL_Z}. In fact, the latter decreases in both contraction and expansion as it is lower than its initial value in all the model. 
As the profile of Poisson's ratio and Young's modulus are not varying in the same direction, in order to monitor the overall material compressibility we compute and provide the variation of the bulk modulus which includes both Young's modulus and Poisson's ratio and considered directly as a measure of the compressibility (and consequently strain-softening or hardening). Figure \ref{fig_KbulkNL_Z} reveals that the solid domain properties variation obeys the given neo-Hookean material model, which could be served as a verification of the FE implementation.

    \begin{figure*}
  \centering
  \begin{subfigure}{5.8cm}
  \includegraphics[width = 5.8cm]{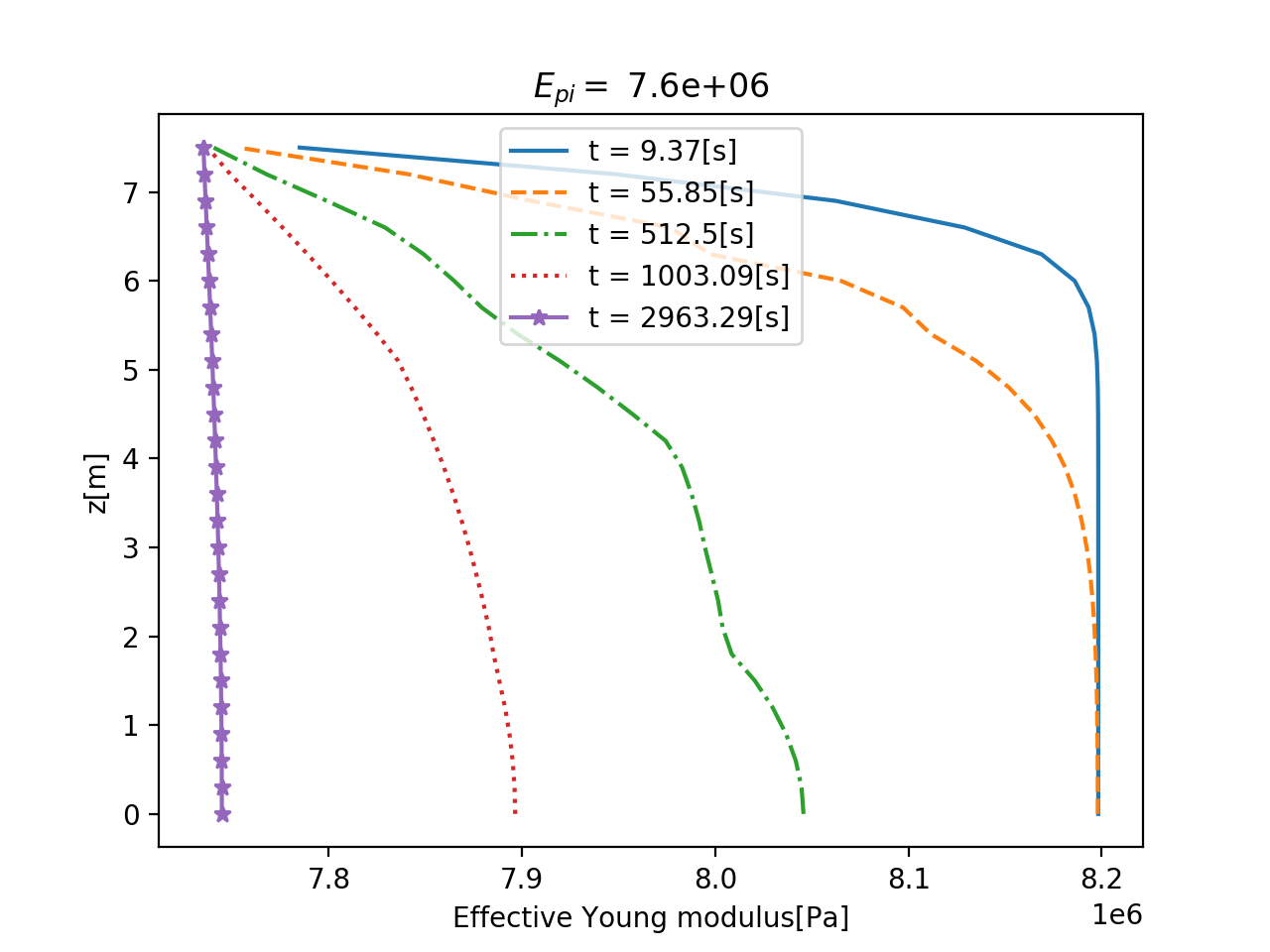}
  \caption{The profile of the effective Young's modulus is a results of the mixed roles of porosity and solid matrix elastic properties. 
  }
  \label{fig_EpNL_Z}
  \end{subfigure}\quad
  \begin{subfigure}{5.8cm}
  \includegraphics[width = 5.8cm]{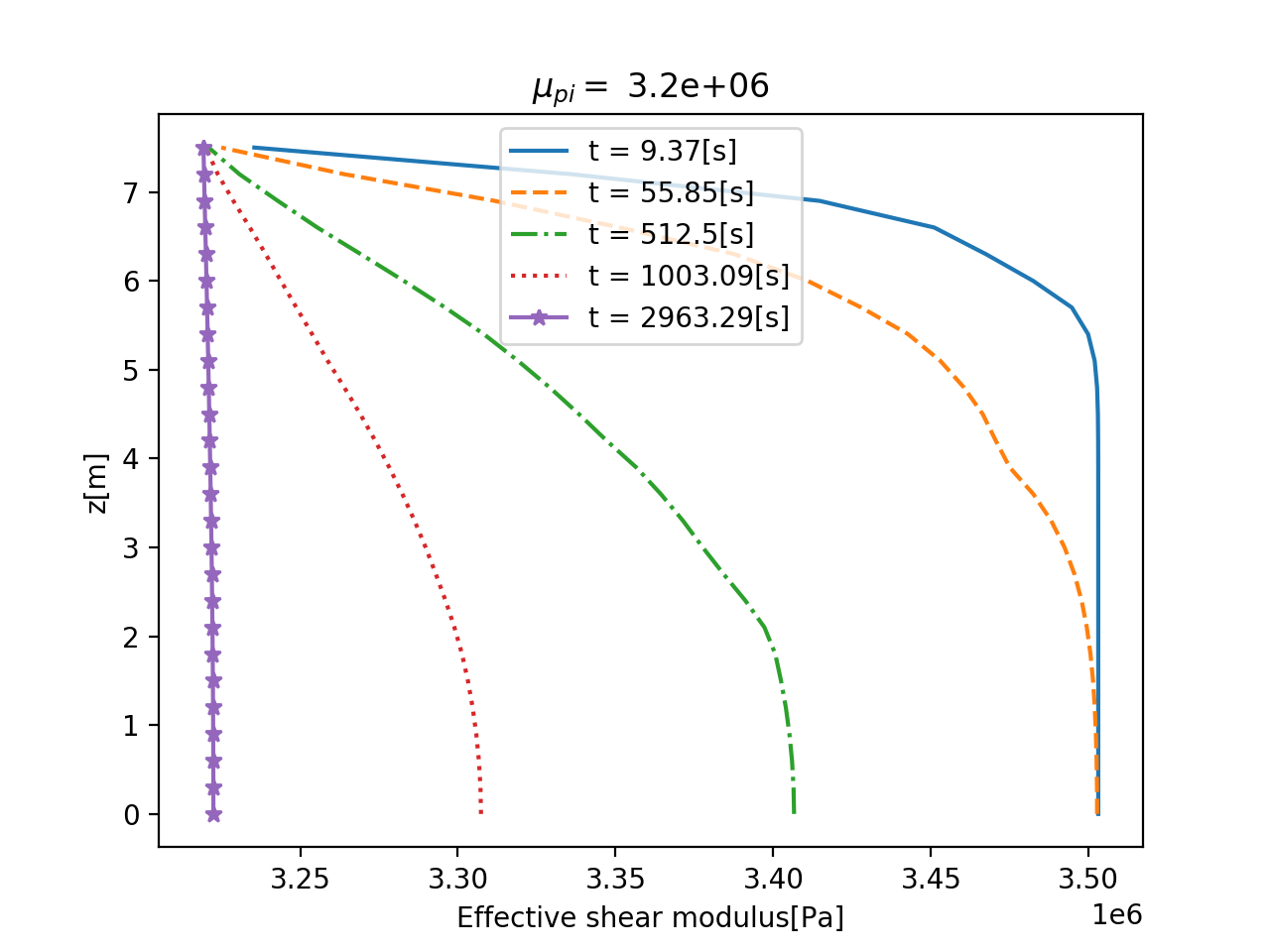}
  \caption{The effective shear modulus varies similar to the effective Young's modulus which is also observed in \cite{Hdehghani}.\vspace{0.4cm}
  }
  \label{fig_MupNL_Z}
  \end{subfigure}
  
  \begin{subfigure}{5.8cm}
  \includegraphics[width = 5.8cm]{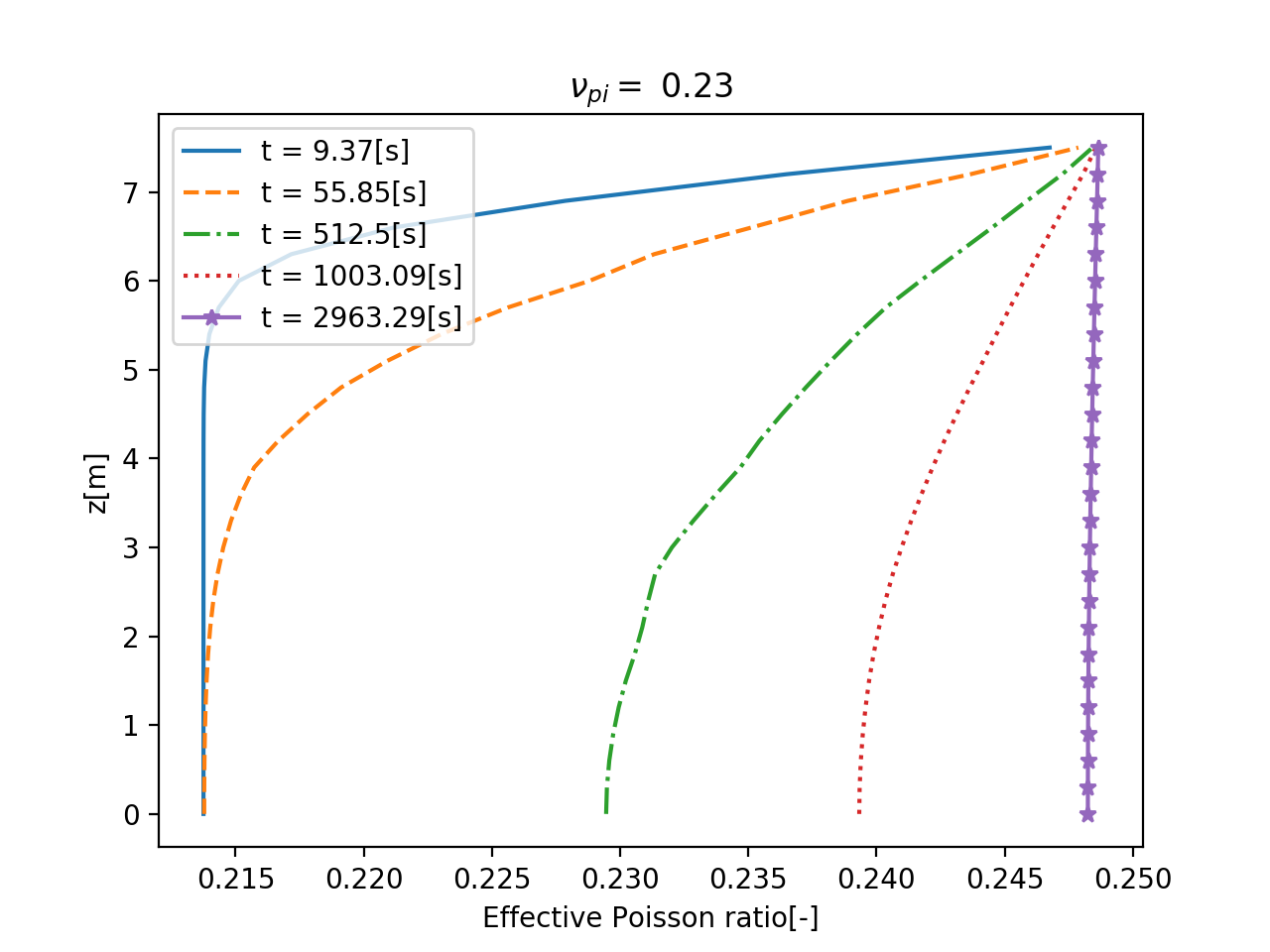}
  \caption{The effective Poisson's ration ranges from values above the initial one to the values below the latter. 
  }
  \label{fig_NupNL_Z}
  \end{subfigure}\quad
  \begin{subfigure}{5.8cm}
  \includegraphics[width = 5.8cm]{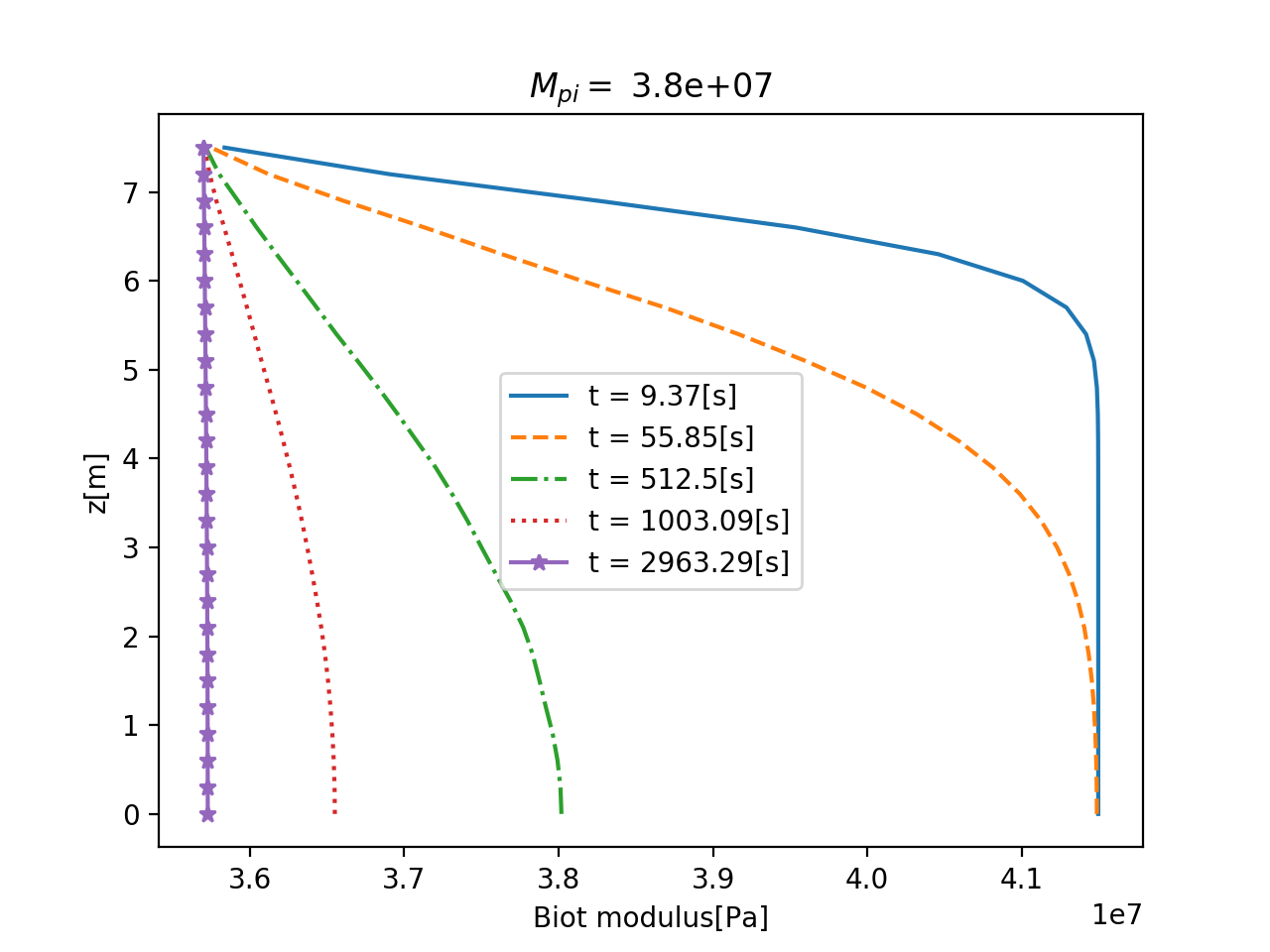}
  \caption{Since the Biot modulus increases at increasing porosity and Solid matrix bulk modulus \cite{Hdehghani}, its variation is reasonable.
  }
  \label{fig_MNL_Z}
  \end{subfigure}
  
  \begin{subfigure}{5.8cm}
  \includegraphics[width = 5.8cm]{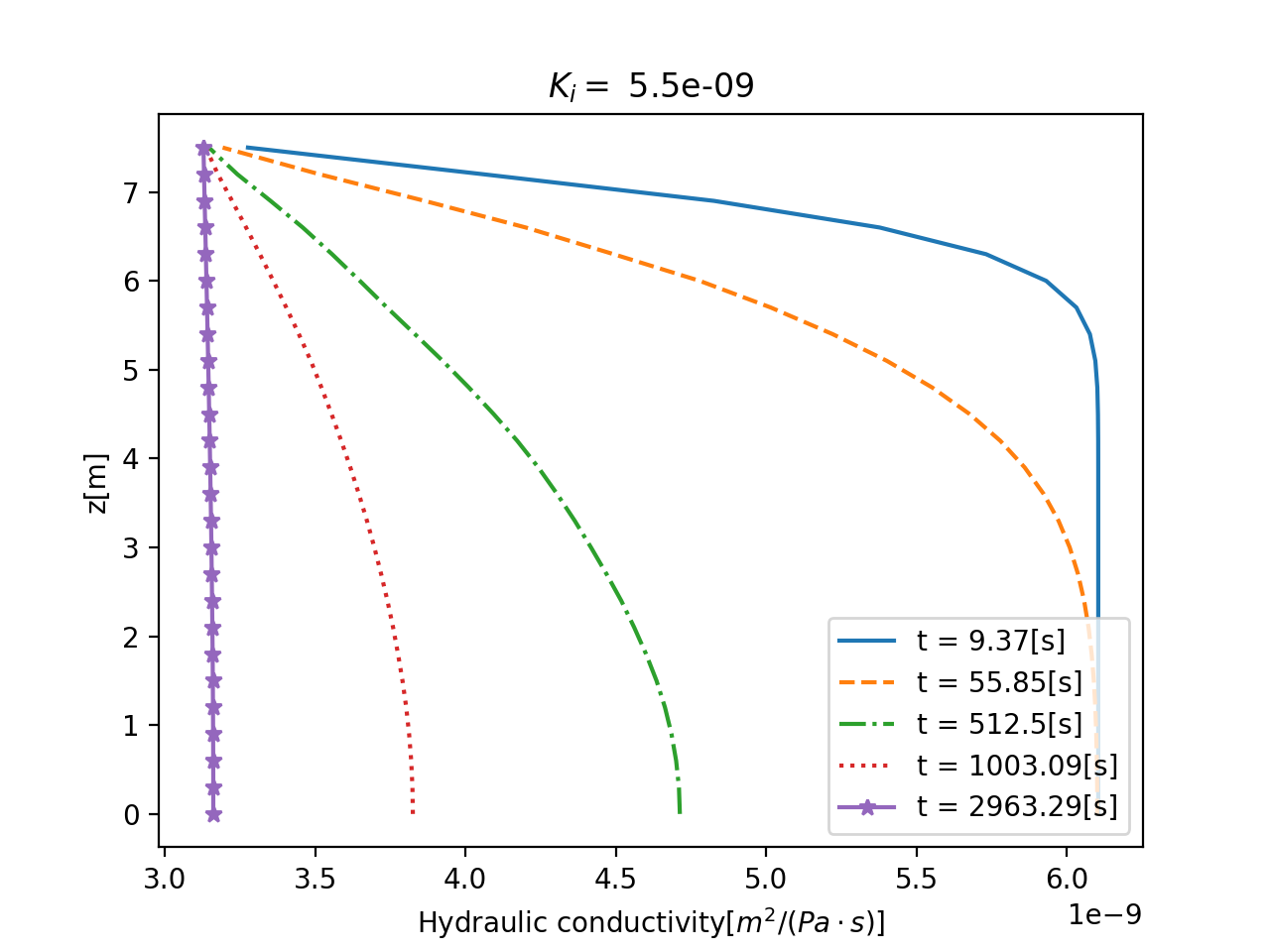}
  \caption{Hydraulic conductivity depends only on porosity, although porosity depends on solid matrix deformation and, consequently, its properties.}
  \label{fig_K11NL_Z}
  \end{subfigure}\quad
  \begin{subfigure}{5.8cm}
  \includegraphics[width = 5.8cm]{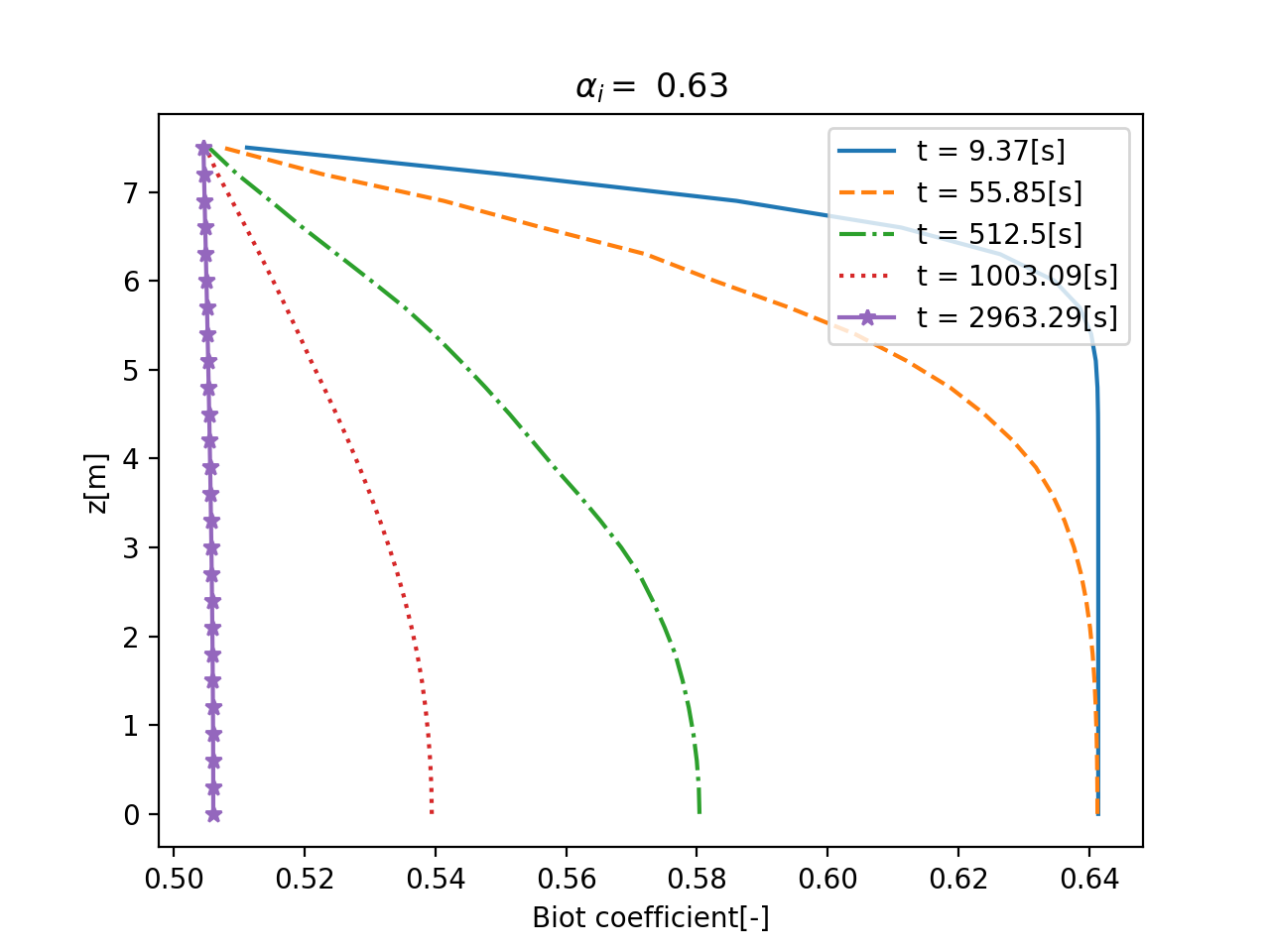}
  \caption{Biot coefficient depends on solid matrix Poisson's ratio and porosity.\vspace{0.38cm}\vspace{0.38cm}}
  \label{fig_AlphaNL_Z}
  \end{subfigure}
  
  \caption{The profiles of the effective poroelastic properties calculated based on the macroscopic mechanical and hydraulic response. We highlight that, due to the nonlinear and asymptotic nature of upscaling, the shown results and conclusions for the problems might vary at different ranges of initial values.}
  \label{fig_macro_prop}
  \end{figure*}
  
At this stage, exploiting the instant upscaling computation via ANN, the fourth rank tensor $\mathbb M$, the second rank tensor $\boldsymbol{\tens Q}$, and the hydraulic conductivity $\boldsymbol{\tens{K}}$ are obtained. The poroelastic properties $\mathbb{\tilde C}$, $\boldsymbol{\tilde{\tens{\alpha}}}$, $\boldsymbol{\tens{K}}$, and $M$ used in the homogenised model (see Equations \eqref{macroequi}-\eqref{eqDarcy}), are calculated via Equations \eqref{eq_TildeC}-\eqref{eq_HydraulicConductivity}. Furthermore, the effective Young's modulus, Poisson's ratio, and shear modulus are calculated based on cubic symmetric $\mathbb{\tilde C}$ via Equations \eqref{effectiveE}-\eqref{effectiveNu} replacing $C_{ij}$ by $\tilde C_{ij}$.
 
 Figure \ref{fig_macro_prop} shows the coefficients of macroscale system of equations that characterise the macroscale mechanical and hydraulic response of the homogenised model. The profile of effective Young's modulus demonstrated in Figure \ref{fig_EpNL_Z} is derived from displacement and pore pressure profiles and the resultant interaction of varying porosity and solid matrix stiffening/softening. In other words, as demonstrated in the literature \cite{Hdehghani, HdehghaniThesis}, increasing porosity at constant solid matrix properties results in a decrease in effective Young's modulus, while, considering remodelling of the solid material properties, it causes higher solid matrix Young's modulus. The latter, in turn, at constant porosity, increases the effective Young's modulus. A decrease in porosity has inverse effects. The result of this complex behaviour can only be determined using the provided AI-assisted multiscale framework. It worth to note that, despite the complex behaviour, effective solid matrix Young's modulus stands at a higher value than its initial value throughout the time and space in the problem.
 
 Effective shear modulus in Figure \ref{fig_MupNL_Z} has a similar complexity as effective Young's modulus although its profile is, somehow, smoother. In fact, similar to effective Young's modulus, it adopts values higher than the initial one resulting in strain stiffening.
Effective Poisson's ratio provided in Figure \ref{fig_NupNL_Z} has a less complex profile as it depends only on the solid matrix Poisson's ratio and porosity.  As an overall description, effective Poisson's ratio decreases, increasing the pore pressure and increases, due to the consolidation at the steady-state.

The fluid drainage and consolidation procedure is considerably influenced by Biot modulus, Hydraulic conductivity, and Biot coefficient.
Biot modulus determines the rate of change in pore pressure due to the macroscopic solid and fluid volume change.  Figure \ref{fig_MNL_Z} shows that this parameter exhibits an increase due to the positive pore pressure while it decreases due to the negative macroscale solid strain.
The considerable decrease in hydraulic conductivity and Biot coefficient in Figures \ref{fig_K11NL_Z} and \ref{fig_AlphaNL_Z} is due to the decreasing porosity at the top of the model as well as, in the case of Biot coefficient, the resultant change in the solid matrix material properties. We highlight that Biot coefficient acts as the coupling term between solid and fluid and hydraulic conductivity relates the spatial gradient of pore pressure to the fluid velocity relative to the solid displacement rate.  The overall effects and importance of the variation in the effective poroelastic properties are reflected in the macroscopic response of the medium.

\begin{figure*}
  \centering
  \begin{subfigure}{5.8cm}
  \includegraphics[width = 5.8cm]{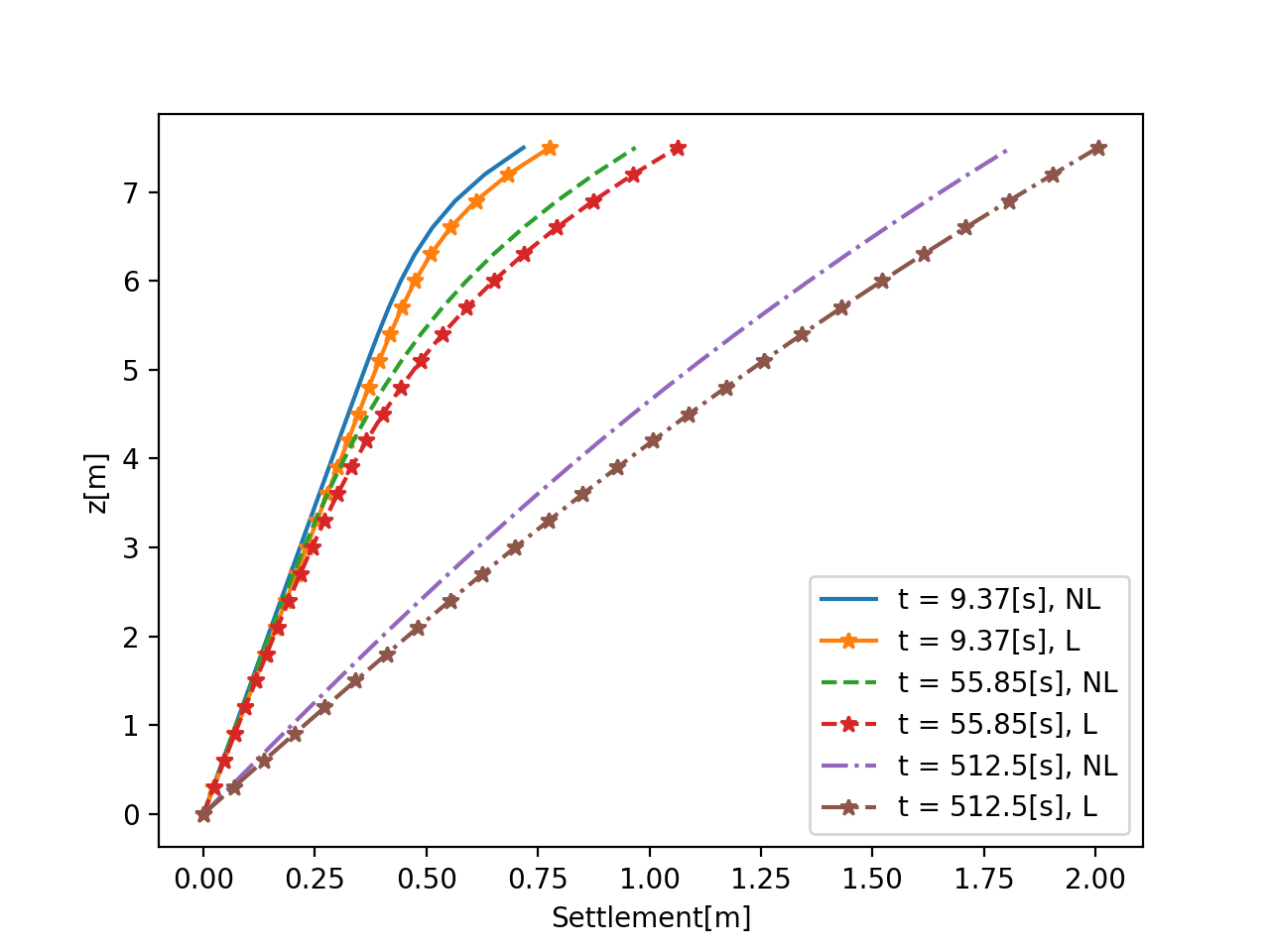}
  \caption{The smaller settlement values calculated via the presented incremental nonlinear poroelasticity ("NL") agrees with the strain stiffening in compression imposed by the chosen material model (neo-Hookean).}
  \label{fig_SettlementLNL_Z}
  \end{subfigure}\quad
  \begin{subfigure}{5.8cm}
  \includegraphics[width = 5.8cm]{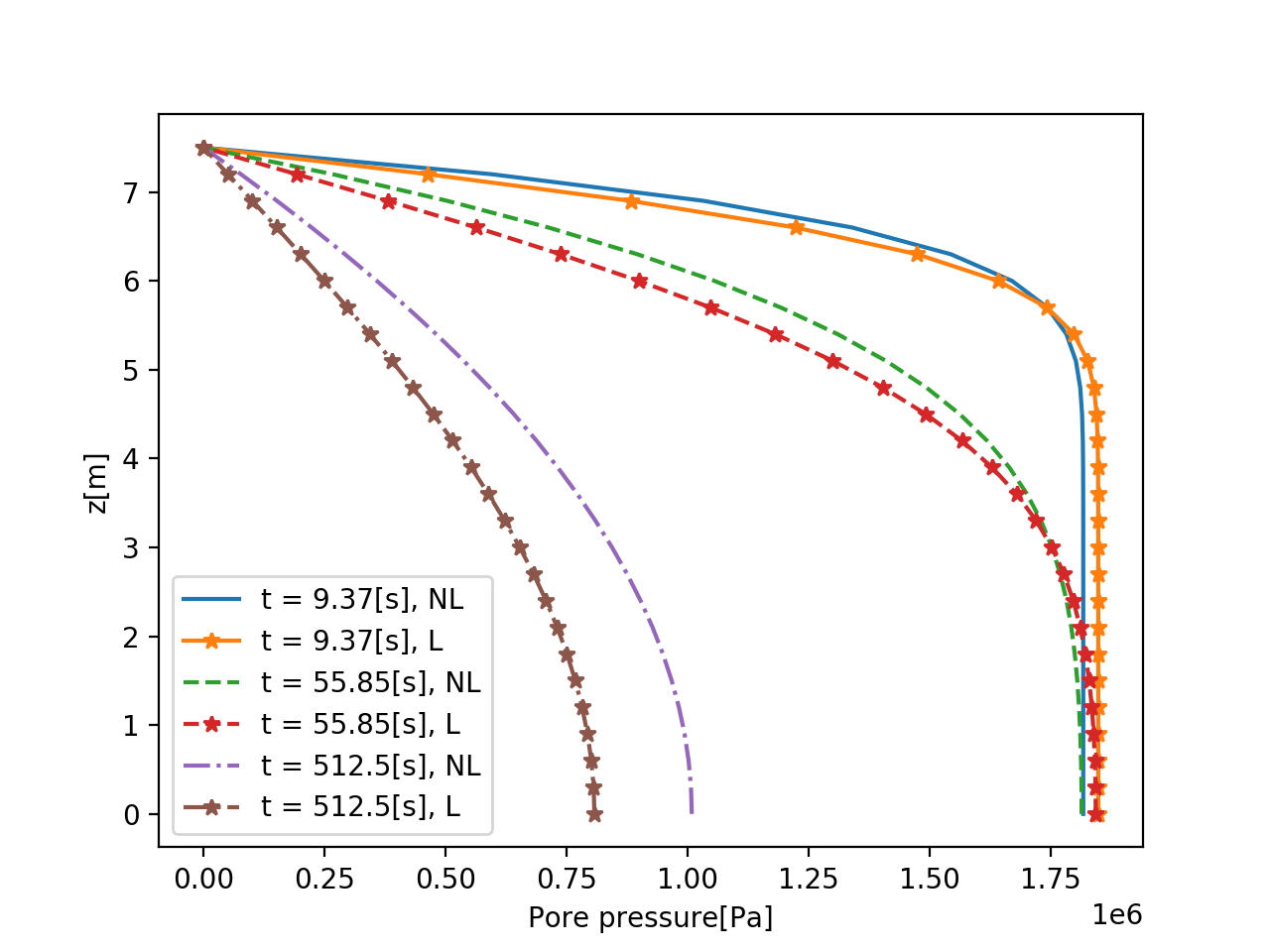}
  \caption{The maximum pore pressure of "NL" case is slightly smaller than the the "L" case shortly after the loading while it is considerably greater than "L" case at longer times.\vspace{0.38cm}}
  \label{fig_PPLNL_Z}
  \end{subfigure}
  
   \begin{subfigure}{5.8cm}
  \includegraphics[width = 5.8cm]{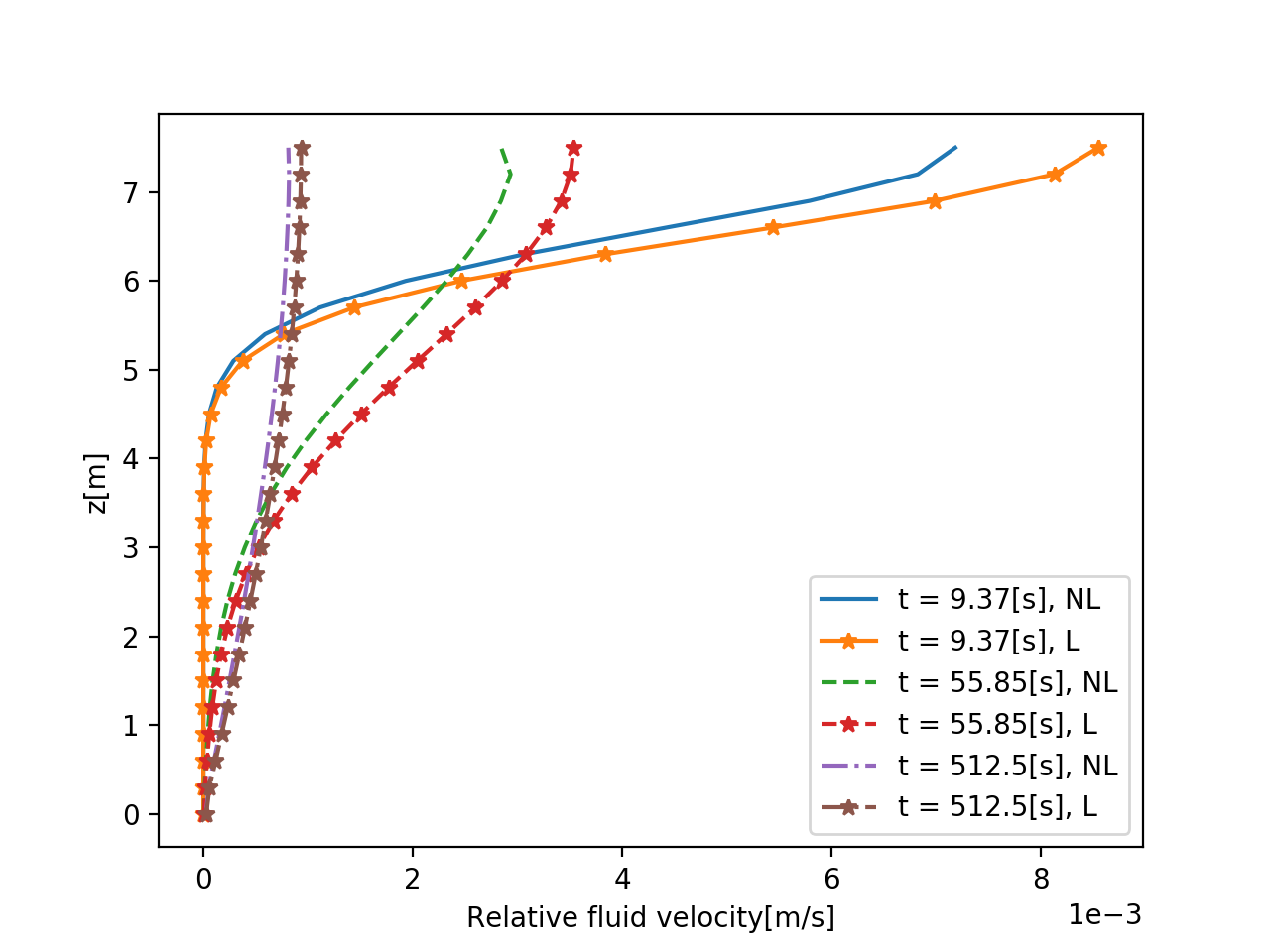}
  \caption{Although the pore pressure of "NL" case is greater than "L" case the relative fluid velocity is slightly smaller than in the "L" case due to the considerable decrease in Hydraulic conductivity shown in Figure \ref{fig_K11NL_Z}.}
  \label{fig_Vrf_Z_LNL}
  \end{subfigure}\quad
  \begin{subfigure}{5.8cm}
  \includegraphics[width = 5.8cm]{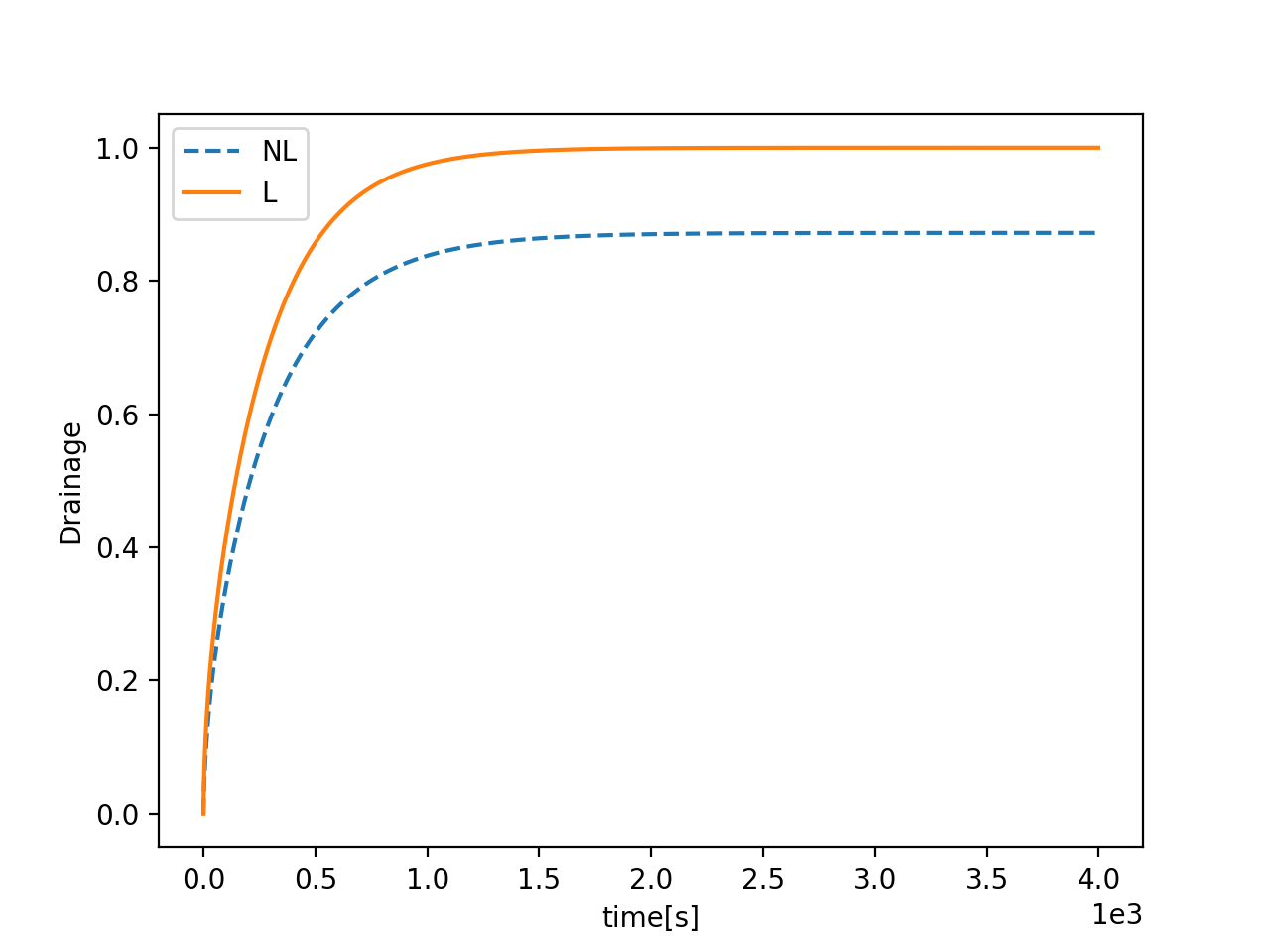}
  \caption{Agreeing with the maximum settlement (Figure \ref{fig_MaxSettlement_T}) and relative fluid velocity profile (Figure \ref{fig_Vrf_Z_LNL}), the fluid drainage in the "NL" case is smaller than the "L" case.\vspace{0.38cm}}
  \label{fig_divUrf_T}
  \end{subfigure}
  \caption{A comparison between the final response of the poroelastic medium calculated via the presented incremental nonlinear method (labeled as "NL") and the outcome of the linear poroelasticity (labeled as "L").}
  \label{fig_macroLNL_response}
  \end{figure*}

As a comparison mean and to understand the influences of the interplay between macroscale response and macroscale/microscale properties, the settlement and pore pressure profiles of both linear Biot consolidation and the present method are provided in Figures \ref{fig_macroLNL_response}. The latter shows a higher polynomial degree of nonlinearity during the transient state while the difference between them is closer to a coefficient change approaching the steady-state.

Applying divergence theorem and having small time incrementation the fluid drainage in one time increment from unit section surface area could be computed via
\begin{equation}
\Delta \varepsilon_{vf} =(\vett v_{rf} \cdot \vett n) \Delta t.
\end{equation}
We can also compute the total fluid drainage until $m$ th increment via
\begin{equation}
\varepsilon_{vf}^m= \sum_{n=1}^m \Delta \varepsilon_{vf}^n
\end{equation}

Figure \ref{fig_divUrf_T} shows the normalised fluid drainage (w.r.t the maximum value of the linear case)

Noteworthy is that all the mentioned properties in both scales are spatially dependent during the transient state while they become homogeneous at the steady-state. This highlights the fact that considering the described material response-properties interdependency is particularly more important in transient problems, e.g. cyclic loading. The effects of spatially dependent solid matrix material properties and porosity is thoroughly discussed in \cite{HDAZ2020}.

\subsection{Deviation from Darcy's law}

 \begin{figure*}
  \centering
  \begin{subfigure}{5.8cm}
  \includegraphics[width = 5.8cm]{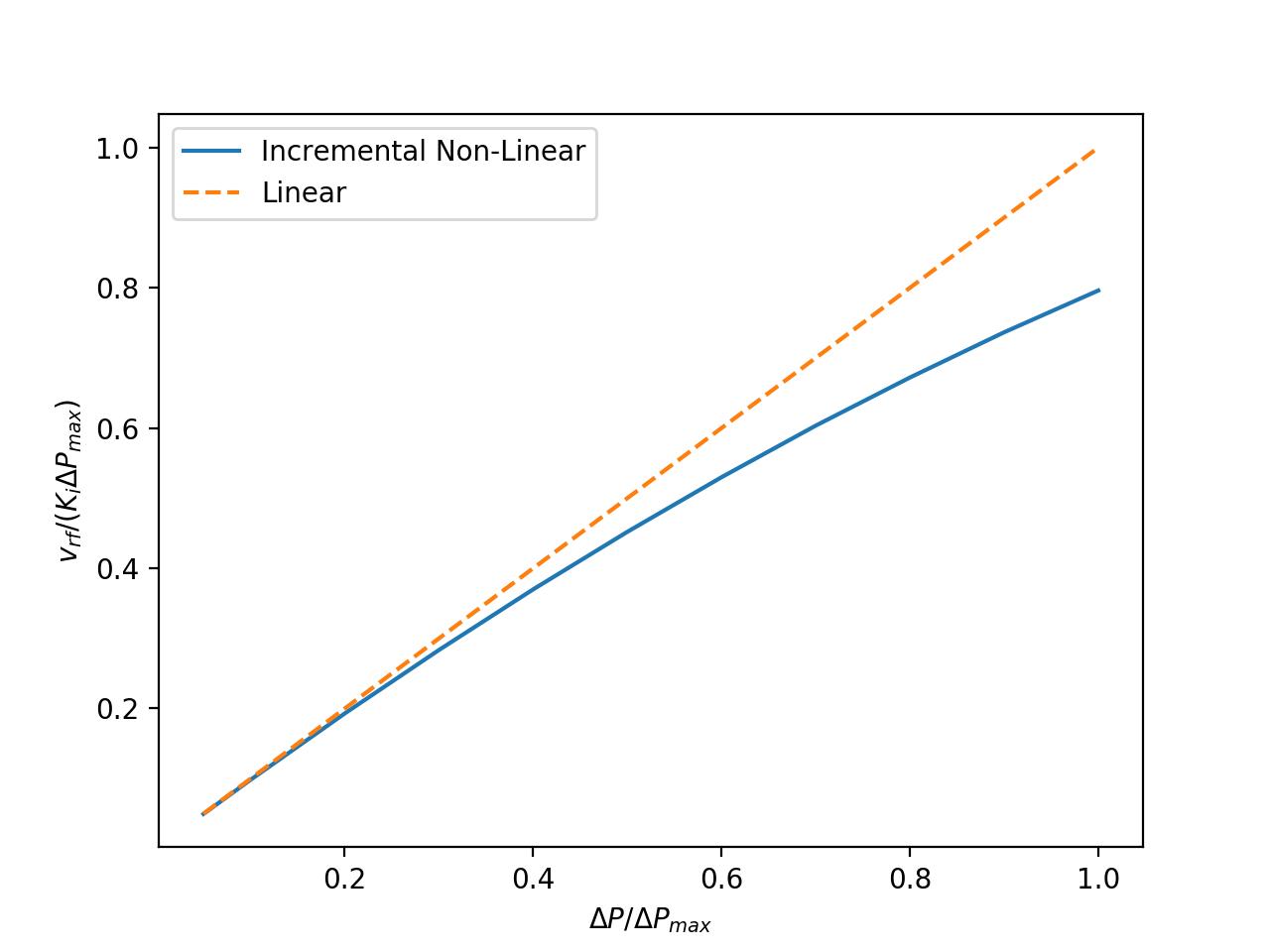}
  \caption{Although linear Darcy's law is used in each time increment, the overall hydraulic response calculated via the present incremental nonlinear method deviates from Darcy's law for linear poroelasticity, which is due to the hydraulic conductivity rearrangement after each increment.}
  \label{fig_Vrf_DeltaP_Normal_PD}
  \end{subfigure}\quad
  \begin{subfigure}{5.8cm}
  \includegraphics[width = 5.8cm]{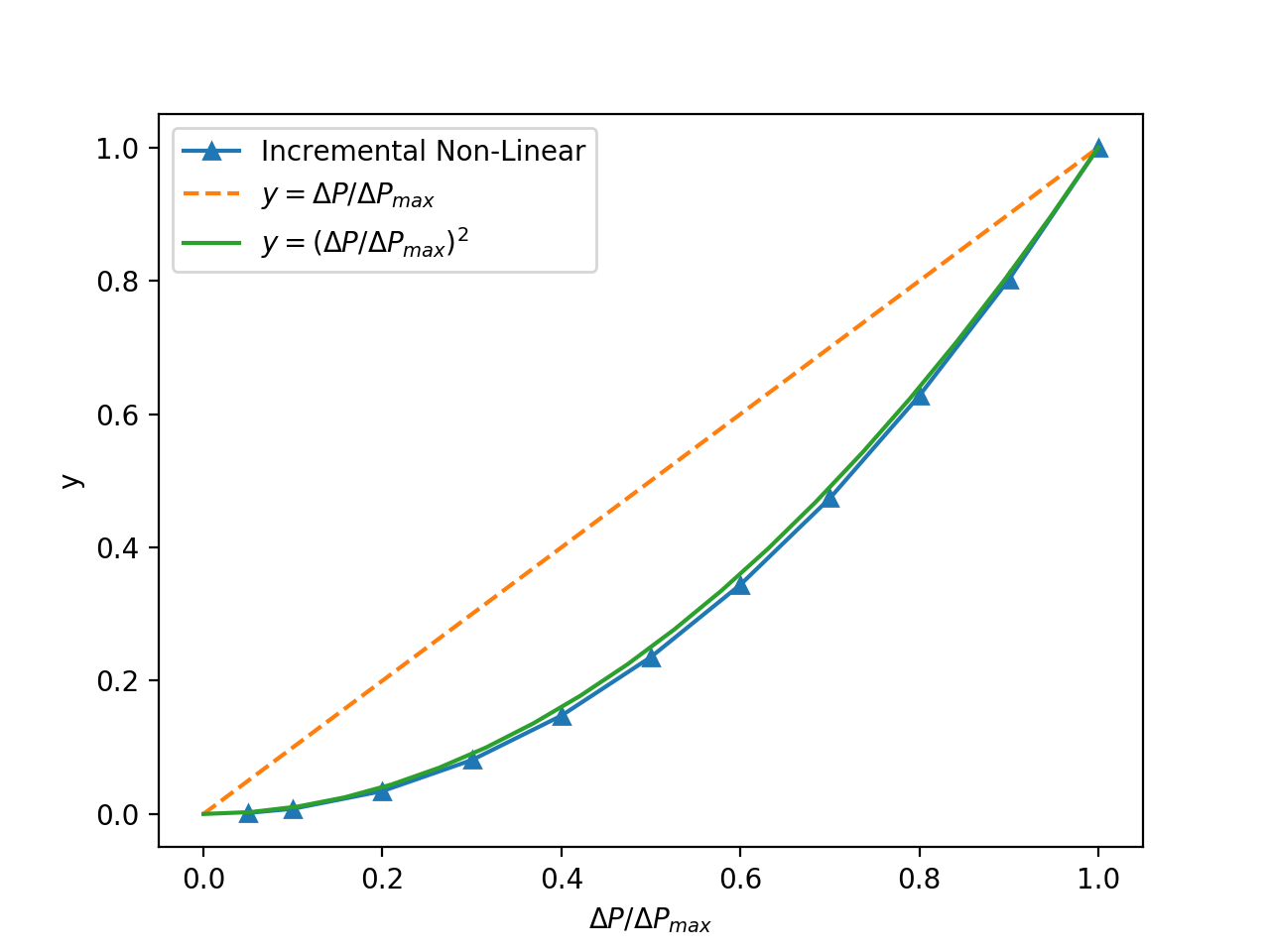}
  \caption{The dimensionless deviation from the linear case (Darcy's law) is close to the quadratic deviation relationship proposed in the literature (e.g. \cite{Firdaouss1997, forchheimer1901}).\vspace{0.38cm}\vspace{0.38cm}\vspace{0.38cm}\vspace{0.38cm}}
  \label{fig_DfD_Re}
  \end{subfigure}
  
  \begin{subfigure}{5.8cm}
  \includegraphics[width = 5.8cm]{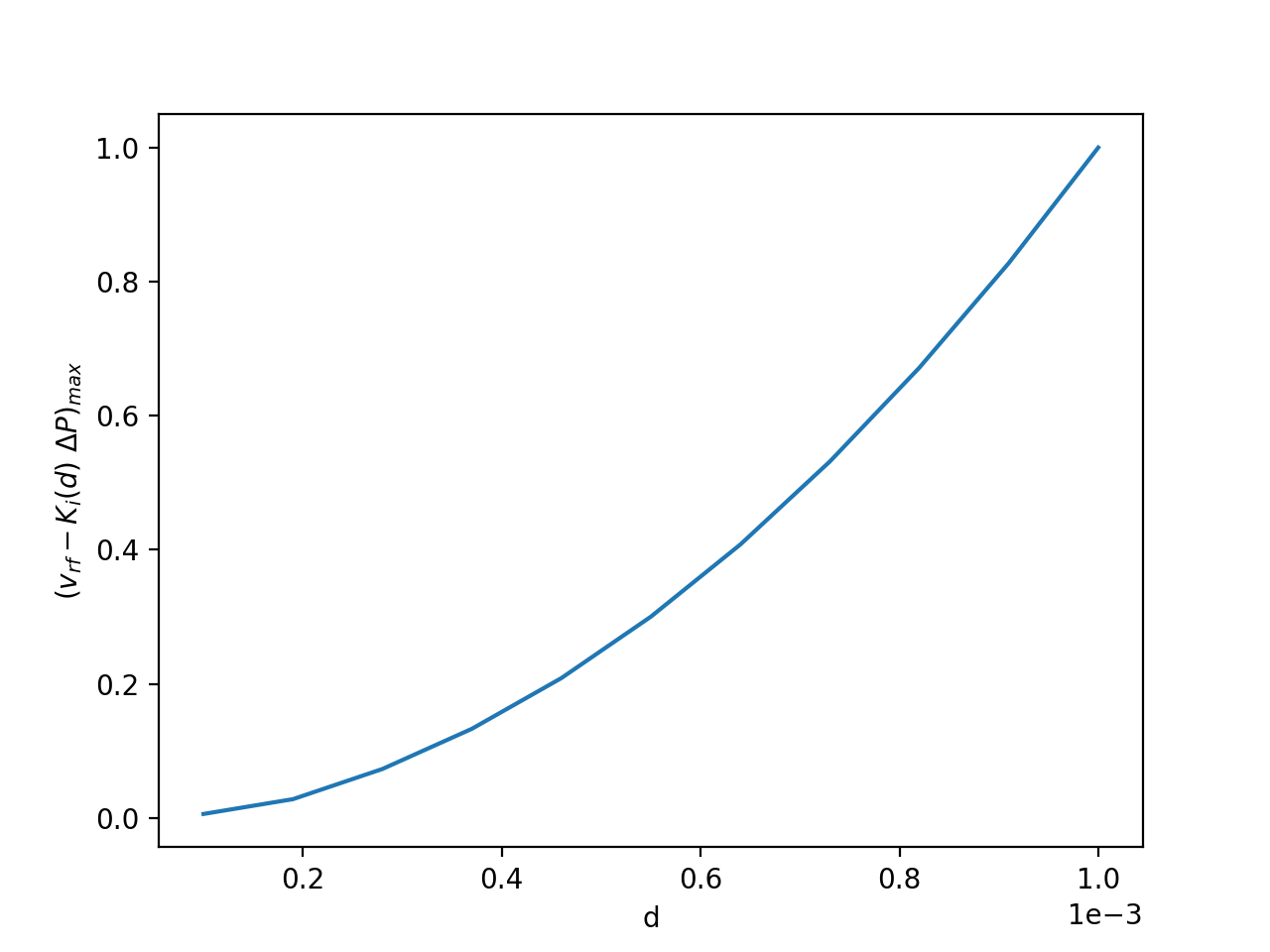}
  \caption{For a fixed $\Delta P$ the average cell dimension nonlinearly affects the deviation from the Darcy's law.\vspace{0.38cm}}
  \label{fig_Dvrf_d_PD}
  \end{subfigure}\quad
  \begin{subfigure}{5.8cm}
  \includegraphics[width = 5.8cm]{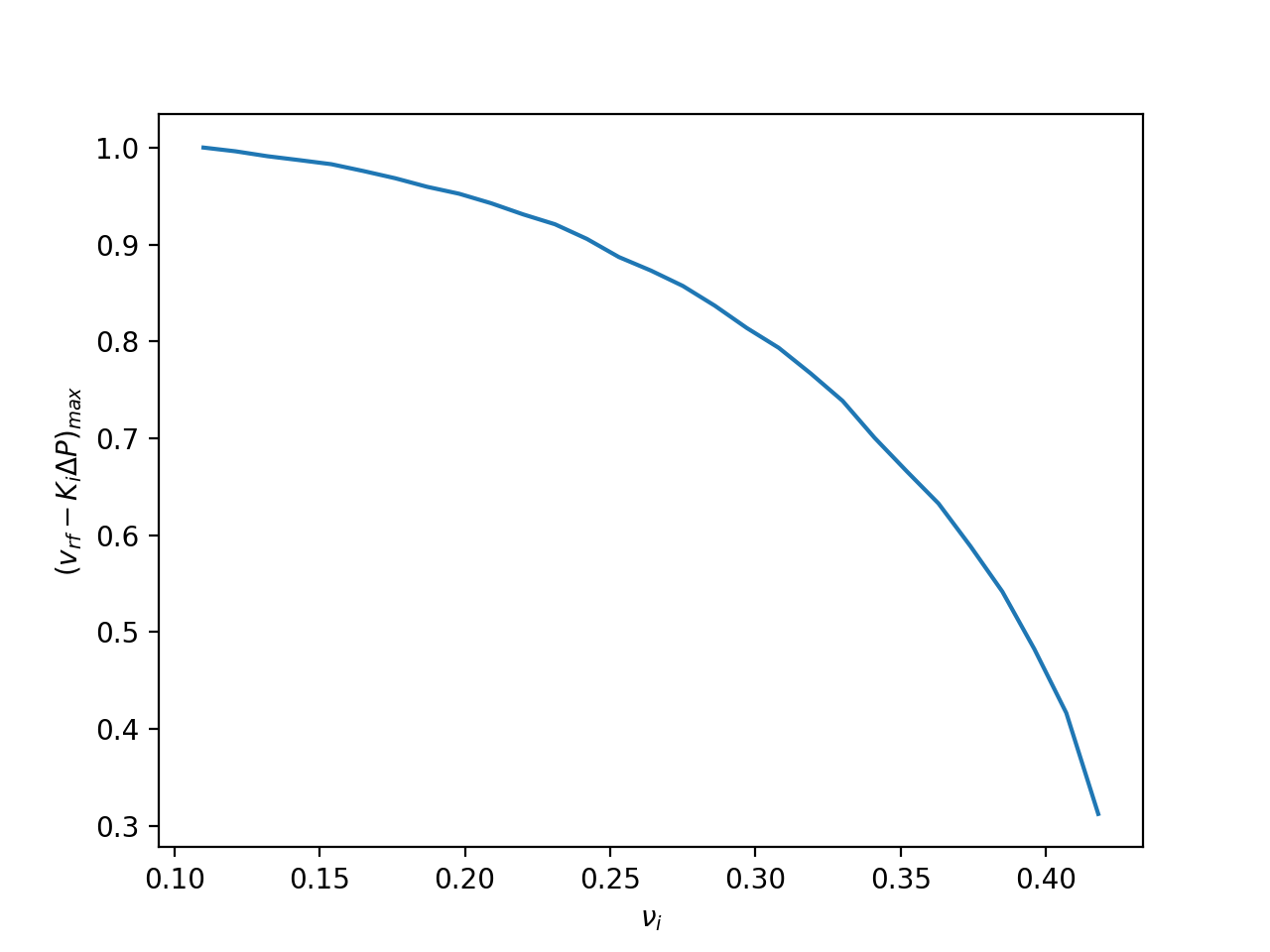}
  \caption{The initial solid matrix Poisson's ratio is another important parameter affecting, nonlinearly, the deviations from Darcy's law. }
  \label{fig_Dvrf_nui}
  \end{subfigure}
  \caption{The deviations from overall linear Darcy's law (without remodelling) is due to poroelastic properties rearrangements. The observed agreement with different experimental results in the literature suggests that, using the present incremental nonlinear method (updating the hydraulic conductivity after each increment), there is no need to any correction to Darcy's law used inside each time increment.}
  \label{fig_Darcy_Ex}
  \end{figure*}

The applicability of Darcy's law \cite{Darcy1856} into poroelastic problems in asymptotic range (at small Reynolds number) has been under debate since its publication. In \cite{Firdaouss1997} it is mentioned that the linear fit to the experimental data yields a negative pressure gradient when the flow rate is zero (e.g. $-\vett{\nabla} p = -0.745773 + 0.640324 q$ where $q$ is the flow rate) which means that there is a need to nonlinear corrections to Darcy's law. In this study, we model the Darcy's filtration experiment for a wide range of pressure differences ($\Delta P$) reaching to a maximum $\Delta P_{max} = E_i/4$ (where $E_i$ is the initial solid matrix Young's modulus). We calculate the relative fluid velocity when the simulation reaches the steady-state with criterion $\frac{v_{rf}^{in} - v_{rf}^{out}}{v_{rf}^{in}} \leq 1e{-3}$. We calculate the dimensionless deviation from Darcy's law via
\begin{equation}
y = \frac{v_{rf}- (K_i \Delta P) }{v_{rf,max}-( K_i \Delta P_{max})}, \label{Eqy}
\end{equation}
where $K_i$ and $\sbt_{max}$ are the initial hydraulic conductivity and the maximum value of $\sbt$ throughout the experiment, respectively.  $\vett v_{rf}$ is the relative fluid velocity computed via the presented incremental nonlinear method in which the solid matrix properties and porosity (consequently hydraulic conductivity) are updated at each increment. 

 Figure \ref{fig_Vrf_DeltaP_Normal_PD} shows that $\vett{v}_{rf}$  deviates considerably from the relative fluid velocity computed via Darcy's law (with the initial hydraulic conductivity) at pressure differences higher than $0.2\Delta P_{max}$ which agrees with the information provided in the literature (see, e.g. \cite{Clark1907}).
The non-dimensional deviation value $y$ in Equation \eqref{Eqy} at different pressure drops 
is provided in Figure \ref{fig_DfD_Re}. The latter provides a mean to compare the results provided via the presented method with the linear and quadratic deviations (i.e. $\vett y = (\Delta P/\Delta P_{max})$ and $\vett y = (\Delta P/\Delta P_{max})^2$, respectively) which shows a great match with the quadratic one. The latter was also concluded in several studies such as in \cite{Firdaouss1997, forchheimer1901}.

The average cell dimension $d=1e{-4}[m]$ was assumed so far. Figure \ref{fig_Dvrf_d_PD} reveals the fact that the average cell dimension plays an important role in the deviation from the linear Darcy's law highlighting that the more the average cell dimension, the more the deviation. One could also conclude that the more the cell average dimension, the less is $\Delta P$ limit at which the nonlinear model starts to, considerably, deviate from the linear. This can be seen in the definition of dimensional hydraulic conductivity which is directly proportional to $d^2$. The results provided in Figure \ref{fig_Dvrf_d_PD} is, qualitatively, in agreement with the effect of the soil grain size on deviation from Darcy's law provided by means of experimental results of \cite{Lopik2017}.

From the microscale cell problems which are solved to determine the hydraulic conductivity and the dimensionalisation procedure we notice that, apart from average cell dimension and viscosity, porosity is another important parameter in this problem which is considerably affected by solid matrix Poisson's ratio ($\nu$) at a specific pore pressure distribution. The reason is that $\nu$ plays a major role in the solid matrix volume change leading to a change in porosity. Figure \ref{fig_Dvrf_nui} shows the influence of solid matrix Poisson's ratio on deviation from linear poroelasticity. 

From our observations, we conclude that considering the medium remodelling after each time increment, there is no need for the formerly proposed corrections to Darcy's law.

\subsection{Uniaxial cyclic test on brain tissue}
Detailed understanding of complex and important mechanical properties and response of brain tissues provides us with vital accurate predictions for the design of treatment protocols and assessment of risk. For this purpose, numerical analysis with appropriate models and accurate parameters can provide crucial information on the effects of external loads. The application of these external loads can be slow or fast and their effects can be short or long term. It is seen that the brain tissue mechanical response differs considerably in each case. The complexity of this tissue behaviour is well studied in \cite{Budday2020} introducing several challenges in this field, however, an appropriate poroelastic modelling of brain tissue was missing.
The mechanical response of brain tissue under uniaxial experimental tests at free drainage from, respectively, sides and top of the model in \cite{FRANCESCHINI20062592} provides the first direct evidence of poroelastic behaviour of brain parenchyma which is proven by comparing the results of uniaxial consolidation experiment with Terzaghi's analytical solution. However, the numerical/analytical models for modelling the uniaxial cyclic experiments are not considered via poroelastic models. According to the latter study, under uniaxial cyclic loading, "brain tissue exhibits a peculiar nonlinear mechanical behaviour, exhibiting hysteresis, Mullins effect and residual strain, qualitatively similar to that observed in filled elastomers". 

In this subsection, we consider the application of the presented framework into the mechanical modelling of biological soft tissues which are typically considered as nonlinear problems. In order to focus on the role of the novel incremental nonlinear poroelastic methodology for mechanical response analysis of the medium, we, again, choose the simple neo-Hookean hyperelastic model for the solid matrix properties rearrangement.
We show, for the first time, that the poroelastic nature of brain tissue considering microscopic properties rearrangement due to the macroscopic mechanical and hydraulic response plays a major role (if not the only reason) in the appearance of the mentioned peculiar behaviour.

 \begin{figure*}
  \centering
  \begin{subfigure}{5.8cm}
  \includegraphics[width = 5.8cm]{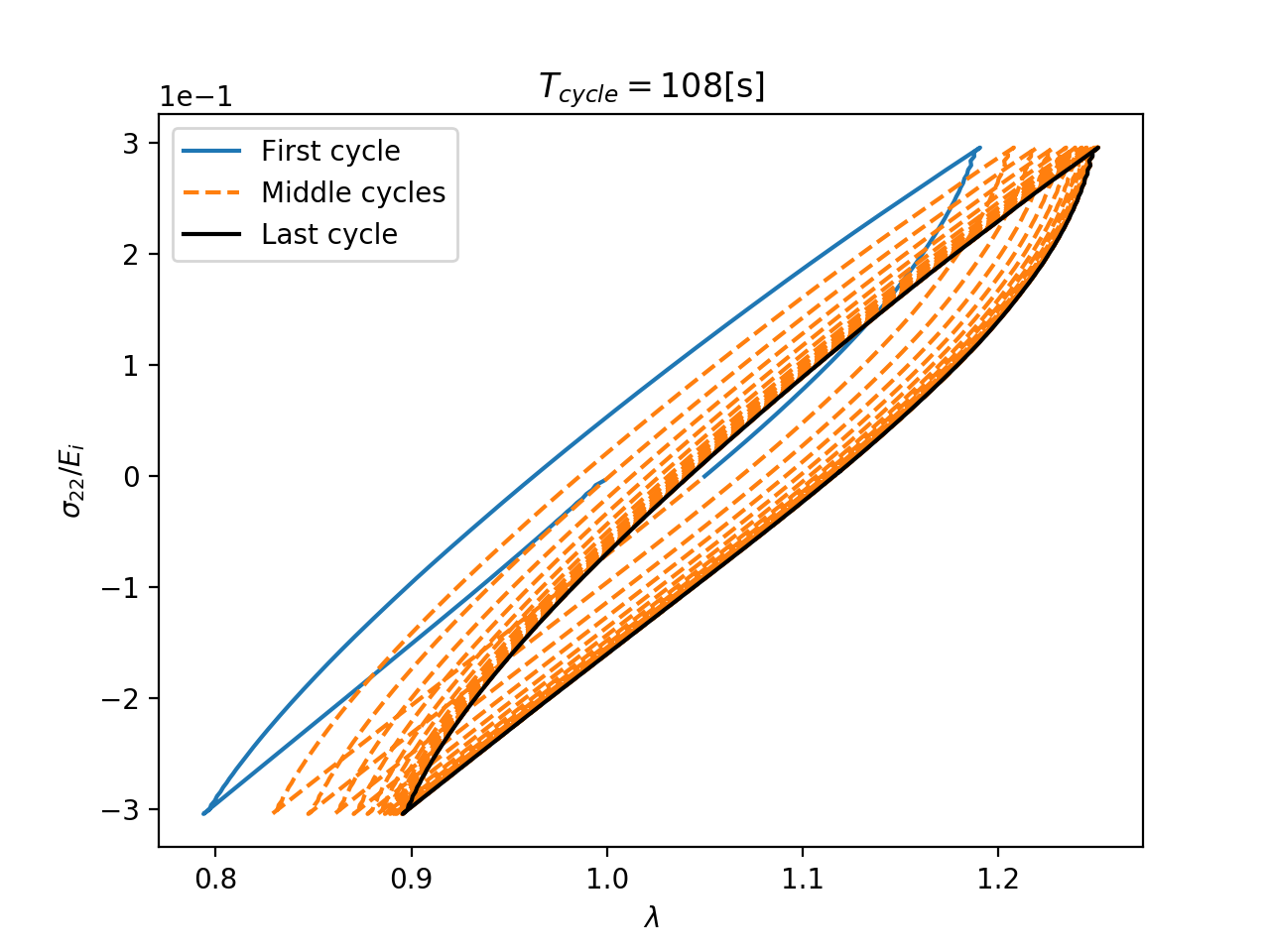}
  \caption{The model response under cyclic uniaxial test for 20 cycles simulated via the present method. The hysteresis behaviour and preconditioning effects are similar to the experimental tests carried out in \cite{FRANCESCHINI20062592}. We consider this model as the reference case.}
  \label{fig_Cyclic_NL}
  \end{subfigure}\quad
  \begin{subfigure}{5.8cm}
  \includegraphics[width = 5.8cm]{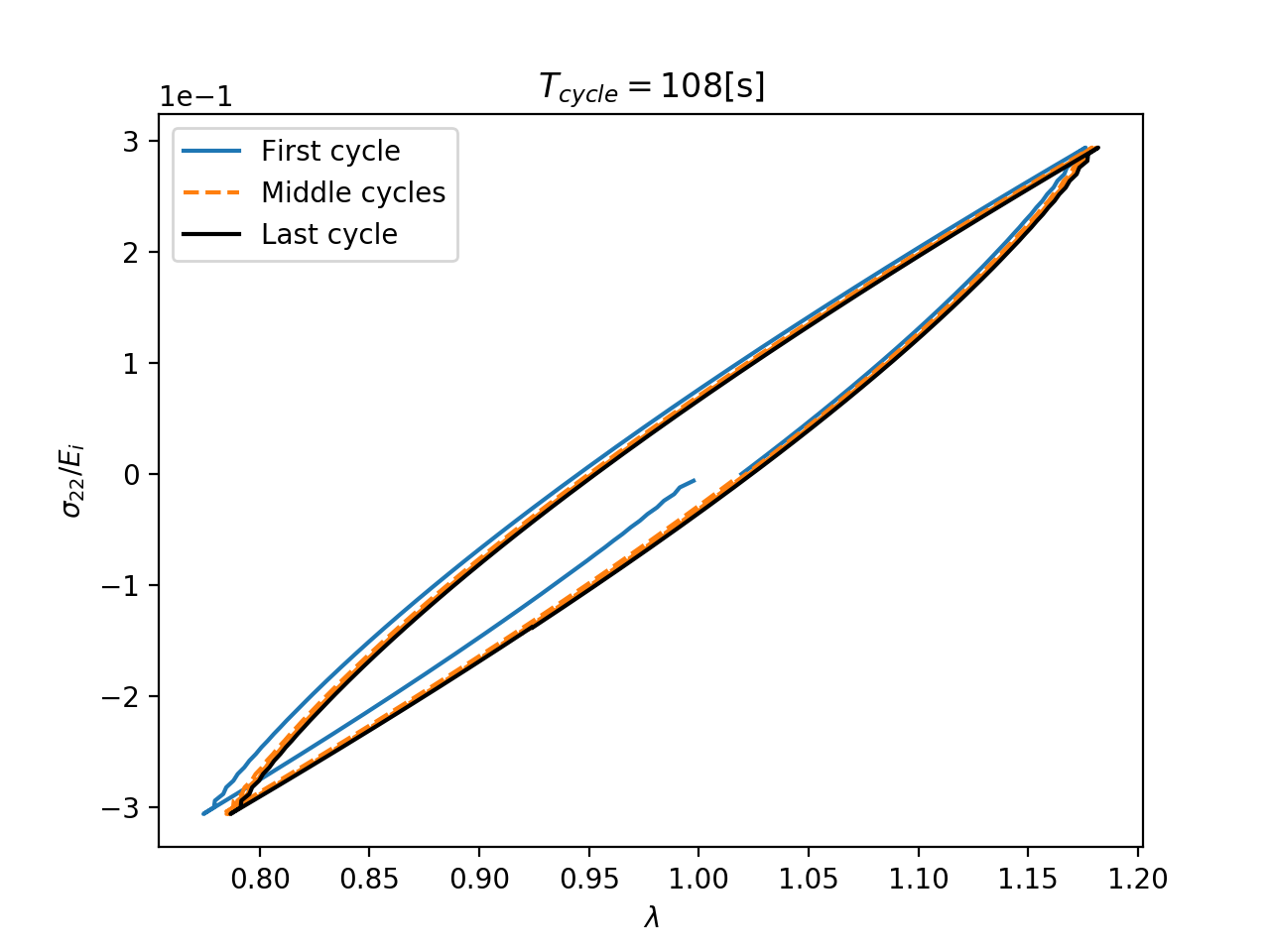}
  \caption{The mechanical response based on the classical linear Biot poroelastic case (without remodelling). Although the hysteresis behaviour is observed the preconditioning effects are negligible. Furthermore, the shape of the cycle is different from the incremental nonlinear case.}
  \label{fig_Cyclic_L}
  \end{subfigure}
  \caption{A comparison between classical Biot poroelastic model and incremental nonlinear method. The model response shows that the latter is more accurate, particularly, in case of soft poroelastic tissues. We highlight that, in the nonlinear case, the material model of the solid phase of the cell is assumed neo-Hookean with initial elastic properties adopted from the linear one.}
  \label{fig_Cyclic_NL_L}
  \end{figure*}

  \begin{figure*}
  \centering
  \begin{subfigure}{5.8cm}
  \includegraphics[width = 5.8cm]{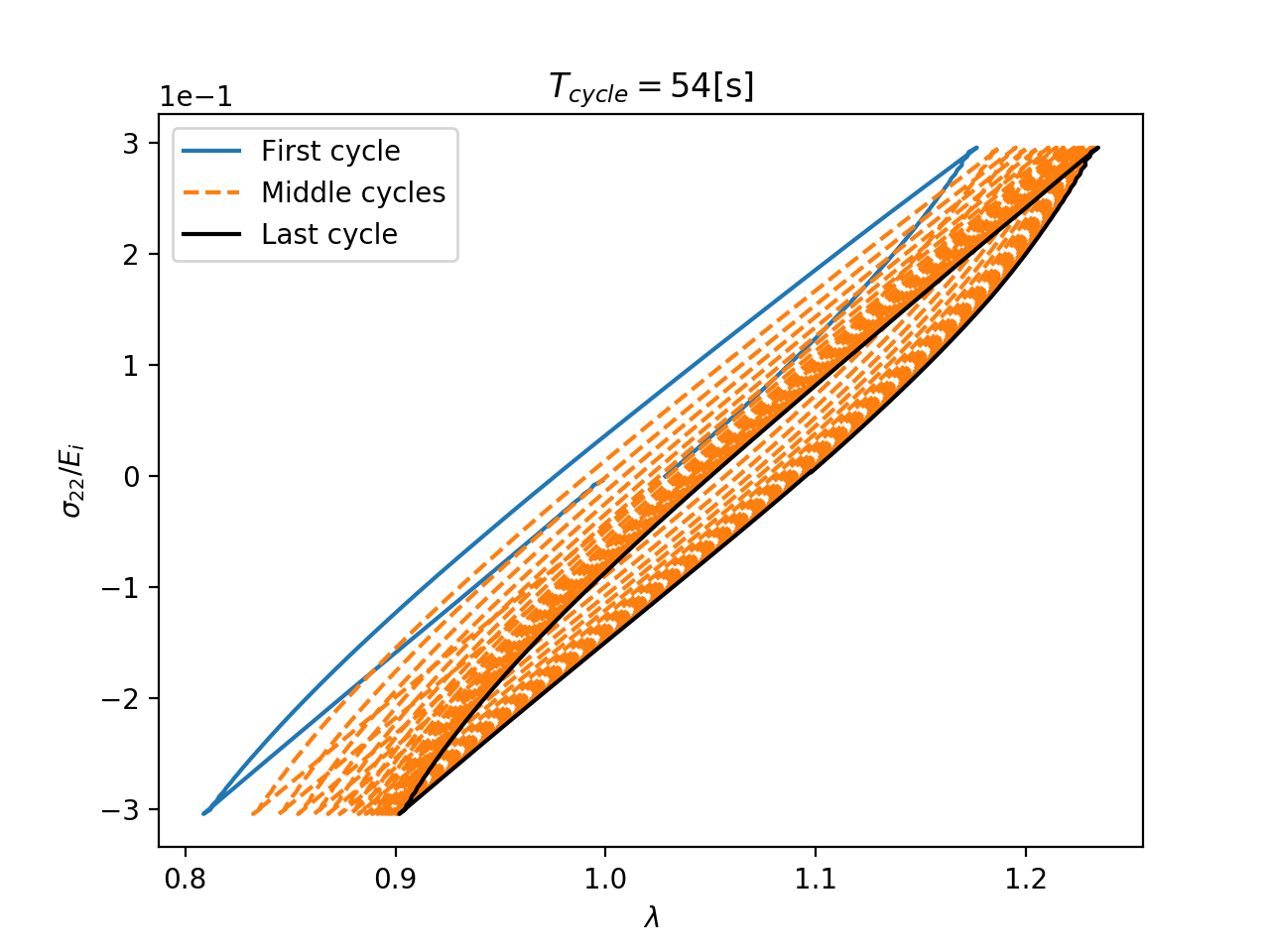}
  \caption{In case of doubled loading rate, the hysteresis area is smaller than the "reference case" (the one in Figure \ref{fig_Cyclic_NL}) while the preconditioning remains more or less similar.}
  \label{fig_CyclicT54}
  \end{subfigure}\quad
  \begin{subfigure}{5.8cm}
  \includegraphics[width = 5.8cm]{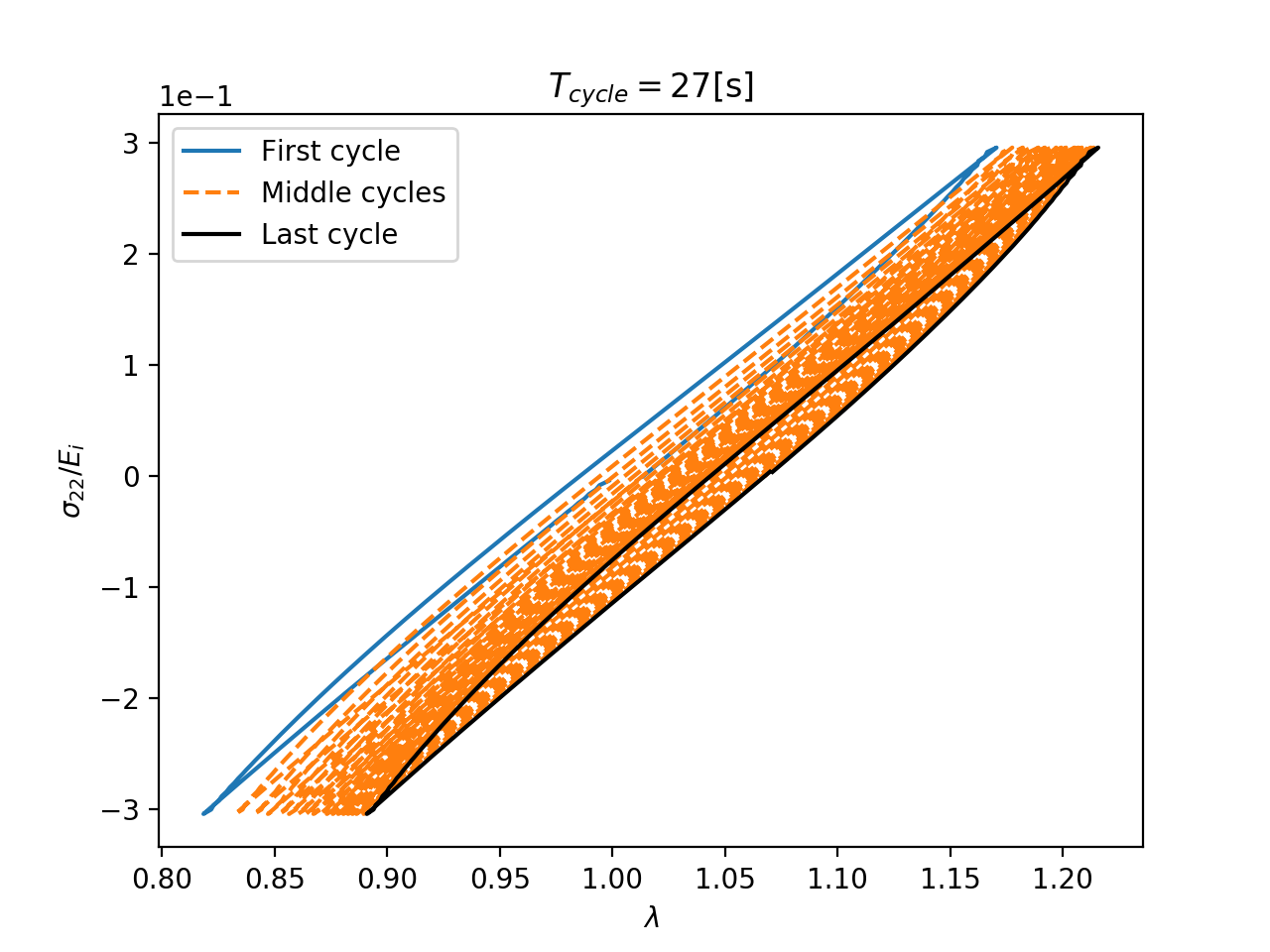}
  \caption{Decreasing the loading time to one fourth of the reference case, the preconditioning magnitude starts decreasing and the hysteresis area continues to drop.\vspace{0.38cm}}
  \label{fig_CyclicT27}
  \end{subfigure}
  
  \begin{subfigure}{5.8cm}
  \includegraphics[width = 5.8cm]{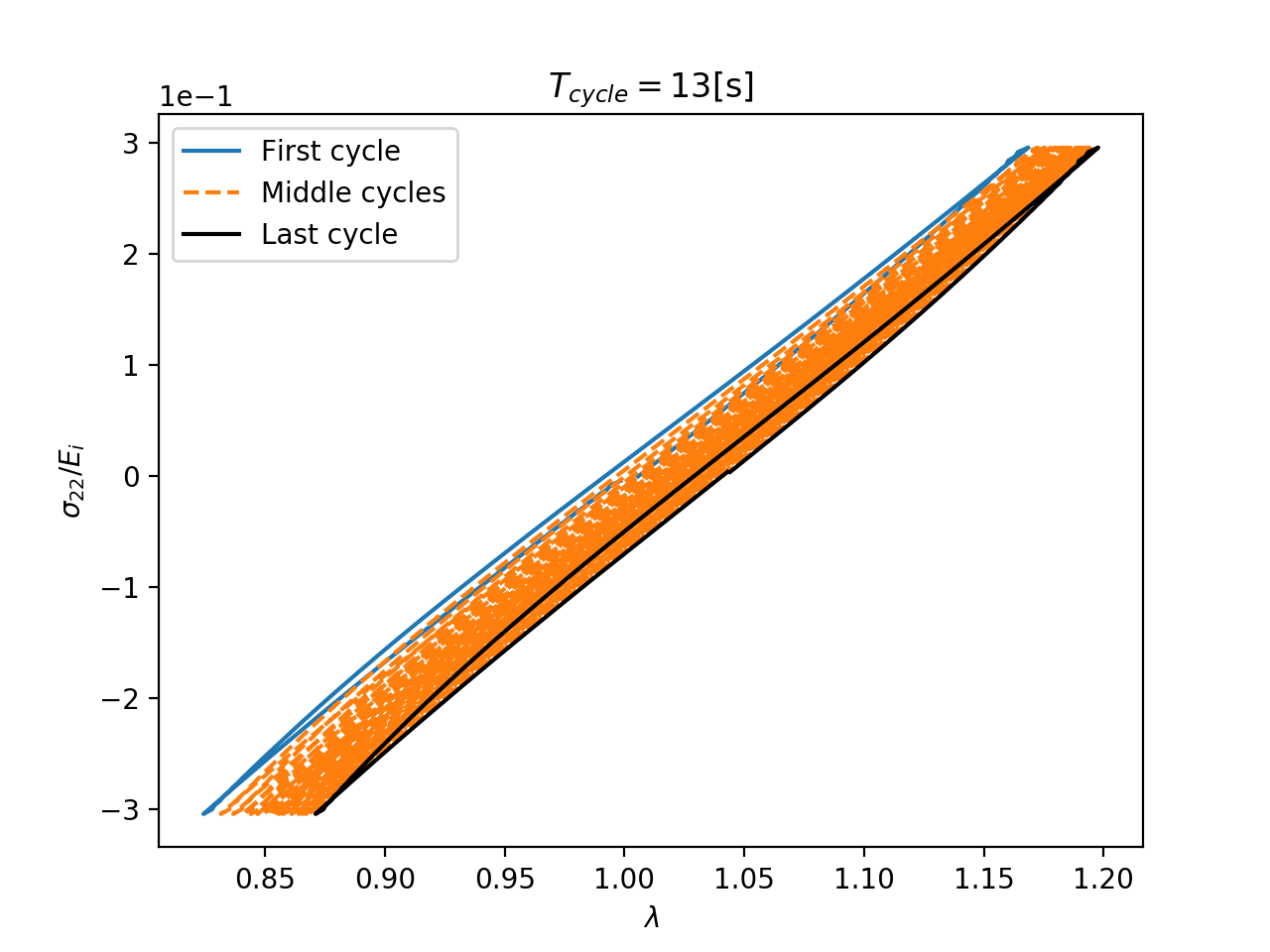}
  \caption{A slight strain-hardening appears more evidently which is expected to be a consequence of insufficient drainage time for the model in which the model approaches to a less compressible elastic type response.\vspace{0.38cm}\vspace{0.38cm}}
  \label{fig_CyclicT13}
  \end{subfigure}\quad
  \begin{subfigure}{5.8cm}
  \includegraphics[width = 5.8cm]{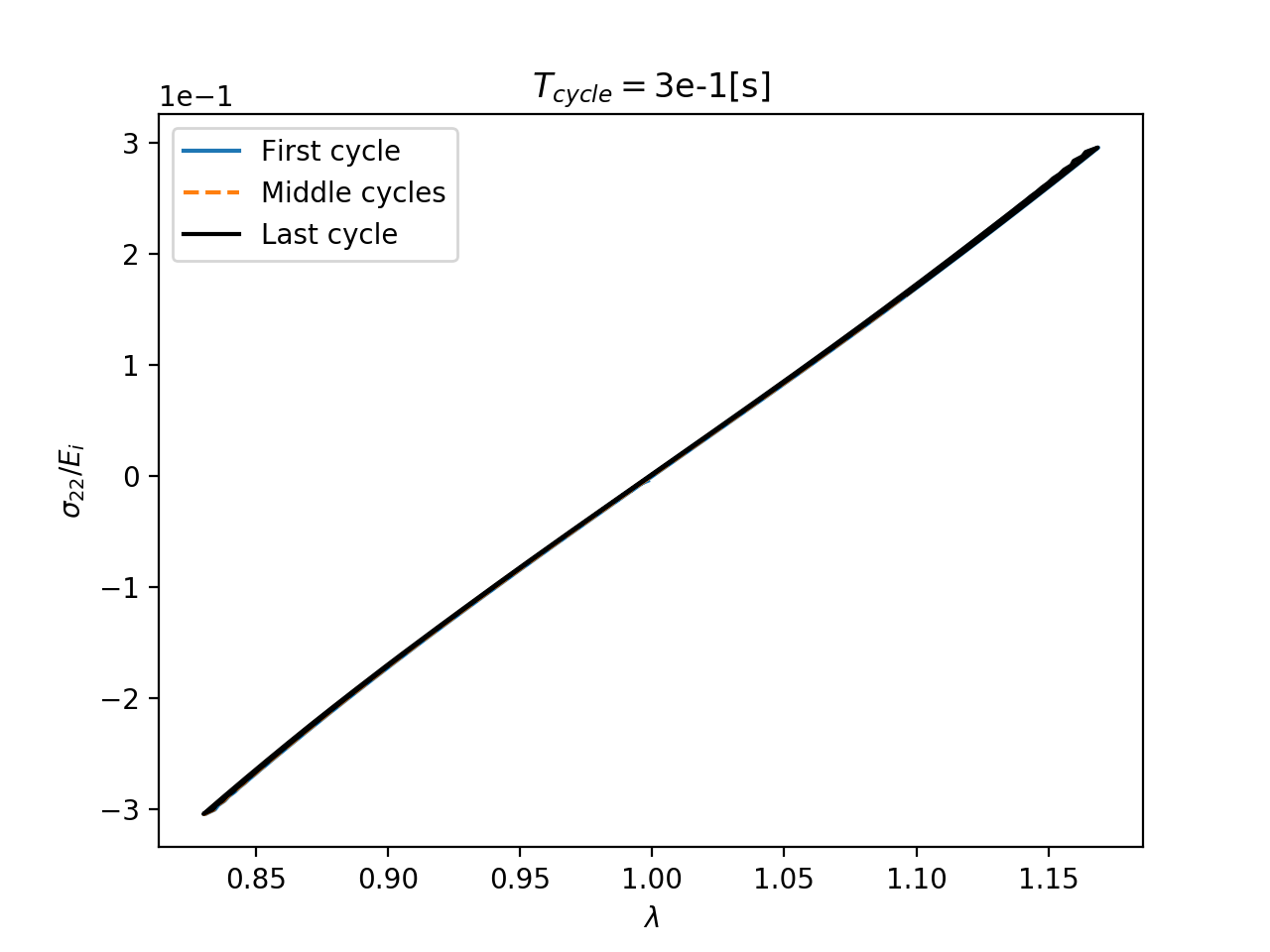}
  \caption{At almost instant loading, there is no hysteresis and preconditioning in the model response as the fluid is trapped inside the solid matrix due to insufficient time. In fact, the material properties is similar to an elastic one with undrained elasticity tensor (see Equation \eqref{Eq_undrained}.)\vspace{0.38cm}}
  \label{fig_CyclicT3}
  \end{subfigure}
  \caption{The effects of increasing loading speed on the model response showing the feasibility of the present method to overcome a source of ambiguity (analysis time) in understanding the brain tissue's mechanical behaviour.}
  \label{fig_Cyclic_Stimes}
  \end{figure*}
  
First, one should pay attention to the characteristic values of the parameters of this application. For example, due to small dimension of the tissue, it is more convenient if the macroscale average length scale is millimetres ($L = 10^{-3}[m]$) so that one can also avoid numerical errors due to very small dimensions of the elements. 
One can perform the calculations based on dimensionless parameters calculated via \eqref{nondim2} in order to avoid ambiguity. 
The dynamic viscosity of cerebrospinal fluid CSF \cite{CSFviscosity} is assumed as the characteristic one $\mu_c = 10^{-3}[Pa.s]$. Average microscopic periodic cell dimension $d = 20^{-6}[m]$ is chosen from the literature \cite{WANG20092371,KimECMsize, Baker2009} and unit force $F_c = 10^{-3}[N]$ is adopted to reach an appropriate unit pressure.
For example, characteristic time and pressure are $t_c = 1[s]$ and $P_c = 10^{3} [Pa]$, respectively,
and, assuming solid matrix initial Poisson's ratio $\nu_i = 0.3[-]$ and Young's modulus $E_i = 13.5\times 10^{3} [Pa]$, the dimensionless initial Young's modulus is $E'_i = 13.5[-]$ which is the same as the one assumed in ANN training. 

We first make use of the conventional linear poroelastic model, providing a comparison mean to understand to what extend the provided incremental nonlinear framework is important in mechanical modelling of brain tissue. Figure \ref{fig_Cyclic_L} shows that, provided appropriate parameters, even the linear poroelastic model shows the hysteresis behaviour (indicating the energy dissipation) of the tissue without introducing viscoelasticity, damage, etc. The preconditioning effect, however, is considerably smaller than the empirically captured values in \cite{FRANCESCHINI20062592}. We highlight that the cycle time $T_{cycle} =108[s]$ is chosen to introduce, more or less, the same loading conditions as in the latter study.
Employing the presented incremental method for this brain tissue deformation example, a considerable difference is observed. The mechanical response of the model shown in Figure \ref{fig_Cyclic_NL} exhibits the effect of preconditioning to a larger extent which agrees with the experimental results provided in \cite{FRANCESCHINI20062592} for brain tissue (and in \cite{MillerPreconditioning,QuinnPrecon} for fibrous connective tissues such as ligament) highlighting the importance of considering remodelling and micro-macro interaction of nonlinear poroelastic media under finite deformation. The preconditioning effects causes the cycles to move to the right hand side of the plot which means that the deformed length of the model increases after each cycle which was again observed in the latter study \cite{FRANCESCHINI20062592}.  Furthermore, the hysteresis behaviour is slightly different from the linear poroelastic one and is closer to the experimentally observed ones.

  \begin{figure*}
  \centering 
  \begin{subfigure}{5.8cm}
  \includegraphics[width = 5.8cm]{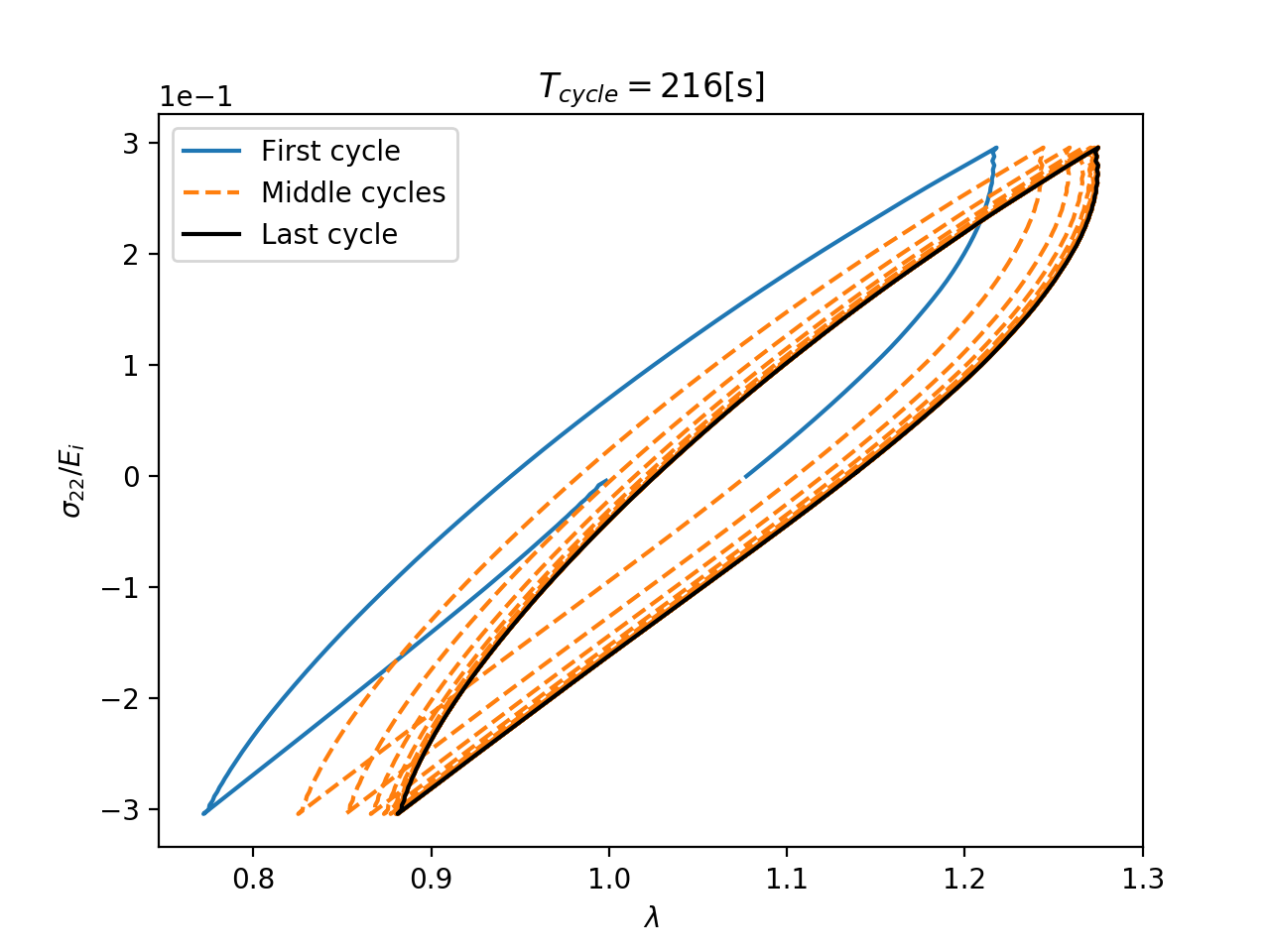}
  \caption{Extending the hysteresis area and decreasing the number of middle cycles (the cycles needed to reach the final cycle state) is observed as the first effects of a greater loading time than the reference one (Figure \ref{fig_Cyclic_NL}).\vspace{0.38cm}\vspace{0.38cm}}
  \label{fig_CyclicT216}
  \end{subfigure}\quad
   \begin{subfigure}{5.8cm}
  \includegraphics[width = 5.8cm]{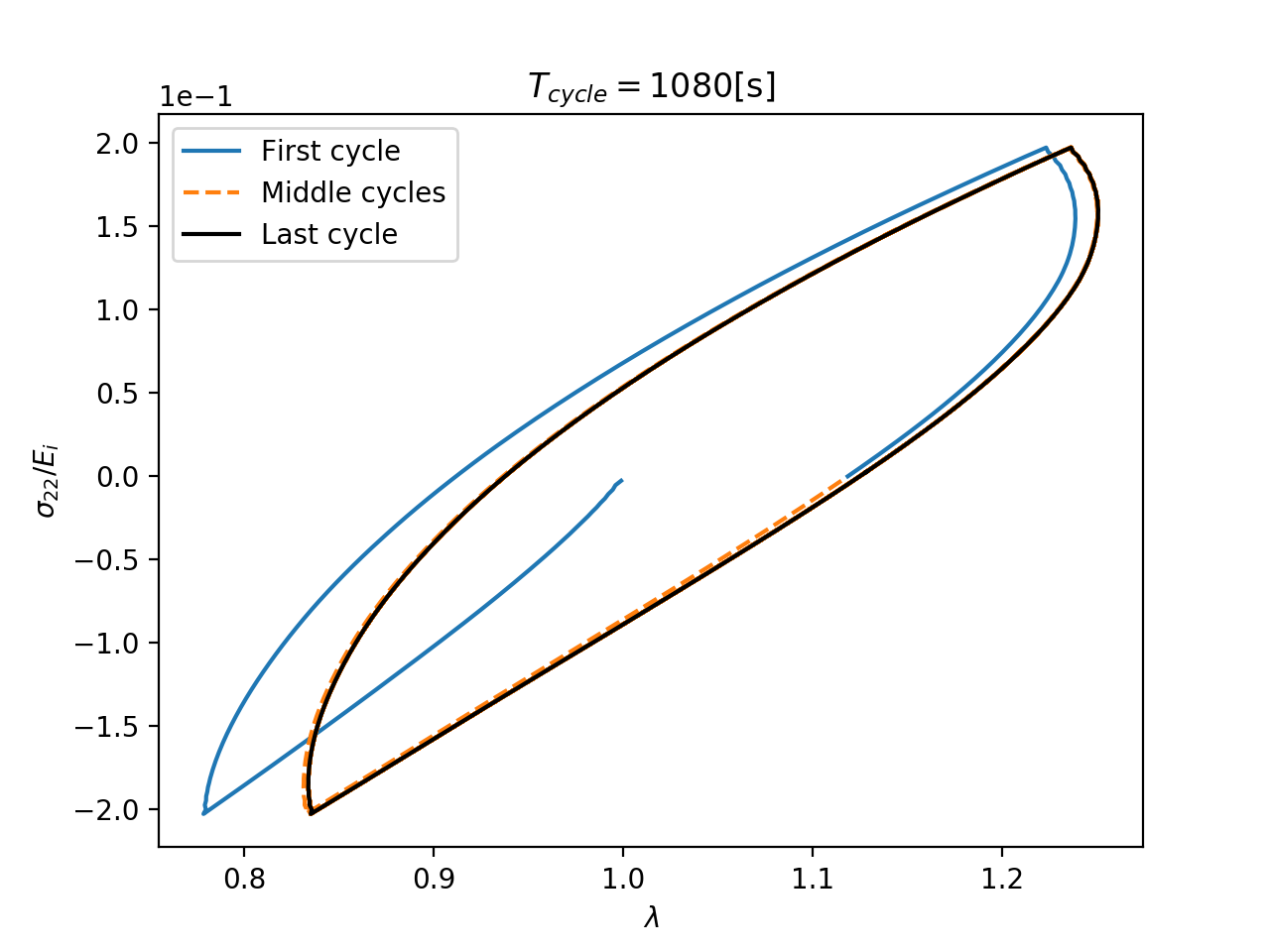}
  \caption{Increasing the cycle time to $1080[s]$ the final state is reached at the second cycle, however, the hysteresis area is considerably greater than the case in Figure \ref{fig_CyclicT216}. Furthermore, a similar stretch value to the reference case is obtained at smaller loads (see the values in vertical direction).}
  \label{fig_CyclicT1080}
  \end{subfigure}
  
   \begin{subfigure}{5.8cm}
  \includegraphics[width = 5.8cm]{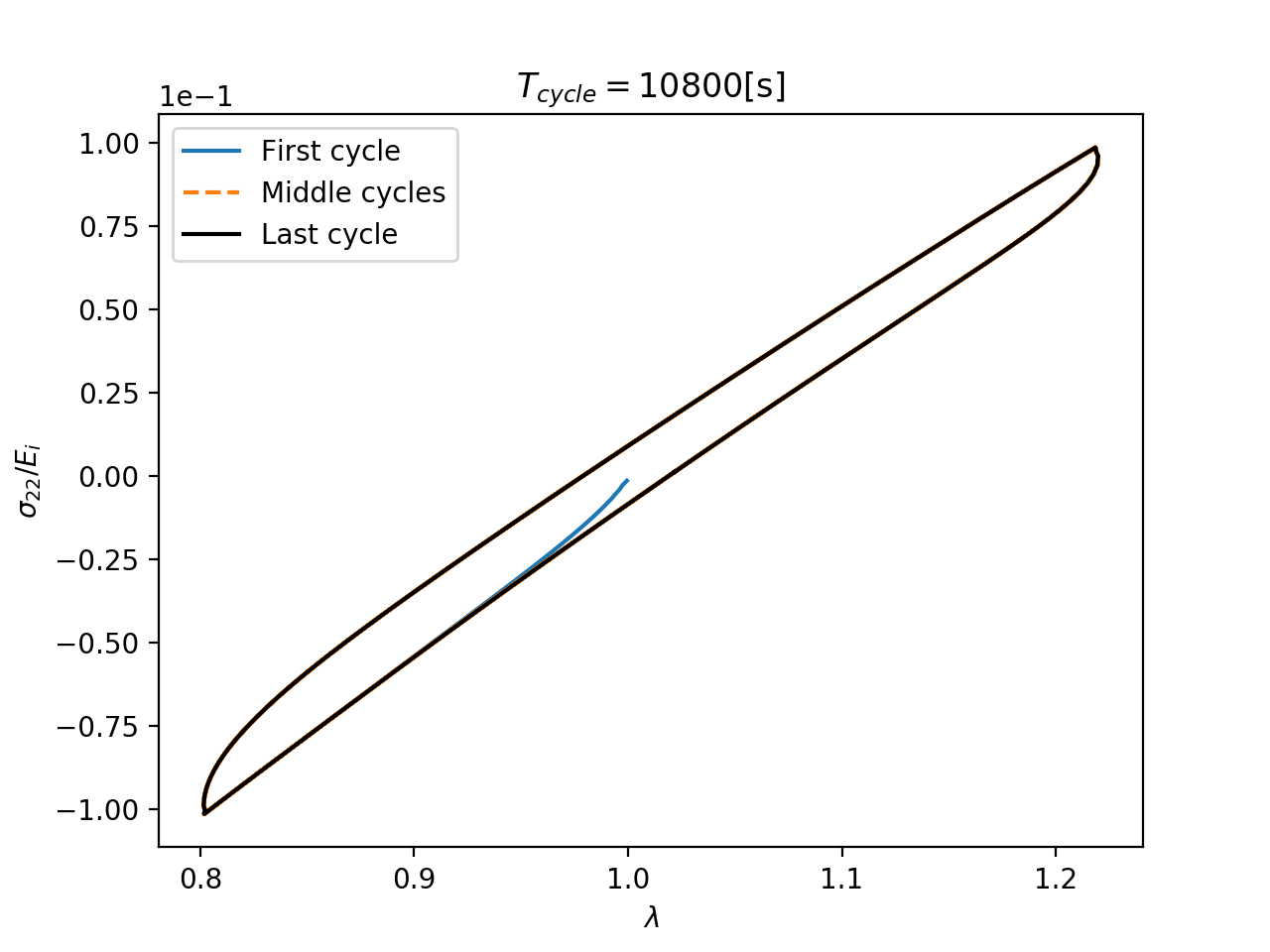}
  \caption{At $T_{cycle}=10800[s]$ the preconditioning effects is negligible and hysteresis area is considerably smaller approaching to an elastic-like mechanical response. Moreover, the model response is softer (more compressible) than the previous cases (see the values in vertical direction).\vspace{0.38cm}}
  \label{fig_CyclicT10800}
  \end{subfigure}\quad
  \begin{subfigure}{5.8cm}
  \includegraphics[width = 5.8cm]{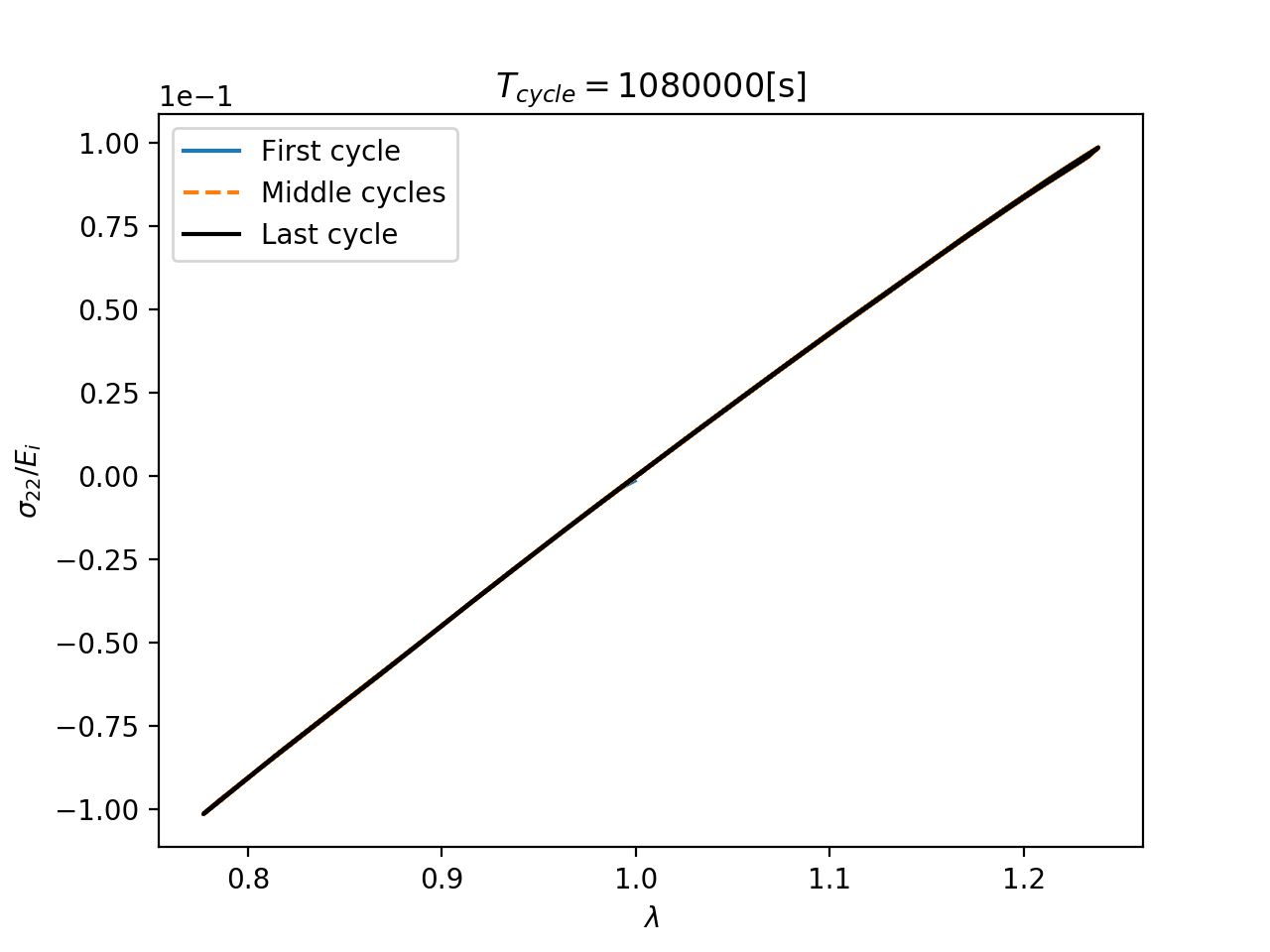}
  \caption{At infinitesimal strain rates the fluid has the sufficient time to drain without considerable interactions with the solid phase. The overall mechanical model response is elastic-like with drained elasticity tensor ($\tilde{\mathbb C}$) which could be considerably more compressible than the undrained one.}
  \label{fig_CyclicT108000}
  \end{subfigure}
  \caption{The effects of increasing $T_{cycle}$. We highlight that all other properties are the same as the reference case shown in Figure \ref{fig_Cyclic_NL}.}
  \label{fig_Cyclic_times}
  \end{figure*}

 After the first cycle, there is a considerable \textit{cycle residual strain} (the gap between the beginning of the loading and the end of the unloading path in one cycle) which is related to the fluid absorption due to the model deformation. As the fluid flow inside the medium requires time, we expect that increasing the loading rate the cycle residual strain and consequently, the preconditioning effects decreases and the medium responds similarly to an elastic one. Figure \ref{fig_Cyclic_Stimes} highlights that not only the preconditioning effects but also the hysteresis behaviour are due to fluid flow and its interaction with the solid phase. Furthermore, it is shown that by increasing the loading rate, the minimum stretch decreases, which means a decrease in the medium compressibility. In this case, the incompressible interstitial fluid does not percolate through the porous medium due to the insufficient time in which case static fluid phase filling the pores could be assumed (undrained case). Considering Equations \eqref{solidCons} and \eqref{eqMassC} the model deformation follows the undrained elasticity tensor (for more details see \cite{Hdehghani, growing, burrigekeller})
 \begin{equation}
 \tilde{\mathbb C} + M \tilde{\alpha} \tilde{\alpha}, \label{Eq_undrained}
 \end{equation}
which expected to be far less compressible than the drained medium \cite{Hdehghani}.

  On the other hand, decreasing the strain rate, the given time for a small deformation increases which, in turn, decreases the resultant pore pressure driven by the solid deformation (see Equation \eqref{eqMassC}) approaching the overall material response to the undrained case. The latter is expected to be considerably more compressible and softer as without Equation \eqref{eqMassC} the pores could be considered as void spaces.
Figure \ref{fig_Cyclic_times} shows that increasing the cycle time, the dissipated energy due to the hysteresis response as well as the preconditioning effects and residual strains increases up to a maximum point (see Figures \ref{fig_CyclicT216} and \ref{fig_CyclicT1080})  from which they decrease approaching zero at very long times shown in Figures \ref{fig_CyclicT10800} and \ref{fig_CyclicT108000}. The latter shows the undrained response of the porous media under cyclic loading which is, compared with Figure \ref{fig_CyclicT3}, considerably softer and more compressible (note the difference in the applied load).
  
\begin{figure*}
  \centering 
   \begin{subfigure}{5.8cm}
  \includegraphics[width = 5.8cm]{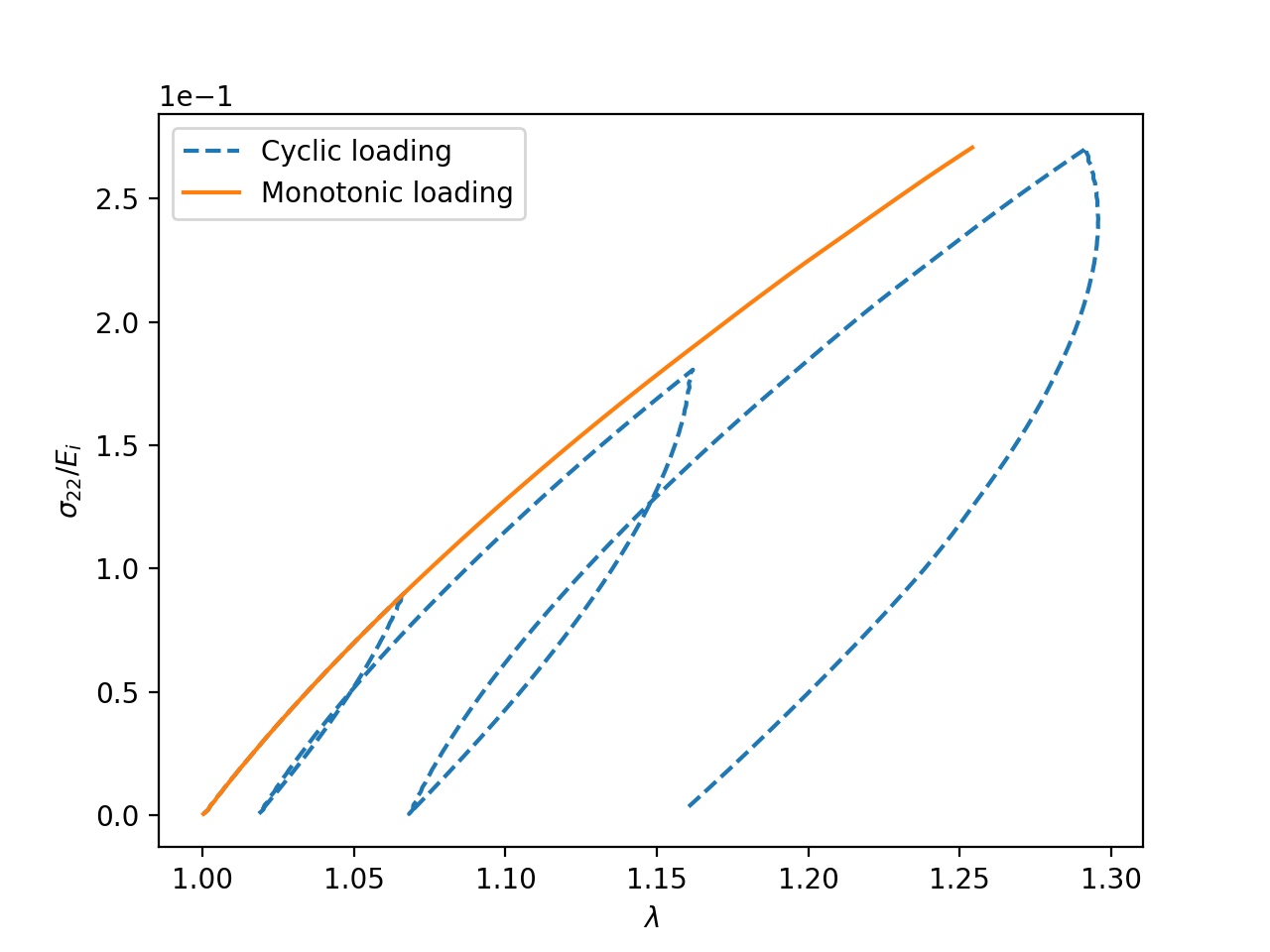}
  \caption{Comparison between monotonic and cyclic loading in a tensile test.}
  \label{fig_Mullins_tensile}
  \end{subfigure}\quad
  \begin{subfigure}{5.8cm}
  \includegraphics[width = 5.8cm]{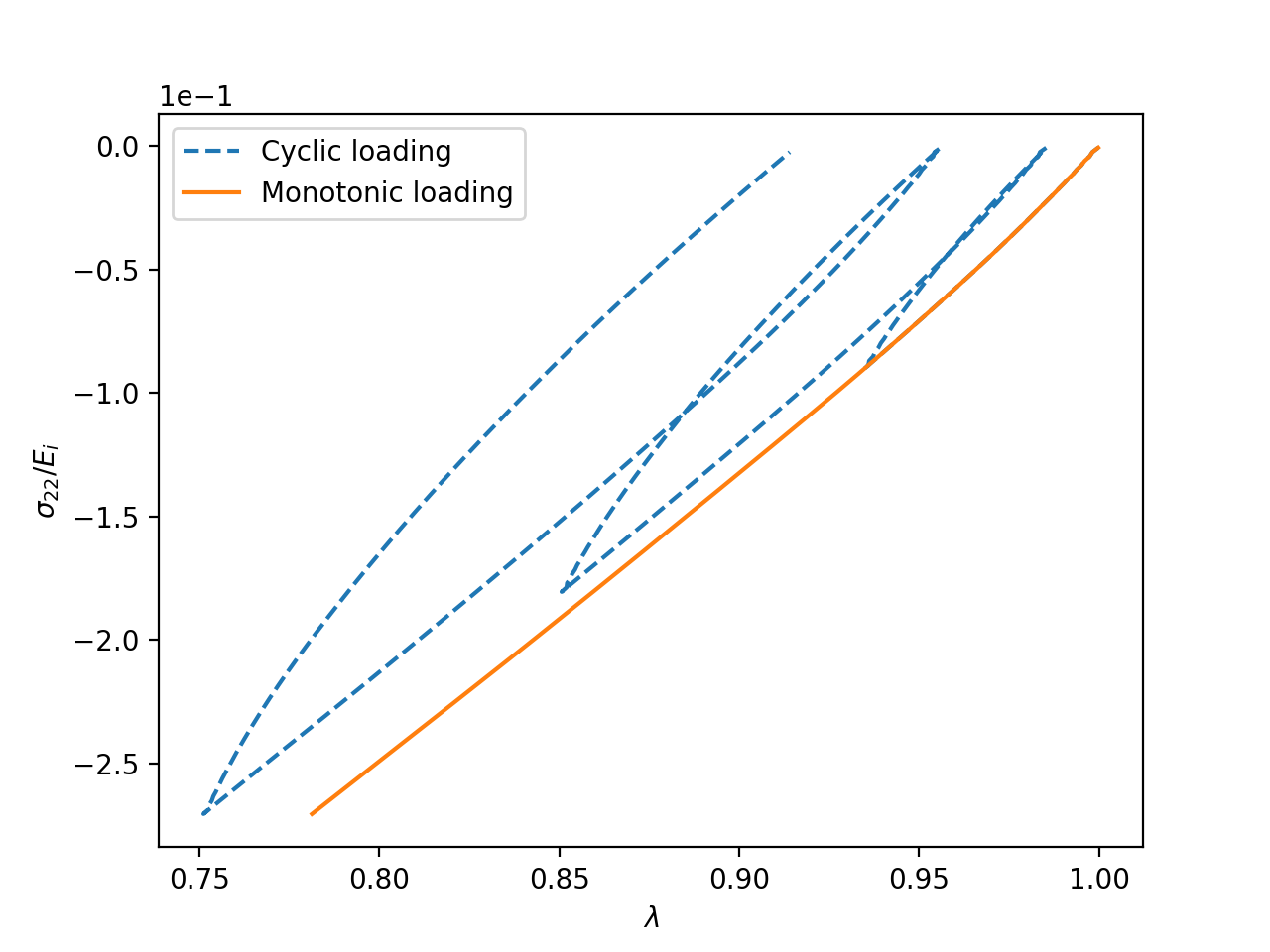}
  \caption{Comparison between monotonic and cyclic loading in a compressive test.}
  \label{fig_Mullins_compressive}
  \end{subfigure}
  \caption{The mechanical response of the poroelastic models with the same properties as in the reference case shows Mullins-like effects with preconditioning (residual strain).}
  \label{fig_Mullins}
  \end{figure*}

In the experiments carried out in \cite{FRANCESCHINI20062592} a behaviour similar to what is called Mullins effect \cite{Mullins} together with residual strain is observed. In \cite{FRANCESCHINI20062592} this behaviour is modelled by means of the pseudo-elastic model provided in \cite{ogden_pseudoelastic} neglecting the solid-fluid interaction, multiscale nature of the problem, and the observed residual strain. Figure \ref{fig_Mullins} shows that the present modelling methodology results in a macroscopic mechanical response similar to what is observed in the experiments, however, based on simple compressible neo-Hookean material employed in the present multiscale and multiphysics methodology. In this figure, the model deformation under both cyclic and monotonic loading is provided so that the dependence of the model response on the previous cycles or the deformation history is clear. As there is no term in the material model accounting for the Mullins effect, we conclude that this type of response is due to the nonlinear poroelastic nature of the medium. Furthermore, due to the chosen material model which imposes strain-softening in tensile deformation and strain-hardening is compressive one, as expected, a greater residual strain is captured under tensile loading (shown in Figure \ref{fig_Mullins_tensile}) compared with the compression test (shown in \ref{fig_Mullins_compressive}). 

The authors highlight that, although several key features and aspects of the brain tissue response to the external loads are addressed, the accurate modelling and simulation of this complex tissue requires further study and effort focused, for example, on the solid phase material model as well as the cell geometry properties.

\section{Conclusions and future work} \label{Sec_conclusions}
In this study, ANNs, a standard multiscale approach (asymptotic or two-scale homogenisation), and an incremental FE analysis framework are integrated to provide a novel and robust computational method for analysis of multiphysics and multiscale poroelastic problems under global deformations that are large enough to invalidate the assumptions of infinitesimal strain theory. This framework considers the full interdependency between the homogenised hydraulic and mechanical response and microscopic (and, consequently, macroscopic) properties rearrangement (multiscale remodelling) which encompasses different sources of nonlinearity, namely, solid-fluid interaction as well as the nonlinear governing equations (due to updating effective coefficients). The whole procedure is based on non-dimesionalised variables and properties so that it is applicable in a wide range of problems in different scales such as soil and biomechanics.

Applied into the soil mechanics (in the sense of the characteristic values), Terzaghi's consolidation and Dar- cy's fluid filtration tests are reconstructed numerically under finite deformation providing detailed information on the combined effects of the solid deformation and pore pressure on the microscopic and effective properties of the medium, as well as its overall response. Agreeing with the experimental results in the literature, for example, the fluid filtration example shows that, employing the presented method, there is no need to the corrections to Darcy's law for fluid flow in poroelastic media. Furthermore, the average cell dimension (grain size) and initial Poisson's ratio nonlinearly increase and decrease the deviation from Darcy's law, respectively. 
In Terzaghi's test, the distribution (in space and time) of the mechanical and hydraulic response, as well as microscopic and effective properties are monitored until the steady-state is reached. In general, we observe spatially and time-dependent distribution of the properties and variables highlighting the complexity of the problem.  Among many other results, we show that, in this problem, the porosity of the medium under mechanical pressure increases instantly after loading (due to the resultant pore pressure) while it decreases during the transient state, as a consequence of the consolidation and fluid drainage, reaching a constant value smaller than the initial one at the steady-state.
Moreover, the captured dependency of the effective properties on the changes in microscopic properties (porosity and solid matrix material properties) due to the solid matrix contraction/expansion and its effects on the overall response of the media highlights the importance of employing the presented method.

Furthermore, we simulate a uniaxial cyclic test on the brain tissue as a poroelastic medium (which is known for its complex and peculiar mechanical response). The test is characterised to mimic the conditions of the experimental tests so that we can evaluate the results. The hysteresis and preconditioning effects together with a behaviour similar to the observed response in Mullins tests are captured for the first time only based on the poroelastic nature of the tissue. The effects of the loading/deformation rate is thoroughly studied as an important source of ambiguity in brain tissue properties identification. Although the numerical results agree with the experimental ones in several aspects, more effort is required for accurate modelling of brain tissue which can be the subject of a future study.

Apart from the development of the mentioned method, several aspects of fundamental issues in the context of poroelasticity has been addressed for the first time in this study. However, there are more challenges to be overcome in the future studies which can be divided into three directions. Firstly, an application-specific and robust nonlinear material law for the solid matrix, able to accurately describe the complex poroelastic media such as brain tissue, is to be chosen/discovered. This, in turn, requires a robust and specific framework for microscopic model parameter identification using AI (similar to poroelastography but considering solid matrix properties). Secondly, the future works could be oriented to remove some simplifying assumptions such as pore section and solid matrix material, respectively, remaining circular and isotropic under the deformation during the analysis etc. 
Last but not least, based on the observations using the presented computational framework, phenomenological homogenised material models can be adopted/introduced considering the mentioned interdependencies and interactions between the micro and macro properties and responses. In other words, as soft poroelastic tissues may not be easily available, one can partially replace the experiments with the numerical simulations using the present method to gain a deep insight of the required elements in the material models for poroelastic media.

\section*{Acknowledgements}
We acknowledge the support of this research work via the framework of DTU DRIVEN, funded by the Luxembourg National Research Fund (PRIDE17/12252781), and the project CDE-HUB, funded by the Luxembourg Ministry of Economy (FEDER 2018-04-024).

The authors would like to thank Prof. Davide Bigoni for providing additional information about experimental test data published in \cite{FRANCESCHINI20062592}.

\appendix
\section{Upscaling} \label{Homo_app}
In this Appendix, for the readers convenience and as it is the basis of the present work, we mention the asymptotic homogenisation procedure (for more details on the specific problem see \cite{HdehghaniThesis} and the references therein).
Applying the transformation \eqref{eq:gradnew} and \eqref{eq:xiTransform} into Equations \eqref{eq:solid}-\eqref{eq:tangente} yields the multiscale equations
\begin{align}
\nabla_{\vett{y}} \cdot \tens{\tau_{\epsilon}}+\epsilon \nabla_{\vett{x}} \cdot \tens{\tau_{\epsilon}} &=0 &\textrm{in}\,\,\Omega_s\label{eq:hstruttura}\\
\frac{1}{\epsilon}\mathbb{C}\xi_{\vett{y}}(\vett{u}_{\epsilon})+ \mathbb{C}\xi_{\vett{x}}(\vett{u}_{\epsilon})&=\tens{\tau_{\epsilon}}&\textrm{in}\,\,\Omega_s\label{eq:hsolidstress}\\ \nonumber\\
\nabla_{\vett{y}} \cdot \tens{\sigma_{\epsilon}}+\epsilon\nabla_{\vett{x}} \cdot \tens{\sigma_{\epsilon}}&=0 &\textrm{in}\,\,\Omega_f \label{eq:hfluido}\\
p_{\epsilon}\tens{I}-\epsilon2\mu \left(\xi_{\vett{y}}( \vett{v}_{\epsilon})\right)-
\epsilon^22\mu \left(\xi_{\vett{x}}( \vett{v}_{\epsilon})\right)&=\tens{\sigma_{\epsilon}} &\textrm{in}\,\,\Omega_f\label{eq:hfluidstress}\\
\nabla_{\vett{y}} \cdot \vett{v}_{\epsilon}+\epsilon\nabla_{\vett{x}} \cdot \vett{v}_{\epsilon}&=0 &\textrm{in}\,\,\Omega_f\label{eq:hinco}\\ \nonumber\\
\tens{\tau_{\epsilon}}\vett{n}&=\tens{\sigma_{\epsilon}}\vett{n} &\textrm{on}\,\,\Gamma\,\,\label{eq:hstress}\\
\dot{\vett{u}}_{\epsilon}&=\vett{v}_{\epsilon} &\textrm{on}\,\,\Gamma\,\,\label{eq:htangente}\\
\vett{u}^s_{\epsilon}&=\vett{u}^f_{\epsilon} &\textrm{on}\,\,\Gamma\,\, \label{eq_dispinterface}
\end{align}
where the subscripts $\epsilon$ indicate the representation in the power series form.

Next, every field, namely, solid and fluid stress, displacement, velocity, and pressure are replaced by their power of series representation, $\psi_{\epsilon}(\vett{x},\vett{y})=\sum_{l=0}^{\infty}\psi^{(l)}(\vett{x},\vett{y})\epsilon^ l$.
Coefficients of $\epsilon^0$ yield
\begin{align}
\nabla_{\vett{y}} \cdot \tens{\tau}^{(0)}&=0 &\textrm{in}\,\,\Omega_s\label{eq:zstruttura}\\
\mathbb{C}\xi_{\vett{y}} (\vett{u}^{(0)})&=0 &\textrm{in}\,\,\Omega_s\label{eq:zsolidstress}\\ \nonumber \\
\nabla_{\vett{y}} \cdot \tens{\sigma}^{(0)}&=0 &\textrm{in}\,\,\Omega_f \label{eq:zfluido}\\
\tens{\sigma}^{(0)}&=-p^{(0)}\tens{I} &\textrm{in}\,\,\Omega_f\label{eq:zfluidstress}\\
\nabla_{\vett{y}} \cdot \vett{v}^{(0)}&=0 &\textrm{in}\,\,\Omega_f\label{eq:zinco}\\ \nonumber \\
\tens{\tau}^{(0)}\vett{n}&=\tens{\sigma}^{(0)}\vett{n} &\textrm{on}\,\,\Gamma\,\,\label{eq:zstress}\\
\dot{\vett{u}}^{(0)}&=\vett{v}^{(0)} &\textrm{on}\,\,\Gamma\,\,\label{eq:ztangente}
\end{align}
while in the case of $\epsilon^1$ it reads
\begin{align}
\nabla_{\vett{y}} \cdot \tens{\tau}^{(1)}+\nabla_{\vett{x}} \cdot \tens{\tau}^{(0)} &=0 &\textrm{in}\,\,\Omega_s\label{eq:ustruttura}\\
\mathbb{C}\left(\xi_{\vett{y}} (\vett{u}^{(1)})+ \xi_{\vett{x}}(\vett{u}^{(0)})\right)&=\tens{\tau}^{(0)} &\textrm{in}\,\,\Omega_s \label{eq:usolidstress} \\ \nonumber \\
\nabla_{\vett{y}} \cdot \tens{\sigma}^{(1)}+\nabla_{\vett{x}} \cdot \tens{\sigma}^{(0)}&=0 &\textrm{in}\,\,\Omega_f \label{eq:ufluido}\\
-p^{(1)}\tens{I}+ \left(2\mu \xi_{\vett{y}}(\vett{v}^{(0)})\right)&=\tens{\sigma}^{(1)} &\textrm{in}\,\,\Omega_f\label{eq:ufluidstress}\\
\nabla_{\vett{y}} \cdot \vett{v}^{(1)}+\nabla_{\vett{x}} \cdot \vett{v}^{(0)}&=0 &\textrm{in}\,\,\Omega_f\label{eq:uinco}\\ \nonumber \\
\tens{\tau}^{(1)}\vett{n}&=\tens{\sigma}^{(1)}\vett{n} &\textrm{on}\,\,\Gamma\,\,\label{eq:ustress}\\
\dot{\vett{u}}^{(1)}&=\vett{v}^{(1)} &\textrm{on}\,\,\Gamma\,\,\label{eq:utangente}
\end{align}
We notice that, according to Equations \eqref{eq:zsolidstress}-\eqref{eq:zfluidstress}, the leading order solid displacement $\vett{u}^{(0)}$ and hydrostatic pressure $p^{(0)}$ are microscale independent (locally constant) and could be referred to as macroscale variables. 

We note that, substituting Equation \eqref{eq:zfluidstress} into Equation \eqref{eq:ufluido}, together with Equation \eqref{eq:ufluidstress} and, Considering the relative fluid velocity as in Equation \eqref{eq_vfr}, with the boundary condition \eqref{eq:ztangente} we reach a Stokes-type boundary value problem which using the Ansatz \eqref{eq:vzero} and \eqref{eq:puno} results in the following fluid auxiliary cell problem.
\begin{align}
\nabla^2_{\vett{y}}\trasp{\boldsymbol{\tens{W}} }-\nabla_{\vett{y}}\vett{P}+\boldsymbol{\tens{I}}&=\boldsymbol 0 &\textrm{in}\,\,\Omega_f\label{eq:stokescellw}\\
\nabla_{\vett{y}}\cdot \trasp{\boldsymbol{\tens{W}} }&=\boldsymbol 0 &\textrm{in}\,\,\Omega_f\label{eq:incellw}\\
\boldsymbol{\tens{W}} &=\boldsymbol 0 &\textrm{on}\,\,\Gamma\,\,\,\label{eq:bccellw}\\
\left\langle \vett{P} \right\rangle_f&=\vett{0} \label{equnique}.
\end{align}
Furthermore, the Equations \eqref{eq:zstruttura}, \eqref{eq:zstress}, and \eqref{eq:usolidstress} and the solid Ansatz \eqref{ansatzsolid} we can construct the following two auxiliary cell problems to be solved in the solid domain which results in 
the fourth rank tensor $\mathbb{M}$ and the second rank tensor $\boldsymbol{\tens{Q}} $, respectively.
\begin{align}
\nabla_{\vett{y}} \cdot (\mathbb{C} \xi_{\vett{y}} (\mathcal{A}))&=\boldsymbol 0  &\textrm{in}\,\,\ \Omega_s \label{6solid21} \\
(\mathbb{C} \xi_{\vett{y}} (\mathcal{A})) \vett{n} + \mathbb{C} \vett{n}&=\boldsymbol 0 &\textrm{on}\,\,\, \Gamma\,\,\, \label{6solid221}\\
\left\langle \mathcal{A}\right\rangle_s&=\boldsymbol 0,  \label{eq:unique6solid21}
\end{align}
\begin{align}
\nabla_{\vett{y}}\cdot\left(\mathbb{C}\xi_{\vett{y}} (\vett{a})\right)&=\boldsymbol 0 &\textrm{in}\,\,\Omega_s \label{eq:unoinside21}\\
\quad (\mathbb{C}\xi_{\vett{y}} (\vett{a}))\vett{n}+\vett{n}&=\boldsymbol 0 &\textrm{on}\,\,\Gamma\,\, \label{eq:unoutside21}\\
\left\langle \vett{a}  \right\rangle_s &=\boldsymbol 0 \label{7thBC},
\end{align}

Where, $\Omega_s$, $\Omega_f$, $\Gamma$, $\mathcal{A}$, and $\vett a$ representing the solid and fluid domains, their interface, a third rank tensor and a vector, respectively. $\vett{n}$ is the inward unit vector normal to the solid-fluid interface $\Gamma$, and
\begin{equation}
\label{eq:mdef21}
\mathbb{M}\coloneqq \xi_{\vett{y}} (\mathcal{A}), \quad\boldsymbol{\tens{Q}} \coloneqq \xi_{\vett{y}} (\vett{a}),
\end{equation}

$\vett{P}$ is an auxiliary vector that encodes microscale information by relating the first order hydrostatic pressure to $\nabla_{\vett{x}} p^{(0)}$ ($p^{(1)}=-\vett{P}\cdot\nabla_{\vett{x}}p^{(0)}$), and $\boldsymbol{\tens{I}}$ is the second rank identity tensor.

\bibliography{Dbib.bib}{}
\bibliographystyle{unsrt} 

\end{document}